\documentclass[aps,prd,amsmath,amssymb,eqsecnum,nofootinbib,notitlepage]{revtex4-1}

\usepackage[retainorgcmds]{IEEEtrantools}

\usepackage{graphicx}
\usepackage{graphics}
\usepackage{color}

\allowdisplaybreaks[1]

\newcommand{\ab}{{\bar{a}}}
\newcommand{\bb}{{\bar{b}}}
\newcommand{\cb}{{\bar{c}}}
\newcommand{\db}{{\bar{d}}}
\newcommand{\eb}{{\bar{e}}}
\newcommand{\fb}{{\bar{f}}}
\newcommand{\hb}{{\bar{h}}}

\newcommand{\xb}{{\bar{x}}}

\newcommand{\xp}{{x'}}

\newcommand{\rb}{r_0}
\newcommand{\rbdot}{\dot{r}_0}

\newcommand{\alphab}{{\bar{a}}}
\newcommand{\betab}{{\bar{b}}}

\newcommand{\Dtau}[1]{\Delta\tau_{(#1)}}

\newcommand{\rbar}{{\bar{r}}}
\newcommand{\sbar}{{\bar{s}}}
\newcommand{\zrho}{{\rho}}
\newcommand{\zrhoz}{{\rho_0}}
\newcommand{\num}{{n}}

\newcommand{\adv}{{(adv)}}
\newcommand{\ret}{{(ret)}}
\newcommand{\sing}{{(S)}}
\newcommand{\reg}{{(R)}}

\newcommand{\PhiS}{\Phi^{\rm \sing}}

\newcommand{\lpow}[1]{[#1]}
\newcommand{\lnpow}[1]{[\scalebox{0.85}{-}#1]}

\DeclareMathOperator{\sgn}{sgn}

\begin{document}
\title{High-order expansions of the Detweiler-Whiting singular field\\ in Schwarzschild spacetime}
\author{Anna Heffernan}
\email{anna.heffernan@ucd.ie}
\affiliation{School of Mathematical Sciences and Complex \& Adaptive Systems Laboratory, University College Dublin, Belfield, Dublin 4, Ireland}

\author{Adrian Ottewill}
\email{adrian.ottewill@ucd.ie}
\affiliation{School of Mathematical Sciences and Complex \& Adaptive Systems Laboratory, University College Dublin, Belfield, Dublin 4, Ireland}

\author{Barry Wardell}
\email{barry.wardell@gmail.com}
\affiliation{School of Mathematical Sciences and Complex \& Adaptive Systems Laboratory, University College Dublin, Belfield, Dublin 4, Ireland}
\affiliation{Max-Planck-Institut f\"{u}r Gravitationphysik, Albert-Einstein-Institut, 14476 Potsdam, Germany}

\begin{abstract}
The self-field of a charged particle has a singular component that diverges at the
particle.  We use both coordinate and covariant approaches to compute an expansion
of this singular field for particles in generic geodesic orbits about a Schwarzschild
black hole for scalar, electromagnetic and gravitational cases.  We check that
both approaches yield identical results and give, as an application, the calculation of previously unknown mode-sum regularisation parameters.

In the so-called
\emph{mode-sum regularization} approach to self-force calculations, each mode of the retarded
field is finite, while their sum
diverges. The sum may be rendered finite and convergent by the subtraction of appropriate
\emph{regularization
parameters}. Higher order parameters lead to faster convergence in the mode-sum. To
demonstrate the significant benefit that they yield, we use our newly derived parameters to calculate a
highly accurate value of $F_r = 0.000013784482575667959(3)$ for the self-force on a scalar particle in a circular orbit around
a Schwarzschild black hole.

Finally, as a second example application of our high-order expansions, we compute
high-order expressions for use in the effective source approach to self-force
calculations.
\end{abstract}

\maketitle

\section{Introduction}
The notion of a self-force has a long history in physics. Early work led to the derivation of the
Abraham-Lorentz-Dirac \cite{Dirac-1938} formula for the radiation reaction force on an
accelerating point electric charge moving in a flat spacetime. In the 1960s, DeWitt and
Brehme \cite{DeWitt:1960} derived the curved spacetime equivalent, and a correction was later
provided by Hobbs \cite{Hobbs:1968a}. For a comprehensive review of the self-force problem,
see Refs.~\cite{Poisson:2003,Detweiler:2005,Barack:2009}.

With the advent of gravitational wave astronomy, the past two decades have seen a surge in interest
in the self-force problem. This has been motivated by the study of so-called Extreme Mass Ratio
Inspiral (EMRI) systems. These binary systems -- consisting of a compact object of $\sim1$ solar mass
spiralling into a massive black hole of $\sim10^6$ solar masses -- are one of the most proimising
candidates for study by planned space-based gravitational wave detectors
\cite{Magorrian:1998,Gair:2004,AmaroSeoane:2007,Gair:2009,Barack:Cutler:2004,Babak:Gair:Sesana:2010,Gair:Sesana:Berti:2010}.
The nature of interferometric gravitational wave detectors is that unlike traditional
electromagnetic spectrum telescopes they are not directional; instead they see all directions at
once. As a result, a crucial component of any observation is the use of data analysis techniques
such as matched filtering to disentangle the desired signal from the noise. This, in turn,
requires an accurate model of the signal one expects to see. This is even more true given the
recent proposals for the eLISA/NGO detector \cite{AmaroSeoane:2012km} (which will have a reduced
sensitivity compared to LISA, the previously proposed space-based gravitational wave detector)
will mean that accurate waveform models are all the more crucial.

The past two decades have seen new derivations of the equations of motion of a point charge moving
in a curved spacetime; this was first done in the gravitational charge case by Mino, Sasaki and
Tanaka \cite{Mino:Sasaki:Tanaka:1996} and Quinn and Wald \cite{Quinn:Wald:1997} and in the
minimally coupled scalar charge case by Quinn \cite{Quinn:2000}. Since then, there have been
significant achievements in putting these derivations on a firmer footing. The use of
distributional sources in Einstein’s equations is known to be problematic \cite{Geroch:1987qn}.
Nevertheless, with sufficient care they have proven to be a useful practical tool in many cases
\cite{Steinbauer:2006qi,Tonita:2011my}. In the case of the self-force problem recent derivations
have avoided the introduction of distributional sources altogether through the use of
matched asymptotic expansions and careful limiting procedures
\cite{Gralla:Wald:2008,Harte-2008,Harte:2009,pound:2009,Gralla:Harte:Wald:2009,Harte:2010}.
At first perturbative order it is satisfying that these
more rigorous derivations reproduce the same equations of motion as one would obtain using
distributional sources. Nevertheless, it is likely that these more rigorous methods
are required to advance to second perturbative order
\cite{Rosenthal:2005ju,Rosenthal:2006iy,Detweiler:2011tt,Pound:2012nt,Gralla:2012db} and beyond
\cite{Galley:2010xn,Galley:2011te,Harte:2011ku}.

Several practical self-force computation strategies have developed from these formal derivations,
all of which are based on the now-justified assumption that the use of a distributional source is
acceptable at first perturbative order:
\begin{itemize}
\item The \emph{mode-sum} approach \cite{Barack:Ori:2000,Barack:Mino:Nakano:Ori:Sasaki:2001}
\item The \emph{effective source} approach \cite{Barack:Golbourn:2007,Vega:Detweiler:2008}
\item The \emph{matched expansion} approach \cite{Anderson:Wiseman:2005,Casals:Dolan:Ottewill:Wardell:2009}
\end{itemize}
The key to all three approaches is the subtraction of an appropriate \emph{singular} component
(which only appears due to the use of a distributional source) from the retarded field to leave
a finite \emph{regular} field which is solely responsible for the self-force. This singular
component must satisfy two criteria:
\begin{enumerate}
\item It has the same singular structure as the full retarded field on the particle's worldline.
\item It does not contribute to the self-force (or its contribution is well known and can be corrected for).
\end{enumerate}
There are many choices for a singular field that satisfies the above criteria, although not all
choices are equal. Detweiler and Whiting \cite{Detweiler-Whiting-2003} identified a particularly
appropriate choice. Through a Green function decomposition, they derived a singular field which not
only satisfies the above two criteria, but also has the property that when it is subtracted from the
full retarted field, it leaves a regularized field which is a solution to the \emph{homogeneous} wave
equation. Extensions of this idea of a singular-regular split to extended charge
distributions \cite{Harte-2008,Harte:2009}, to second perturbative
order \cite{Rosenthal:2005ju,Rosenthal:2006iy,Detweiler:2011tt,Pound:2012nt,Gralla:2012db}
and fully non-perturbative contexts \cite{Harte:2011ku} have recently been developed.

In this paper, we develop highly accurate approximations to the Detweiler-Whiting singular field of
point scalar and electromagnetic charges as well as that of a point mass. This is achieved through high-order
series expansions in a parameter $\epsilon$, which acts as a measure of distance from the particle's
world-line. As examples, we derive explicit expressions for the case of geodesic motion in
Schwarzschild spacetime and show how these may be applied to both the mode-sum and effective source
approaches described above. Similar expansions may also be used for the quasi-local component 
in the matched expansion approach \cite{Ottewill:Wardell:2008,Ottewill:Wardell:2009,QL}
whereby the Green function
is matched onto a quasi-normal mode sum and
the retarded field is then computed as the integral of this matched
retarded Green function along the worldline of the particle.

While the primary focus of this paper is on computing the singular field for Schwarzschild
spacetime, many of the expressions we give are valid in more general spacetimes. In particular,
where space allows, we do not make any assumptions about the spacetime being Ricci-flat.
To make this distinction explicit, we use the Weyl tensor, $C_{abcd}$, in expressions
which are valid only in vacuum and the Riemann tensor, $R_{abcd}$ in expressions which are also
valid for non-vacuum spacetimes. Note that this is done only for space reasons\footnote{The
notable exception is the case of the gravitational singular field, as in that case the
equations of motion have not yet been derived for non-Ricci-flat spacetimes.}; our raw
calculations include all non-vacuum terms in addition to those given in this paper and we
have made the full expressions available in electronic form \cite{BarryWardell.net}.

The layout of this paper is as follows. In Sec.~\ref{sec:SingularField} we give exact formal
expressions for the singular field in scalar, electromagnetic and gravitational cases. In
Sec.~\ref{sec:SingularFieldExpansion} we describe covariant and coordinate approaches to the
computation of expansions of the key fundamental bitensors appearing in the formal expressions
for the singular field. In Sec.~\ref{sec:SingularFieldExplicit} we give explicit expressions for
the singular field in the form of covariant series expansions. Our equivalent coordinate
expressions are too large to be of use in print; instead we have made them available
electronically \cite{BarryWardell.net}. In Sec.~\ref{sec:ModeSum}, we use our coordinate expansions
to derive high-order regularization parameters for use in the mode-sum method. In doing so, we
give the next two previously unknown non-zero regularization parameters in scalar, electromagnetic
and gravitational cases. In
Sec.~\ref{sec:EffectiveSource} we apply our coordinate expansions to the effective source method
and compute an effective source for our high order singular field. This improves on previous
effective source calculations by adding an additional four orders, bringing the approximation to
eight-from-leading order. In Sec.~\ref{sec:Discussion} we summarize our results and
discuss further prospects for their application. Finally, in Appendices
\ref{sec:CoordinateExpansions} and \ref{sec:ExpansionCoefficients} we give explicit
expressions for many of the expansions in coordinate and covariant form, respectively,
along with a more in-depth derivation of our expressions.

Throughout this paper, we use units in which $G=c=1$ and adopt the sign conventions of
\cite{Misner:Thorne:Wheeler:1974}. We denote symmetrization of indices using parenthesis
(e.g. $(ab)$), anti-symmetrization using square brackets (e.g. $[ab]$), and exclude indices
from (anti-)symmetrization by surrounding them by vertical
bars (e.g. $(a | b | c)$, $[a | b | c]$). We denote pairwise (anti-)symmetrization using an
overbar, e.g. $R_{(\overline{ab}\,\overline{cd})} = \frac12 (R_{abcd}+R_{cdab})$.
Capital letters are used to denote the
spinorial/tensorial indices appropriate to the field being considered.

In many of our calculations, we have several spacetime points to be considered. Our convention
is that (see Fig.~\ref{fig:SingularField}):
\begin{itemize}
\item the point $x$ refers to the point where the field is evaluated
\item the point $x'$ refers to an arbitrary spacetime point
\item the point $\xb$ refers to an arbitrary point on the world-line
\item the point $x_{\rm \adv}$ refers to the advanced point of $x$ on the world-line
\item the point $x_{\rm \ret}$ refers to the retarded point of $x$ on the world-line
\end{itemize}
In computing expansions, we use $\epsilon$ as an expansion parameter to denote the fundamental
scale of separation, so that $x-x' \approx x-\xb \approx x'-x = \mathcal{O}(\epsilon)$. Where
tensors are to be evaluated at one of these points, we decorate their indices appropriately using
$'$ and $\bar{ }$, e.g. $T^a$, $T^{a'}$ and $T^\ab$ refer to tensors at $x$, $x'$ and $\xb$,
respectively.

\section{Singular field and self-force of a point particle} \label{sec:SingularField}

In an appropriate gauge, the retarded field, $\varphi^{A}(x)$, of an arbitrary point particle
satisfies the inhomogeneous wave equation with a distributional source,
\begin{equation}
\label{eq:Wave}
\mathcal{D}^{A}{}_B \varphi^{B} = - 4\pi \mathcal{Q} \int u^{A} \delta_4 \left( x,z(\tau') \right) d\tau',
\end{equation}
where
\begin{equation}
\label{eq:Wave-operator}
\mathcal{D}^{A}{}_B  = \delta^{A}{}_B (\square - m^2) - P^{A}{}_B
\end{equation}
is the wave operator, $\square \equiv g^{ab}\nabla_{a}\nabla_{b}$, 
$g^{ab}$ is the (contravariant) metric tensor, $\nabla_{a}$ is the covariant
derivative defined by a connection $\mathcal{A}^{A}{}_{Ba}$:
$\nabla_{a}\varphi^{A}= \partial_{a}\varphi^{A}+  \mathcal{A}^{A}{}_{Ba} \varphi^{B}$,
$m$ is the mass of the field, $\mathcal{Q}$ is the charge
of the particle, $P^{A}{}_B(x)$ is a potential term, $u^{A}(x)$ is an appropriate tensor
constructed from products of the four-velocity on the world-line parallel transported to $x$ and $\delta_4 \left( x,z(\tau') \right)$ is an invariant Dirac functional in a four-dimensional curved spacetime as defined in Eq.~(13.1) of Ref.~\cite{Poisson:2003}.
%$\delta_4 \left( x, x' \right)$ is defined by the relations
%\begin{equation}
%\int_V f\left(x \right) \delta_4 \left( x, x' \right) \sqrt{-g} d^4 x = f \left(x'\right), \quad \quad \int_{V'} f\left(x' \right) \delta_4 %\left( x, x' \right) \sqrt{-g'} d^4 x' = f \left(x\right),
%\end{equation}
%where $f\left(x \right)$ is a smooth test function, $V$ any four-dimensional region that contains $x'$, and $V'$ any four-dimensional region that contains $x$.  
The retarded solutions to this equation give rise to a field which one might na\"ively
expect to exert a self-force
\begin{equation}
F^a = p^a{}_A \varphi^{A}_{\rm{\ret}}
\end{equation}
on the particle, where $p^a{}_A(x)$ is a tensor at $x$ and depends on the type of charge. The
distributional nature of the source leads to this self-force being divergent at the location of the
particle and a regularisation scheme must be employed. We require a \emph{singular} field, 
$\varphi^{A}_{(\mathrm{S})}$, that captures the singular behaviour of $\varphi^{A}$ and that, when
subtracted from $\varphi^{A}$, leaves a finite \emph{physical} self-force.

Detweiler and Whiting \cite{Detweiler-Whiting-2003} showed how such a singular field can be
constructed through a Green function decomposition. In four spacetime dimensions and within a normal
neighborhood, the Green function for the retarded/advanced solutions to Eq.~\eqref{eq:Wave}
may be given in Hadamard form,
\begin{equation}
\label{eq:Hadamard}
G_{\mathrm{\ret/\adv}}{}^{A}{}_{B'}\left( x,x' \right) = \theta_{-/+} \left( x,x' \right) \left\lbrace U^{A}{}_{B'} \left( x,x' \right) \delta \left[ \sigma \left( x,x' \right) \right] - V^{A}{}_{B'} \left( x,x' \right) \theta \left[ - \sigma \left( x,x' \right) \right] \right\rbrace ,
\end{equation}
where $\delta\left[ \sigma\left(x,x'\right)\right]$ is the covariant form of the Dirac delta
function, $\theta\left[ \sigma\left(x,x'\right)\right]$ is the Heaviside step function, while 
$U^{A}{}_{B'}\left( x,x' \right)$ and $V^{A}{}_{B'}\left( x,x' \right)$ are symmetric bi-spinors/tensors
which are regular for $x' \rightarrow x$. The bi-scalar $\sigma \left( x,x' \right)$ is the
Synge~\cite{Poisson:2003} world function; it is equal to one half of the squared geodesic distance
between $x$ and $x'$. The first term here, involving $U^{A}{}_{B'} \left( x,x' \right)$,
represents the \emph{direct} part of the Green function 
while the second term, involving $V^{A}{}_{B'} \left( x,x' \right)$, is known as the
\emph{tail} part of the Green function.

Detweiler and Whiting proposed to define a \emph{singular} Green function by taking the symmetric
Green function, $G_{\mathrm{(sym)}}{}^{A}{}_{B'} = \frac12(G_{\mathrm{\ret}}{}^{A}{}_{B'} + G_{\mathrm{\adv}}{}^{A}{}_{B'})$
and adding $V^{A}{}_{B'}\left( x,x' \right)$ (a homogeneous solution to Eq.~\eqref{eq:Wave}). This
leads to the singular Green function,
\begin{equation}
\label{eq:Hadamard-singular}
G_{\mathrm{\sing}}{}^{A}{}_{B'}\left( x,x' \right) =
  \frac12\left\lbrace U^{A}{}_{B'} \left( x,x' \right) \delta \left[ \sigma \left( x,x' \right) \right]
  + V^{A}{}_{B'} \left( x,x' \right) \theta \left[\sigma \left( x,x' \right) \right] \right\rbrace .
\end{equation}
Note that this has support on and \emph{outside} the past and future light-cone 
(i.e. for points $x$ and $x'$ spatially separated)
and is only uniquely defined provided $x$ and $x'$ are within a convex normal neighbourhood. 
Given this singular Green function, we may define the Detweiler-Whiting singular field,
\begin{equation} \label{eq:SingularField}
\varphi^{A}_{\rm \sing} = \int_{\tau_{\rm \adv}}^{\tau_{\rm \ret}} G_{\mathrm{\sing}}{}^{A}{}_{B'}\left( x, z(\tau') \right) u^{B'} d\tau',
\end{equation}
which also satisfies Eq.~\eqref{eq:Wave}. 
Subtracting this singular field from the retarded field, we obtain the \emph{regularized} field,
\begin{equation}
\varphi^{A}_{\rm{\reg}} = \varphi^{A}_{\rm{\ret}} - \varphi^{A}_{\rm{\sing}},
\end{equation}
which Detweiler and Whiting showed gives the correct finite physical self-force,
\begin{equation} \label{eqn:SelfForce}
F^a = p^a{}_A \varphi^{A}_{\rm{\reg}}.
\end{equation}
Moreover, this regularized field is a solution of the homogeneous wave equation,
\begin{equation}
\mathcal{D}^{A}{}_B \varphi^{B}_{\rm{\reg}} = 0.
\end{equation}
This holds independently of whether one is considering a scalar or electromagnetically charged
point particle or a  point mass. To make this more explicit, in the following subsections
we give the form these expressions take in each of scalar, electromagnetic and gravitational cases.

\subsection{Scalar Case}
In the scalar case the singular field, $\PhiS$, is
a solution of the inhomogeneous scalar wave equation,
\begin{equation}
(\Box - \xi R - m^2) \PhiS = q \int \sqrt{-g} \delta_4(x,z(\tau)) d\tau,
\end{equation}
where $q$ is the scalar charge and $\xi$ is the coupling to the background scalar curvature.
An expression for $\PhiS$ may be found by considering the scalar Green function [obtained
by taking $U^{A}{}_{B'} = U(x,x')$ in Eq.~\eqref{eq:Hadamard-singular}],
\begin{equation}\label{eqn:Gs}
G^{\rm \sing} = \frac{1}{2}\left\{U(x,x') \delta[\sigma(x,x')] + V(x,x') \theta[\sigma(x,x')]\right\},
\end{equation}
with $U(x,x') = \Delta^{1/2}(x,x')$, where $\Delta^{1/2}(x,x')$ is the Van Vleck-Morette determinant as defined in Eq.~(7.1) of Ref.~\cite{Poisson:2003}.  This Green function is a solution of the equation
\begin{equation}
(\Box - \xi R - m^2)  G^{\rm \sing} = -4\pi \delta(x,x').
\end{equation}
Given this expression for the Green function, the scalar singular
field is
\begin{eqnarray} \label{eq:PhiS}
\PhiS(x) &=& q\int_\gamma G^{\rm \sing} (x,z(\tau)) d\tau \nonumber \\
 &=& \frac{q}{2} \Bigg[ \frac{U(x,x')}{\sigma_{c'} u^{c'}} \Bigg]_{x'=x_{\rm \ret}}^{x'=x_{\rm \adv}}
   + \frac{q}{2} \int_{\tau_{\rm \ret}}^{\tau_{\rm \adv}} V(x,z(\tau)) d\tau
\end{eqnarray}
and one computes the scalar self-force from the regularized scalar field
$\Phi^{\rm \reg} = \Phi^{\rm{\ret}} - \PhiS$ as
\begin{equation} \label{eqn:SelfForceScalar}
F^a = q \, g^{ab} \Phi^{\rm{\reg}}_{, b}.
\end{equation}

\subsection{Electromagnetic Case}
In Lorenz gauge, the electromagnetic singular field satisfies the equation
\begin{equation}
\Box A_a^{\rm \sing} - R_a{}^b A_b^{\rm{\sing}}= -4\pi e \int g_{aa'} u^{a'} \sqrt{-g} \delta_4(x,z(\tau)) d\tau,
\end{equation}
where $e$ is the electric charge.
An expression for $A_a^{\rm \sing}$ may be found by considering the electromagnetic Green
function [obtained by taking $U^{A}{}_{B'} = U(x,x')_{aa'}$ in Eq.~\eqref{eq:Hadamard-singular}],
\begin{equation}
G^{\mathrm{\sing}}_{aa'}(x,x') =
\frac{1}{2} \left\lbrace U(x,x')_{aa'} \delta \left( \sigma (x,x') \right) + V(x,x')_{aa'} \theta \left( \sigma (x,x') \right) \right\rbrace,
\end{equation}
with $U^a{}_{a'} = \Delta^{1/2} g^a{}_{a'}$, where $g^a{}_{a'}$ is the bi-vector of parallel transport as defined in Eq.~(5.11) of Ref.~\cite{Poisson:2003}.  This Green function is a solution of the equation
\begin{equation}
\Box G_{aa'}^{\rm \sing} - R_a{}^b G_{ba'}^{\rm{\sing}}= - 4\pi g_{aa'} \delta_4(x,x').
\end{equation}
Given this expression for the Green function, the electromagnetic singular field is
\begin{align} \label{eq:AS}
A_{a}^{\rm \sing} &= e \int_\gamma G^{\rm \sing}_{aa'} (x, z(\tau')) u^{a'} d\tau' \nonumber \\
&= \frac{e}{2} \Bigg[\frac{u^{a'} U_{aa'}(x,x')}{\sigma_{c'} u^{c'}} \Bigg]_{x'=x_{\rm \ret}}^{x'=x_{\rm \adv}}
  + \frac{e}{2} \int_{\tau_{\rm \ret}}^{\tau_{\rm \adv}} V_{aa'}(x,z(\tau)) u^{a'} d\tau.
\end{align}
One computes the electromagnetic self-force from the electromagnetic regular
field,
$A_a^{\rm \reg} = A_a^{\rm{\ret}} - A_a^{\rm{\sing}}$, as
\begin{equation} \label{eqn:SelfForceEM}
F^a = e \, g^{ab} u^c A^{\rm{\reg}}_{[c, b]}.
\end{equation}

\subsection{Gravitational Case}
In Lorenz gauge, the trace-reversed singular first order metric perturbation satisfies the equation
\begin{equation}
\Box \hb_{ab}^{\rm \sing} + 2 C_a{}^c{}_b{}^d \hb_{cd}^{\rm{\sing}}= -16\pi \mu \int g_{a'(a} u^{a'} g_{b)b'} u^{b'} \sqrt{-g} \delta_4(x,z(\tau)) d\tau,
\end{equation}
where $\mu$ is the mass of the particle and the trace-reversed singular field is related to the
non-trace-reversed version by $\hb_{ab}^{\rm \sing} = h_{ab}^{\rm \sing} - \frac12 h^{\rm \sing} g_{ab}$
with $h^{\rm \sing} =  g^{ab} h^{\rm \sing}_{ab}$.
An expression for $\hb_{ab}^{\rm \sing}$ may be found by considering the gravitational Green
function [obtained by taking $U^{A}{}_{B'} = U(x,x')_{a b a' b'}$ in
Eq.~\eqref{eq:Hadamard-singular}],
\begin{equation}
G^{\mathrm{\sing}}_{a b a' b'}(x,x') =
\frac{1}{2} \left\lbrace U(x,x')_{a b a' b'} \delta \left[ \sigma (x,x') \right] + V(x,x')_{a b a' b'} \theta \left[ \sigma (x,x') \right] \right\rbrace,
\end{equation}
with $U^{ab}{}_{a'b'} = \Delta^{1/2} g^{(a}{}_{a'} g^{b)}{}_{b'}$. 
This Green function is a solution of the equation
\begin{equation}
\Box G_{aba'b'}^{\rm \sing} + 2 C_a{}^p{}_b{}^q G_{pqa'b'}^{\rm{\sing}}= - 4\pi g_{a'(a}g_{b)b'} \delta_4(x,x').
\end{equation}
Given this expression for the Green function, the trace-reversed singular first order metric
perturbation is
\begin{align} \label{eq:hS}
\hb_{a b}^{\rm \sing} &= 4 \mu \int_\gamma G^{\rm \sing}_{a b a' b'} (x, z(\tau')) u^{a'} u^{b'} d\tau' \nonumber \\
&= 2 \mu \Bigg[\frac{u^{a'} u^{b'} U_{aba'b'}(x,x')}{\sigma_{c'} u^{c'}} \Bigg]_{x'=x_{\rm \ret}}^{x'=x_{\rm \adv}}
  + 2 \mu \int_{\tau_{\rm \ret}}^{\tau_{\rm \adv}} V_{aba'b'}(x,z(\tau)) u^{a'} u^{b'} d\tau.
\end{align}
One computes the gravitational self-force from the regularized trace-reversed singular first order metric
perturbation,
$\hb_{ab}^{\rm \reg} = \hb_{ab}^{\rm{\ret}} - \hb_{ab}^{\rm{\sing}}$, as
\begin{equation}\label{eqn:SelfForceGravityBasic}
F^a = \mu \, k^{abcd} \hb^{\rm{\reg}}_{bc; d},
\end{equation}
where
\begin{equation}
k^{abcd} \equiv \frac12 g^{ad} u^b u^c - g^{ab} u^c u^d - 
  \frac12 u^a u^b u^c u^d + \frac14 u^a g^{b c} u^d + 
 \frac14 g^{a d} g^{b c}.
\end{equation}

\section{Covariant and Coordinate Expansion of Fundamental Bitensors} \label{sec:SingularFieldExpansion}
In the previous section, we gave expressions for the singular field in terms of the bitensors $U^{A}{}_{B'} \left( x,x' \right)$ and $V^{A}{}_{B'} \left( x,x' \right)$. The first of these is given by 
\begin{equation} \label{eq: U}
U^{AB'} \left( x,x' \right) = \Delta^{1/2} \left( x,x' \right) g^{AB'}\left( x,x' \right),
\end{equation}
where $\Delta \left( x,x' \right)$ is the Van Vleck-Morette determinant \cite{Poisson:2003},
\begin{equation} \label{eq:vanVleckDefinition}
\Delta \left( x,x' \right) = - \left[ -g \left( x \right) \right] ^{-1/2} \det \left( -\sigma _{;\alpha \beta '} \left( x,x' \right) \right) \left[ -g \left( x' \right) \right] ^{-1/2} 
 = \det \left( -g^{\alpha'}{}_\alpha \left( x,x' \right) \sigma ^{;\alpha}{}_{ \beta '} \left( x,x' \right) \right),
\end{equation}
$g^{AB'}$ is the bi-tensor of parallel transport appropriate to the tensorial nature of the field, eg.
\begin{equation}
 g^{A B'} = \begin{cases}
             1 & \text{(scalar)}\\
             g^{a b'} & \text{(electromagnetic)} \\
             g^{a' (a} g^{b) b'} & \text{(gravitational)},
            \end{cases}
\end{equation}
and where the higher spin fields are taken in Lorentz gauge. Here, $g^{\alpha'}{}_\alpha\left( x,x' \right)$ is the bi-vector of parallel transport
defined by the transport equation
\begin{equation}
\sigma^{\alpha} g_{a b' ;\alpha}  = 0 = \sigma^{\alpha'} g_{a b' ;\alpha'}.
\end{equation}
The bitensor $V^{AB'}\left( x,x' \right)$ may be expressed in terms of a formal expansion in increasing powers of $\sigma$ \cite{Decanini:Folacci:2005a}:
\begin{equation}
\label{eq:V}
V^{AB'}\left( x,x' \right) = \sum_{\num =0}^{\infty} V_{\num}{}^{AB'}\left( x,x' \right) \sigma ^{\num}\left( x,x' \right),
\end{equation}
where the coefficients $V_{\num}^{AB'}\left( x,x' \right)$ satisfy the recursion relations
\begin{subequations}
\label{eq:RecursionV}
\begin{align}
\label{eq:recursionVn}
  \sigma ^{;\alpha'} (\Delta ^{-1/2} V^{AB'}_{\num})_{;\alpha'}  
 + \left( \num +1 \right)  \Delta ^{-1/2}  V_{\num}^{AB'} + {\frac{1}{2 \num}} \Delta ^{-1/2}  \mathcal{D}^{B'}{}_{C'} V^{AC'}_{\num -1} = 0 ,
\end{align}
for $\num \in \mathbb{N}$, along with the `initial condition'
\begin{eqnarray}
\label{eq:recursionV0}
\sigma ^{;\alpha'} (\Delta ^{-1/2} V_0^{AB'}){}_{;\alpha'} 
+ \Delta ^{-1/2} V^{AB'}_0 + {\frac{1}{2}}\Delta ^{-1/2} \mathcal{D}^{B'}{}_{C'} ( \Delta ^{1/2} g^{AC'}) &=& 0.
\end{eqnarray}
\end{subequations}

Looking at the above equations for $U^{AB'}\left( x,x' \right)$ and $V_{\num}{}^{AB'}\left( x,x' \right)$, we see that a key component of the present work involves the computation of several fundamental
bitensors, in particular, the world function $\sigma (x, x')$,
Van Vleck determinant $\Delta^{1/2}(x,x')$, four-velocity $u^{a}(x)$, and bivector of parallel
transport $g_a{}^{b'}(x,x')$. This may be achieved by expressing them as expansions about some arbitrary point $\xb$ which is close to $x$ and $x'$.
We derive these here using both covariant and coordinate methods, each of which has its own 
advantages and disadvantages.  The covariant expression is more elegant, allowing for compact 
formulas; however these formulas hide complex terms such as high order derivatives of the Weyl
tensor that quickly become extremely time consuming to compute, even using computer tensor algebra 
packages such as GRTensorII \cite{GRTensor} or xAct \cite{xTensorOnline}.  The coordinate
approach is less elegant but more practical 
for explicit calculations and it avoids the need to use tensor algebra. Independently of the approach taken, these expansions may be used to compute expansions of $U^{AB'}\left( x,x' \right)$ and $V_{\num}{}^{AB'}\left( x,x' \right)$ (by substituting into the above equations), and hence of the singular field. In the case
of covariant expansions, for explicit calculations one must further expand the covariant
expressions in coordinates, yielding an expression which may be directly compared with
those obtained from the coordinate approach. The resulting expressions are long
but are explicit functions of the coordinates, enabling them to be transformed directly into,
for example, C functions, indeed we give them in such form online \cite{BarryWardell.net}.

\begin{figure}
\includegraphics[scale=0.5]{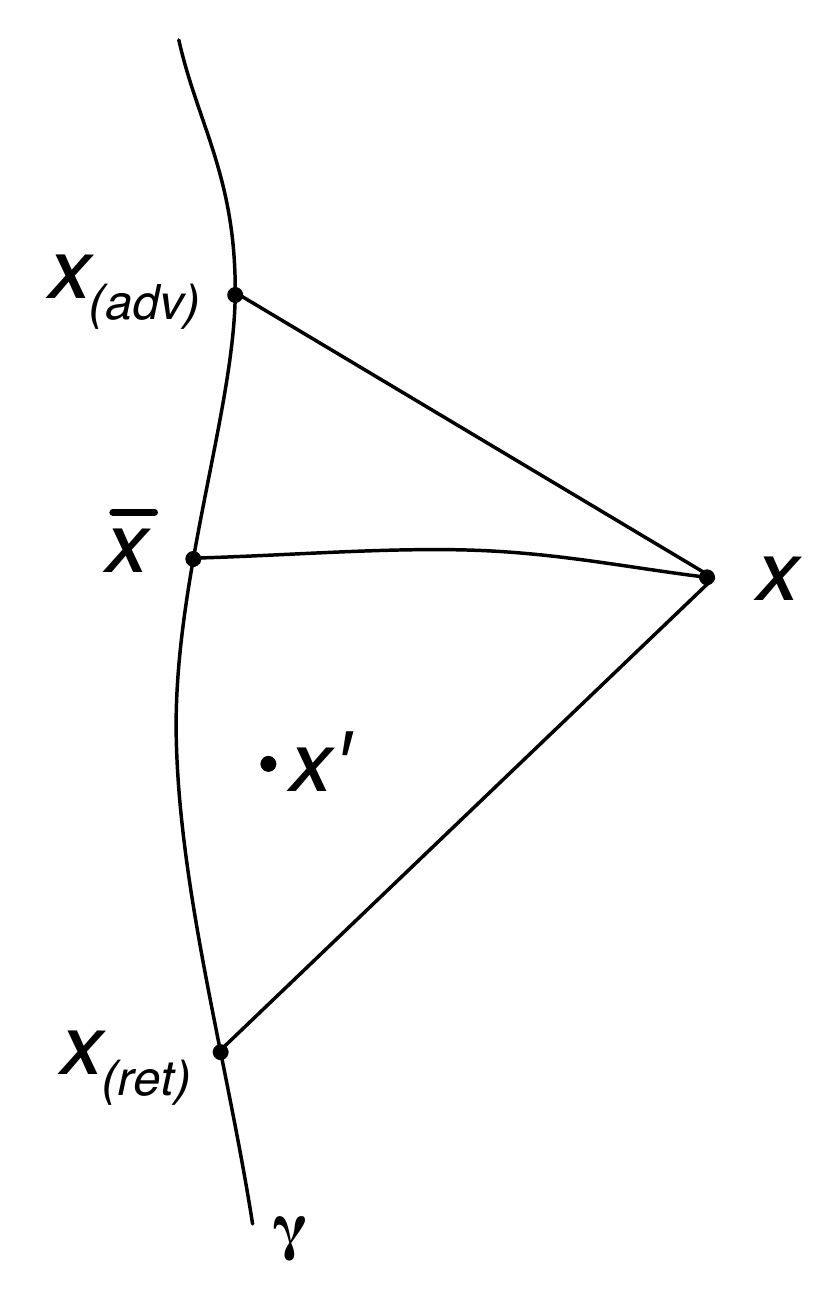}
\caption{We expand all bitensors (which are functions of $x$ and $x'$)
about the arbitrary point $\xb$ on the worldine.}
\label{fig:SingularField}
\end{figure}

\subsection{Covariant Approach} \label{sec:Covariant}
In this subsection, we briefly discuss our method for obtaining covariant
expansions for the biscalars
appearing in Eqs.~\eqref{eq:PhiS}, \eqref{eq:AS} and \eqref{eq:hS}. We eventually seek expansions about a point $\bar{x}$
on the worldline (which we may treat as fixed in the majority of this paper). In doing
so, we follow the strategy of Haas and Poisson \cite{Poisson:2003,Haas:Poisson:2006}:
\begin{itemize}
\item For the generic biscalar $A(x,z(\tau))$, write it as $A(\tau) \equiv A(x,z(\tau))$.
\item Compute the expansion about $\tau = \bar{\tau}$. This takes the form
\begin{equation}
A(\tau) = A(\bar{\tau}) + \dot{A}(\bar{\tau}) (\tau-\bar{\tau}) + \frac{1}{2}\ddot{A}(\bar{\tau}) (\tau-\bar{\tau})^2 + \cdots,
\end{equation}
where $\dot{A}(\bar{\tau}) = A_{;\ab} u^{\ab}$,
$\ddot{A}(\bar{\tau}) = A_{;\ab \bb} u^{\ab} u^{\bb}$, $\cdots$.
\item Compute the covariant expansions of the coefficients
$\dot{A}(\bar{\tau})$, $\ddot{A}(\bar{\tau})$, $\cdots$ about $\bar{\tau}$.
\item Evaluate the expansion at the desired point, eg. $A(x') = A(x,x')$.
\item The resulting expansion depends on $\tau$ through the powers of $\tau-\bar{\tau}$.
Replace these by their expansion in $\epsilon$ (about $\bar{x}$), the distance between $x$ and
the world-line.
\end{itemize}

A key ingredient of this calculation is the expansion of $\Delta\tau \equiv \tau-\bar{\tau}$ in
$\epsilon$. The leading orders in this expansion were developed by Haas and Poisson \cite{Haas:Poisson:2006} for the particular choices $\Delta\tau_+ \equiv v-\bar{\tau}$
and $\Delta\tau_- \equiv u-\bar{\tau}$. They found
\begin{equation}
\label{eq:Delta}
\Delta\tau_\pm = (\rbar \pm \sbar) \mp \frac{(\rbar \pm \sbar)^2}{6\,\sbar} R_{u \sigma u \sigma} \mp
  \frac{(\rbar \pm \sbar)^2}{24\,\sbar} \left[
    (\rbar \pm \sbar)R_{u \sigma u \sigma | u}
    - R_{u \sigma u \sigma | \sigma}\right]
  + \mathcal{O}(\epsilon^5),
\end{equation}
where $\rbar \equiv \sigma_\ab u^\ab$ and $\sbar \equiv (g^{\ab \bb} + u^\ab u^\bb) \sigma_\ab \sigma_\bb$. In Appendix \ref{sec:ExpansionCoefficients} we
extend their calculation to the higher orders required in the present work.
In the same Appendix, we also apply the above method to compute covariant expansions of all quantities appearing in the expression for the singular field.

In order to obtain explicit expressions, we substitute in the coordinate expansion for
$\sigma_{\ab}$ (as discussed in Sec.~\ref{sec:CoordinateExpansions}) along with the metric, Riemann tensor and
4-velocity (all evaluated at $\xb$). In doing so, we only have to keep terms that contribute up to
the required order and truncate any higher order terms.

\subsection{Coordinate Approach} \label{sec:Coordinate}
In this subsection we follow a similar approach as described above, but in coordinates.  
We will start by considering two arbitrary points $x$ and $x'$ near $\xb$.  We will seek expansion where the
coefficients are evaluated at $\xb$, so we introduce the notation
\begin{equation} \label{eqn:deltaxprime}
\Delta x^a = x^a - x^\ab , \qquad \delta x^{a'} = x^{a'} - x^a = x^{a'} - \Delta x^a - x^{\bar{a}},
 \end{equation}
where we use the convention that the index carries the information about the point: $\xb^a = x^\ab$.  
In the calculations below $\Delta x^{a}$ and $\delta x^{a'}$
are both assumed to be small, of order $\epsilon$.

The first item we require for our calculations is a coordinate expansion of the biscalar $\sigma \left( x,x' \right)$, 
 the Synge world function.  We start with a standard coordinate series expansion about $x$, see for example \cite{Ottewill:Wardell:2009} (note the difference in convention for $\Delta x^a$ in that paper), to get
\begin{equation} \label{eqn: sigma}
\sigma(x,x') = \tfrac{1}{2} g_{a b}(x) \delta x^{a'} \delta x^{b'} + A_{abc}(x) \delta x^{a'} \delta x^{b'} \delta x^{c'} + B_{abcd}(x) \delta x^{a'} \delta x^{b'} \delta x^{c'} \delta x^{d'} 
+ \cdots.
\end{equation}
The coefficients are readily determined in terms of derivatives of the metric at $x$ by use of the defining identity $2 \sigma = \sigma_{a'} \sigma^{a'}$,
see \cite{Ottewill:Wardell:2009}. To be explicit, the first few are given by
\begin{align*}
A_{a b c}(x) 	&= \tfrac{1}{4}g_{(a b ,c)}(x) \\
 B_{a b c d}(x)	&= \tfrac{1}{12}g_{(a b, c d)}(x) - \tfrac{1}{24}  g^{p q}(x) \left(g_{(a b ,|p|}(x) g_{c d) ,q}(x)
			-12   g_{(a b ,|p|} (x)g_{|q| c , d)}(x)+36    g_{p (a, b} (x)g_{|q| c , d)}(x) \right) .
\end{align*}

We now go one step further by expanding the coefficients about $\xb$  to give a double expansion in $\Delta x^a$ and $\delta x^{a'}$ with coefficients at $\xb$. The first few terms are
\begin{align}
\sigma(x,x') &=  \tfrac{1}{2} g_{\ab \bb}(\xb) \delta x^{a'}  \delta x^{b'} +
\left[\tfrac{1}{2} g_{\ab \bb,\cb }(\xb) \delta x^{a'}  \delta x^{b'}  \Delta x^c+A_{\ab \bb \cb}(\xb)\delta x^{a'} \delta x^{b'}  \delta x^{c'}  \right]\nonumber \\
&+
 \left[ \tfrac{1}{4} g_{\ab \bb,\cb\db }(\xb) \delta x^{a'}  \delta x^{b'} \Delta x^c \Delta x^d +A_{\ab \bb \cb,\db}(\xb)\delta x^{a'} \delta x^{b'}  \delta x^{c'}  \Delta x^{d}+B_{\ab \bb \cb \db}(\xb) \delta x^{a'} \delta x^{b'} \delta x^{c'}  \delta x^{d'}  \right]+ O(\epsilon^5),
\end{align}
where now we interpret $\delta x^{a'}$ as   $x^{a'} - \Delta x^a - x^{\bar{a}}$ and we use square brackets to distinguish terms of different order in $\epsilon$.
Rather than disturb the flow here and throughout this section we just give the first few terms of each expansion for a general metric to make the structure clear and give explicit expressions in Schwarzschild space-time to much higher order in Appendix \ref{sec:CoordinateExpansions}.

Now that the coefficients are at the fixed point $\xb$, it is straightforward to take derivatives of $\sigma$ at $x$ and $\xp$, for example,
\begin{align}
\sigma_{{a'}} &=   g_{\ab \bb} \delta x^{b'}  +
\left[ g_{\ab \bb,\cb } \delta x^{b'}  \Delta x^c
 +3 A_{\ab \bb \cb}\delta x^{b'}  \delta x^{c'} \right] \nonumber\\
&\qquad+\left[ \tfrac{1}{2} g_{\ab \bb,\cb\db }\delta x^{b'} \Delta x^c \Delta x^d +3 A_{\ab \bb \cb,\db}\delta x^{b'} \delta x^{c'} \Delta x^{d} + 4B_{\ab \bb \cb \db}\delta x^{b'} \delta x^{c'}  \delta x^{d'} \right]+O(\epsilon^4),
 \label{eqn: sigmaaprime}
\\ \sigma_{a} &= -\sigma_{{a'}} +\tfrac{1}{2} g_{ \bb \cb ,\ab } \delta x^{b'}  \delta x^{c'}  + \left[ \tfrac{1}{2} g_{ \bb\cb,\db\ab  } \delta x^{b'} \delta x^{c'} \delta x^{d'} +A_{ \bb \cb \db,\ab}\delta x^{b'} \delta x^{c'} \delta x^{d'} \right]+O(\epsilon^4), \\
\sigma_{{a'}b} &=  - g_{\ab \bb} -
g_{\ab\bb,\cb}  \Delta x^c +\left[ ( 3 A_{\ab \cb \db, \bb}-12 B_{\ab \bb \cb \db})\delta x^{c'}  \delta x^{d'}  -\tfrac{1}{2} g_{\ab\bb,\cb\db}\Delta x^{c}  \Delta x^{d}\right] +O(\epsilon^3).
\label{eqn: sigmaalphaprime}
\end{align}
Likewise we can calculate the Van Vleck-Morette determinant directly from its definition,
\begin{equation} \label{eqn: Van Vleck}
\Delta^{\frac{1}{2}} (x,x') = \left(-\left[-g(x)\right]^{- \frac{1}{2}} \left|- \sigma_{a' b} (x,x')\right| \left[-g(x')\right]^{- \frac{1}{2}}\right)^{\frac{1}{2}},
\end{equation}
giving 
\begin{align} \label{eqn: Van Vleck 2}
\Delta^{\frac{1}{2}} (x,x') =& 1 + \Bigl[\tfrac{1}{2}g^{\cb\db}(2g_{\ab\cb,\bb\db}-g_{\ab\bb,\cb\db}-g_{\cb\db,\ab\bb} )+ \tfrac{1}{4}g^{\cb\db}g^{\eb\fb}\bigl(g_{\cb\eb,\ab}g_{\db\fb,\bb}+2g_{\cb\ab,\bb}(g_{\eb\fb,\db}-2g_{\db\eb,\fb}) \nonumber\\ 
&
\quad+2g_{\ab\cb,\eb}(g_{\bb\db,\fb}-g_{\bb\fb,\db})
-g_{\ab\bb,\cb}(g_{\eb\fb,\db}-2g_{\db\eb,\fb}) \bigr)\Bigr]\delta x^a\delta x^b + O(\epsilon^3).
\end{align}

To obtain an expressions at $x_\textrm{\adv}$ and $x_\textrm{\ret}$, we allow $x^{a'}$ to
be on the worldline and again give it as an expansion around the point $\bar{x}$, as
shown in Fig.~\ref{fig:SingularField}.  Writing $x^{a'}$ in
terms of proper time $\tau$ gives\footnote{In principal this expression
is valid for an arbitrary worldline. However, here we restrict
ourselves to the case of geodesic motion and derive the higher derivative terms from the
geodesic equations; we will address the case of accelerated motion
in a follow-up work \cite{NonGeodesic:InPreparation}.} 
\begin{equation} \label{eqn: xtilde}
x^{{a'}}(\tau) = x^{\ab} + u^{\ab} \Delta \tau + \tfrac{1}{2!} \dot{u}^{\ab} \Delta \tau^2 + \tfrac{1}{3!} \ddot{u}^{\ab} \Delta \tau^3 +  \cdots,
\end{equation}
where $u^{\ab}$ is the four velocity at the point $x^{\ab}$, $\Delta \tau = \tau - \bar{\tau}$,
and an overdot denotes differentiation with respect to $\tau$.

We are interested in determining the points on the worldline that are connected to $x$ by a null geodesic, that is we want to solve
\begin{align} \label{eqn: nullseparated}
\sigma\bigl( x^a,x^{{a'}}(\tau) \bigr) = 0 =& 
\tfrac{1}{2} g_{\ab \bb}(u^{\ab} \Delta \tau  -\Delta x^a) (u^{\bb} \Delta \tau  -\Delta x^b) \nonumber\\
&
+\bigl[\tfrac{1}{2} g_{\ab \bb }( u^{\ab}  \Delta \tau -\Delta x^a) \dot{u}^{\bb} \Delta \tau^2+\tfrac{1}{2} g_{\ab \bb,\cb }( u^{\ab} \Delta \tau  -\Delta x^a) (u^{\bb} \Delta \tau   -\Delta x^b)   \Delta x^c
\nonumber\\
&\quad+\tfrac{1}{4}g_{\ab \bb , \cb}(\xb) (u^{\ab} \Delta \tau  -\Delta x^a) (u^{\bb} \Delta \tau  -\Delta x^b) 
 (u^{\cb} \Delta \tau  -\Delta x^c)   \bigr]+ O(\epsilon^4).
\end{align}
By writing 
$
\Delta \tau = \tau_1 \epsilon + \tau_2 \epsilon^2 + \tau_3 \epsilon^3  + \cdots
$
and explicitly inserting an $\epsilon$ in front of $\Delta x^a$, we may equate coefficients of powers of $\epsilon$ to obtain
\begin{align}\label{eqn: tau1}
 \tau_1{}^2  +2 g_{\ab \bb} u^{\ab} \Delta x^b\tau_1 -  g_{\ab \bb} \Delta x^a \Delta x^b =&~ 0,\\
\label{eqn: tau2}
 g_{\ab \bb} u^{\ab}(u^{\ab} \tau_1  -\Delta x^a) \tau_2 =& -\bigl[
\tfrac{1}{2} g_{\ab \bb }( u^{\ab}   \tau_1 -\Delta x^a) \dot{u}^{\bb}  \tau_1{}^2+\tfrac{1}{2} g_{\ab \bb,\cb }( u^{\ab}  \tau_1  -\Delta x^a) (u^{\bb}  \tau_1   -\Delta x^b)   \Delta x^c
\nonumber\\
&\quad+\tfrac{1}{4}g_{\ab \bb , \cb}(\xb) (u^{\ab}  \tau_1  -\Delta x^a) (u^{\bb}  \tau_1  -\Delta x^b) 
 (u^{\cb}  \tau_1  -\Delta x^c) \bigr].
\end{align}
Equation \eqref{eqn: tau1} is a quadratic with two real roots of opposite sign (for $x$ spacelike separated from $\xb$) corresponding to the 
first approximation to our points $x_\textrm{\adv}$ and $x_\textrm{\ret}$,  
\begin{align}
\tau_{1\pm} = g_{\ab \bb} u^{\ab} \Delta x^b \pm \sqrt{(g_{\ab \bb} u^{\ab} \Delta x^b)^2 +g_{\ab \bb} \Delta x^a \Delta x^b}
\equiv \rbar_{(1)} \pm \rho,
\end{align}
where $\rbar_{(1)}$ is the leading order term in the coordinate expansion of the
quantity $\rbar$ appearing in our covariant expansions.
Equation \eqref{eqn: tau2} is typical of the higher order equations giving $\tau_n$ in 
terms of lower order terms.

\section{Expansions of the Singular Field} \label{sec:SingularFieldExplicit}

In this section we list the covariant form of the singular field to order $\epsilon^4$, where $\epsilon$ is the fundamental 
scale of separation, so, for example,  $\rbar$, $\sbar$ and $\sigma(x,\xb)^{\ab}$ are all of leading order $\epsilon$. 
The cooordinate forms of these expansions are too long to be useful in print form so instead they are available to 
download~\cite{BarryWardell.net}.  

\subsection{Scalar singular field}
To $\mathcal{O}(\epsilon^4)$, the scalar singular field is
\begin{equation} \label{eq:phiS-approx}
\PhiS =
q \Big\{
    \frac{1}{\sbar}
  + \frac{\rbar^2-\sbar^2}{6 \sbar^3} C_{u \sigma u \sigma}
  + \frac{1}{24 \sbar^3} \Big[
      (\rbar^2 - 3 \sbar^2) \rbar C_{u \sigma u \sigma ; u}
    - (\rbar^2-\sbar^2) C_{u \sigma u \sigma ; \sigma} \Big]
  + \frac{1}{360 \sbar^5} \left[\PhiS\right]_{(3)}
  + \frac{1}{4320 \sbar^5} \left[\PhiS\right]_{(4)}
  + \mathcal{O}(\epsilon^5)
\Big\},
\end{equation}
where
\begin{IEEEeqnarray}{rCl}
\left[\PhiS\right]_{(3)} &=&
    15 \Big[\rbar^2 - \sbar^2 \Big]^2 C_{u\sigma u\sigma }C_{u\sigma u\sigma }
  + \sbar^2 \Big[
        (\rbar^2 -  \sbar^2)(3 C_{u\sigma u\sigma; \sigma \sigma } + 4 C_{u\sigma \sigma \ab } C_{u\sigma \sigma }{}^{\ab })
	  + (\rbar^4 - 6 \rbar^2 \sbar^2 - 3 \sbar^4)(4 C_{u\sigma u\ab } C_{u\sigma u}{}^{\ab }
\nonumber \\
&&      + 3 C_{u\sigma u\sigma; uu})
      + \rbar (\rbar^2 - 3 \sbar^2) (16 C_{u\sigma u}{}^{\ab } C_{u\sigma \sigma \ab } - 3 C_{u\sigma u\sigma; u\sigma })
    \Big]
    + \sbar^4 \Big[
	    2 C_{u}{}^{\ab }{}_{u}{}^{\bb } [ (\rbar^2 + \sbar^2) C_{\sigma \ab \sigma \bb }\nonumber \\
&&      + 2 \rbar (\rbar^2 + 3 \sbar^2) C_{u\ab \sigma \bb }]
      + 2 C_{u}{}^{\ab }{}_{\sigma }{}^{\bb } [2 \rbar C_{\sigma \ab \sigma \bb } + (\rbar^2 + \sbar^2) C_{u\bb \sigma \ab }]
      + [\rbar^4 + 6 \rbar^2 \sbar^2 + \sbar^4] C_{u\ab u\bb } C_{u}{}^{\ab }{}_{u}{}^{\bb }
\nonumber \\
&&    + [\rbar^2 + \sbar^2] 2 C_{u\ab \sigma \bb } C_{u}{}^{\ab }{}_{\sigma }{}^{\bb }
      + C_{\sigma \ab \sigma \bb } C_{\sigma }{}^{\ab }{}_{\sigma }{}^{\bb }
    \Big]
\end{IEEEeqnarray}
and
\begin{IEEEeqnarray}{rCl}
\left[\PhiS\right]_{(4)} &=&
      30 C_{u\sigma u\sigma } \Big[
          \rbar (3 \rbar^4 - 10 \rbar^2 \sbar^2 + 15 \sbar^4) C_{u\sigma u\sigma; u}
        - 30 (\rbar^2 - \sbar^2)^2 C_{u\sigma u\sigma ;\sigma }
        \Big]
    + 2 \sbar^2 \Big[
        3 C_{u\sigma u\sigma; uuu} \rbar [\rbar^4 - 10 \rbar^2 \sbar^2 - 15 \sbar^4]
\nonumber \\
&&     +\: 3  \rbar [\rbar^2 - 3 \sbar^2] C_{u\sigma u\sigma; u\sigma \sigma }
      - 3 [\rbar^2 -  \sbar^2] C_{u\sigma u\sigma; \sigma \sigma \sigma }
      - 3 [\rbar^4 - 6 \rbar^2 \sbar^2 - 3 \sbar^4] C_{u\sigma u\sigma; uu\sigma }
      - 9 \rbar [\rbar^4 - 10 \rbar^2 \sbar^2
\nonumber \\
&&      - 15 \sbar^4] C_{u\sigma u\ab; u}
      - C_{u\sigma \sigma }{}^{\ab } [
          18 C_{u\sigma \sigma \ab; \sigma } (\rbar^2 - \sbar^2)
        - \rbar (\rbar^2 - 3 \sbar^2) (
              10 C_{u\sigma \sigma \ab; u}
            - 16 C_{u\sigma u\ab; \sigma }
            - 5 C_{u\sigma u\sigma; \ab })
\nonumber \\
&&      - 30 C_{u\sigma u\ab; u} (\rbar^4 - 6 \rbar^2 \sbar^2 - 3 \sbar^4)]
      - C_{u\sigma u}{}^{\ab } [
          36 \rbar (\rbar^2 - 3 \sbar^2) C_{u\sigma \sigma \ab; \sigma }
        + (\rbar^4 - 6 \rbar^2 \sbar^2 - 3 \sbar^4) (13 C_{u\sigma u\ab; \sigma } + 5 C_{u\sigma u\sigma; \ab }
\nonumber \\
&&        - 25 C_{u\sigma \sigma \ab; u} )]
    \Big]
    - 12 \sbar^4 \Big[
        C_{\sigma }{}^{\ab }{}_{\sigma }{}^{\bb } [
          C_{\sigma \ab \sigma \bb; \sigma }
        - \rbar ( C_{\sigma \ab \sigma \bb; u} - 2 C_{u\ab \sigma \bb ;\sigma })
        - (\rbar^2 + \sbar^2) (2 C_{u\ab \sigma \bb; u} - C_{u\ab u\bb; \sigma })
\nonumber \\
&&      - \rbar (\rbar^2 + 3 \sbar^2) C_{u\ab u\bb; u}]
      + 2 C_{u}{}^{\ab }{}_{\sigma }{}^{\bb } [
          C_{\sigma \ab \sigma \bb; \sigma } \rbar
        + (\rbar^2 + \sbar^2)(C_{u\ab \sigma \bb ;\sigma } + C_{u\bb \sigma \ab ;\sigma } - C_{\sigma \ab \sigma \bb; u})
\nonumber \\
&&      + \rbar (\rbar^2 + 3 \sbar^2)(C_{u\ab u\bb ;\sigma } - C_{u\ab \sigma \bb; u} - C_{u\bb \sigma \ab ;u})
        - C_{u\ab u\bb u} (\rbar^4 + 6 \rbar^2 \sbar^2 + \sbar^4)]
      + C_{u}{}^{\ab }{}_{u}{}^{\bb } [
          (\rbar^2 + \sbar^2) C_{\sigma \ab \sigma \bb; \sigma }
 \nonumber \\
&&       - \rbar (\rbar^2 + 3 \sbar^2) (C_{\sigma \ab \sigma \bb; u} - 2 C_{u\ab \sigma \bb; \sigma })
        + (\rbar^4 + 6 \rbar^2 \sbar^2 + \sbar^4) (C_{u\ab u\bb; \sigma } - 2 C_{u\ab \sigma \bb; u})
\nonumber \\
&&      - C_{u\ab u\bb; u} \rbar (\rbar^4 + 10 \rbar^2 \sbar^2 + 5 \sbar^4)]
    \Big].
\end{IEEEeqnarray}

\subsection{Electromagnetic singular field}
To $\mathcal{O}(\epsilon^4)$, the electromagnetic singular field is
\begin{IEEEeqnarray}{rCl} \label{eq:AS-approx}
A_{a}^{\rm \sing} &=&
e g_{a}{}^\ab \Big\{
    \frac{u_\ab}{\sbar}
  + \frac{1}{6 \sbar^3} \Big[
    3 \rbar \sbar^2 C_{\ab u u \sigma}
    + C_{u \sigma u \sigma} (\rbar^2 - \sbar^2) u_\ab
    \Big]
  + \frac{1}{24 \sbar^3} \Big[
      4 \sbar^2 (C_{\ab uu \sigma ;u} - \rbar C_{\ab uu \sigma ; \sigma})
\nonumber \\
&&  +\: \Big( \rbar (\rbar^2-3\sbar^2) C_{u \sigma u \sigma ; u} - (\rbar^2 - \sbar^2) C_{u \sigma u \sigma \sigma}\Big) u_\ab \Big]
  + \frac{1}{2880 \sbar^5} \left[A_\ab^{\rm \sing}\right]_{(3)}
  + \frac{1}{25920 \sbar^5} \left[A_\ab^{\rm \sing}\right]_{(4)}
  + \mathcal{O}(\epsilon^5)
\Big\},
\end{IEEEeqnarray}
where
\begin{IEEEeqnarray}{rCl}
[A_{\ab}^{\rm \sing}]_{(3)} &=& 
120 C_{u\sigma u\ab }{}_{;}{}_{u\sigma } \sbar^4 (\rbar^2 + \sbar^2) - 120 C_{u\sigma u\ab }{}_{;}{}_{\sigma \sigma } \rbar \sbar^4 + 120 C_{u}{}^{\cb }{}_{u}{}^{\db } C_{\sigma \cb \ab \db } \rbar \sbar^6 - 240 C_{u\sigma u\ab } C_{u\sigma u\sigma } \rbar \sbar^2 (\rbar^2 - 3 \sbar^2)\nonumber \\ 
&& -\: 360 C_{u\sigma \ab \cb } C_{u\sigma u}{}^{\cb } \sbar^4 (\rbar^2 + \sbar^2) - 120 C_{u\sigma u\ab }{}_{;}{}_{uu} \rbar \sbar^4 (\rbar^2 + 3 \sbar^2) + 120 \sbar^6 C_{u}{}^{\cb }{}_{\sigma }{}^{\db } (C_{u\db \ab \cb } \rbar + C_{\sigma \db \ab \cb }) \nonumber \\ 
&& +\: 40 C_{u\cb \ab \db } C_{u}{}^{\cb }{}_{u}{}^{\db } \sbar^6 (3 \rbar^2 + \sbar^2) - 120 C_{u\ab \sigma \cb } [C_{u\sigma \sigma }{}^{\cb } \rbar \sbar^4 + 2 C_{u\sigma u}{}^{\cb } \sbar^4 (\rbar^2 + \sbar^2)] - 120 C_{u\ab u\cb } [ C_{u\sigma \sigma }{}^{\cb } \sbar^4 (\rbar^2 + \sbar^2) \nonumber \\ 
&& +\: C_{u\sigma u}{}^{\cb } \rbar \sbar^4 (\rbar^2 + 3 \sbar^2)] + \{8 C_{\sigma \cb \sigma \db } C_{\sigma }{}^{\cb }{}_{\sigma }{}^{\db } \sbar^4 - 5 C_{\cb \db \eb \fb } C^{\cb \db \eb \fb } \sbar^8 - 24 C_{u\sigma u\sigma }{}_{;}{}_{u\sigma } \rbar \sbar^2 (\rbar^2 - 3 \sbar^2) \nonumber \\ 
&& +\: 128 C_{u\sigma u}{}^{\cb } C_{u\sigma \sigma \cb } \rbar \sbar^2 (\rbar^2 - 3 \sbar^2) + 24 C_{u\sigma u\sigma }{}_{;}{}_{\sigma \sigma } \sbar^2 (\rbar^2 -  \sbar^2) + 32 C_{u\sigma \sigma \cb } C_{u\sigma \sigma }{}^{\cb } \sbar^2 (\rbar^2 -  \sbar^2) \nonumber \\ 
&& +\: 120 C_{u\sigma u\sigma } C_{u\sigma u\sigma } (\rbar^2 -  \sbar^2)^2 + 16 C_{u}{}^{\cb }{}_{u}{}^{\db } C_{\sigma \cb \sigma \db } \sbar^4 (\rbar^2 + \sbar^2) + 24 C_{u\sigma u\sigma }{}_{;}{}_{uu} \sbar^2 (\rbar^4 - 6 \rbar^2 \sbar^2 - 3 \sbar^4) \nonumber \\ 
&& +\: 32 C_{u\sigma u\cb } C_{u\sigma u}{}^{\cb } \sbar^2 (\rbar^4 - 6 \rbar^2 \sbar^2 - 3 \sbar^4) + 8 C_{u\cb u\db } C_{u}{}^{\cb }{}_{u}{}^{\db } \sbar^4 (\rbar^4 + 6 \rbar^2 \sbar^2 + \sbar^4) + 16 \sbar^4 C_{u}{}^{\cb }{}_{\sigma }{}^{\db } [2 C_{\sigma \cb \sigma \db } \rbar \nonumber \\ 
&& +\: C_{u\db \sigma \cb } (\rbar^2 + \sbar^2)] + 16 \sbar^4 C_{u\cb \sigma \db } [C_{u}{}^{\cb }{}_{\sigma }{}^{\db } (\rbar^2 + \sbar^2) + 2 C_{u}{}^{\cb }{}_{u}{}^{\db } \rbar (\rbar^2 + 3 \sbar^2)]\} u_{\ab }
\end{IEEEeqnarray}
and 
\begin{IEEEeqnarray}{rCl}
[A_\ab^{\rm \sing}]_{(4)} &=&
216 C_{u\sigma u\ab }{}_{;}{}_{\sigma \sigma \sigma } \rbar \sbar^4 + 540 C_{u\sigma u\sigma }{}_{;}{}_{\sigma } C_{u\sigma u\ab } \rbar \sbar^2 (\rbar^2 - 3 \sbar^2) + 720 C_{u\sigma u\ab }{}_{;}{}_{\sigma } C_{u\sigma u\sigma } \rbar \sbar^2 (\rbar^2 - 3 \sbar^2) \nonumber \\ 
&& -\: 216 C_{u\sigma u\ab }{}_{;}{}_{u\sigma \sigma } \sbar^4 (\rbar^2 + \sbar^2) + 1512 C_{u\sigma \ab }{}^{\cb }{}_{;}{}_{\sigma } C_{u\sigma u\cb } \sbar^4 (\rbar^2 + \sbar^2) + 216 C_{u\sigma u\ab }{}_{;}{}_{uu\sigma } \rbar \sbar^4 (\rbar^2 + 3 \sbar^2) \nonumber \\ 
&& -\: 864 C_{u\sigma \ab }{}^{\cb }{}_{;}{}_{u} C_{u\sigma u\cb } \rbar \sbar^4 (\rbar^2 + 3 \sbar^2) - 216 C_{u\sigma u\ab }{}_{;}{}_{uuu} \sbar^4 (\rbar^4 + 6 \rbar^2 \sbar^2 + \sbar^4) - 540 C_{u\sigma u\sigma }{}_{;}{}_{u} C_{u\sigma u\ab } \sbar^2 (\rbar^4 \nonumber \\ 
&& -\: 6 \rbar^2 \sbar^2 - 3 \sbar^4) + 720 C_{u\sigma u\ab }{}_{;}{}_{u} C_{u\sigma u\sigma } \sbar^2 (- \rbar^4 + 6 \rbar^2 \sbar^2 + 3 \sbar^4) - 432 \sbar^6 C_{\sigma }{}^{\cb }{}_{\ab }{}^{\db }{}_{;}{}_{\sigma } ( C_{u\cb u\db } \rbar + C_{u\db \sigma \cb })\nonumber \\ 
&& -\: 648 \sbar^6 C_{u}{}^{\cb }{}_{\sigma }{}^{\db }{}_{;}{}_{\sigma } ( C_{u\db \ab \cb } \rbar + C_{\sigma \db \ab \cb } ) + 16 \sbar^8 C_{u\ab }{}^{\cb \db }{}^{;}{}^{\eb } ( C_{u\cb \db \eb } \rbar - C_{u\eb \cb \db } \rbar + C_{\sigma \cb \db \eb } - C_{\sigma \eb \cb \db }) \nonumber \\ 
&& +\: 432 \sbar^4 C_{u\sigma \sigma }{}^{\cb }{}_{;}{}_{\sigma } [ C_{u\ab \sigma \cb } \rbar + C_{u\ab u\cb } (\rbar^2 + \sbar^2)] + 504 \sbar^4 C_{u\ab \sigma }{}^{\cb }{}_{;}{}_{\sigma } [ C_{u\sigma \sigma \cb } \rbar + 2 C_{u\sigma u\cb } (\rbar^2 + \sbar^2)] \nonumber \\ 
&& -\: 324 \sbar^4 C_{u\sigma \sigma }{}^{\cb }{}_{;}{}_{u} [ C_{u\ab \sigma \cb } (\rbar^2 + \sbar^2) + C_{u\ab u\cb } \rbar (\rbar^2 + 3 \sbar^2)] + 108 \sbar^4 C_{u\sigma u}{}^{\cb }{}_{;}{}_{\sigma } [5 C_{u\ab \sigma \cb } (\rbar^2 + \sbar^2) + 6 C_{u\sigma \ab \cb } (\rbar^2 + \sbar^2) \nonumber \\ 
&& +\: 3 C_{u\ab u\cb } \rbar (\rbar^2 + 3 \sbar^2)] - 96 \sbar^4 C_{u\ab \sigma }{}^{\cb }{}_{;}{}_{u} [3 C_{u\sigma \sigma \cb } (\rbar^2 + \sbar^2) + 8 C_{u\sigma u\cb } \rbar (\rbar^2 + 3 \sbar^2)] + 96 \sbar^4 C_{u\sigma u\ab }{}^{;}{}^{\cb } [3 C_{u\sigma \sigma \cb } (\rbar^2 + \sbar^2) \nonumber \\ 
&& +\: 4 C_{u\sigma u\cb } \rbar (\rbar^2 + 3 \sbar^2)] + 24 \sbar^4 C_{u\ab u}{}^{\cb }{}_{;}{}_{\sigma } [9 C_{u\sigma \sigma \cb } (\rbar^2 + \sbar^2) + 17 C_{u\sigma u\cb } \rbar (\rbar^2 + 3 \sbar^2)] - 216 \sbar^4 C_{u\sigma u}{}^{\cb }{}_{;}{}_{u} [3 C_{u\ab \sigma \cb } \rbar (\rbar^2 \nonumber \\ 
&& +\: 3 \sbar^2) + 6 C_{u\sigma \ab \cb } \rbar (\rbar^2 + 3 \sbar^2) + 2 C_{u\ab u\cb } \sbar^4 (\rbar^4 + 6 \rbar^2 \sbar^2 + \sbar^4)]- 72 \sbar^4 C_{u\ab u}{}^{\cb }{}_{;}{}_{u} [8 C_{u\sigma \sigma \cb } \rbar (\rbar^2 + 3 \sbar^2) \nonumber \\ 
&& +\: 7 C_{u\sigma u\cb } (\rbar^4 + 6 \rbar^2 \sbar^2 + \sbar^4)] - 216 \sbar^6 C_{u}{}^{\cb }{}_{u}{}^{\db }{}_{;}{}_{\sigma } [3 C_{\sigma \cb \ab \db } \rbar + C_{u\cb \ab \db } (3 \rbar^2 + \sbar^2)] - 144 \sbar^6 C_{u}{}^{\cb }{}_{\ab }{}^{\db }{}_{;}{}_{\sigma } [3 C_{u\db \sigma \cb } \rbar \nonumber \\
&& +\: C_{u\cb u\db } (3 \rbar^2 + \sbar^2)] + 48 \sbar^6 C_{u\ab \sigma }{}^{\cb }{}^{;}{}^{\db } [3 C_{u\cb \sigma \db } \rbar + 3 C_{u\db \sigma \cb } \rbar + 3 C_{\sigma \cb \sigma \db } + C_{u\cb u\db } (3 \rbar^2 + \sbar^2)] \nonumber \\ 
&& +\: 216 \sbar^6 C_{\sigma }{}^{\cb }{}_{\ab }{}^{\db }{}_{;}{}_{u} [3 C_{u\db \sigma \cb } \rbar + C_{u\cb u\db } (3 \rbar^2 + \sbar^2)] + 144 \sbar^6 C_{u}{}^{\cb }{}_{\sigma }{}^{\db }{}_{;}{}_{u} [3 C_{\sigma \db \ab \cb } \rbar + C_{u\db \ab \cb } (3 \rbar^2 + \sbar^2)] \nonumber \\ 
&& +\: 48 \sbar^6 C_{u\ab u}{}^{\cb }{}^{;}{}^{\db } [3 \rbar C_{\sigma \cb \sigma \db } + 3 \rbar C_{u\cb u\db } (\rbar^2 + \sbar^2) + C_{u\cb \sigma \db } (3 \rbar^2 + \sbar^2) + C_{u\db \sigma \cb } (3 \rbar^2 + \sbar^2)] \nonumber \\ 
&& +\: 216 \sbar^6 C_{u}{}^{\cb }{}_{\ab }{}^{\db }{}_{;}{}_{u} [3 C_{u\cb u\db } \rbar (\rbar^2 + \sbar^2) + C_{u\db \sigma \cb } (3 \rbar^2 + \sbar^2)] + 144 \sbar^6 C_{u}{}^{\cb }{}_{u}{}^{\db }{}_{;}{}_{u} [3 C_{u\cb \ab \db } \rbar (\rbar^2 + \sbar^2) + C_{\sigma \cb \ab \db } (3 \rbar^2 + \sbar^2)] \nonumber \\ 
&& +\: \{45 C^{\cb \db \eb \fb }{}_{\sigma } C_{\cb \db \eb \fb } \sbar^8 - 45 C^{\cb \db \eb \fb }{}_{u} C_{\cb \db \eb \fb } \rbar \sbar^8 + 36 C_{u\sigma u\sigma }{}_{;}{}_{u\sigma \sigma } \rbar \sbar^2 (\rbar^2 - 3 \sbar^2) - 540 C_{u\sigma u\sigma }{}_{;}{}_{\sigma } C_{u\sigma u\sigma } (\rbar^2 -  \sbar^2)^2 \nonumber \\ 
&& -\: 36 C_{u\sigma u\sigma }{}_{;}{}_{\sigma \sigma \sigma } \sbar^2 (\rbar^2 - \sbar^2) + 36 C_{u\sigma u\sigma }{}_{;}{}_{uuu} \rbar \sbar^2 (\rbar^4 - 10 \rbar^2 \sbar^2 - 15 \sbar^4) - 36 C_{u\sigma u\sigma }{}_{;}{}_{uu\sigma } \sbar^2 (\rbar^4 - 6 \rbar^2 \sbar^2 - 3 \sbar^4) \nonumber \\ 
&& +\: 180 C_{u\sigma u\sigma }{}_{;}{}_{u} C_{u\sigma u\sigma } (3 \rbar^5 - 10 \rbar^3 \sbar^2 + 15 \rbar \sbar^4) - 216 \sbar^2 C_{u\sigma \sigma }{}^{\cb }{}_{;}{}_{\sigma } [2 \rbar C_{u\sigma u\cb } (\rbar^2 - 3 \sbar^2) - 2 C_{u\sigma \sigma \cb } (- \rbar^2 + \sbar^2)] \nonumber \\ 
&& -\: 72 \sbar^4 C_{\sigma }{}^{\cb }{}_{\sigma }{}^{\db }{}_{;}{}_{\sigma } [2 C_{u\cb \sigma \db } \rbar + C_{\sigma \cb \sigma \db } + C_{u\cb u\db } (\rbar^2 + \sbar^2)] - 144 \sbar^4 C_{u}{}^{\cb }{}_{\sigma }{}^{\db }{}_{;}{}_{\sigma } [C_{\sigma \cb \sigma \db } \rbar + C_{u\cb \sigma \db } (\rbar^2 + \sbar^2) \nonumber \\ 
&& +\: C_{u\db \sigma \cb } (\rbar^2 + \sbar^2) + C_{u\cb u\db } \rbar (\rbar^2 + 3 \sbar^2)] + 72 \sbar^4 C_{\sigma }{}^{\cb }{}_{\sigma }{}^{\db }{}_{;}{}_{u} [C_{\sigma \cb \sigma \db } \rbar + 2 C_{u\cb \sigma \db } (\rbar^2 + \sbar^2) + C_{u\cb u\db } \rbar (\rbar^2 + 3 \sbar^2)] \nonumber \\ 
&& +\: 60 \sbar^2 C_{u\sigma \sigma }{}^{\cb }{}_{;}{}_{u} [2 C_{u\sigma \sigma \cb } \rbar (\rbar^2 - 3 \sbar^2) + 5 C_{u\sigma u\cb } (\rbar^4 - 6 \rbar^2 \sbar^2 - 3 \sbar^4)] + 72 \sbar^2 C_{u\sigma u}{}^{\cb }{}_{;}{}_{u} [3 C_{u\sigma u\cb } \rbar (\rbar^4 - 10 \rbar^2 \sbar^2 \nonumber \\ 
&& -\: 15 \sbar^4) + 5 C_{u\sigma \sigma \cb } (\rbar^4 - 6 \rbar^2 \sbar^2 - 3 \sbar^4)] - 72 \sbar^4 C_{u}{}^{\cb }{}_{u}{}^{\db }{}_{;}{}_{\sigma } [C_{\sigma \cb \sigma \db } (\rbar^2 + \sbar^2) + 2 C_{u\cb \sigma \db } \rbar (\rbar^2 + 3 \sbar^2) \nonumber \\ 
&& +\: C_{u\cb u\db } (\rbar^4 + 6 \rbar^2 \sbar^2 + \sbar^4)] + 144 \sbar^4 C_{u}{}^{\cb }{}_{\sigma }{}^{\db }{}_{;}{}_{u} [C_{\sigma \cb \sigma \db } (\rbar^2 + \sbar^2) + C_{u\cb \sigma \db } \rbar (\rbar^2 + 3 \sbar^2) + C_{u\db \sigma \cb } \rbar (\rbar^2 + 3 \sbar^2) \nonumber \\ 
&& +\: C_{u\cb u\db } (\rbar^4 + 6 \rbar^2 \sbar^2 + \sbar^4)] - 60 \sbar^2 C_{u\sigma u\sigma }{}^{;}{}^{\cb } [C_{u\sigma \sigma \cb } \rbar (\rbar^2 - 3 \sbar^2) + C_{u\sigma u\cb } (\rbar^4 - 6 \rbar^2 \sbar^2 - 3 \sbar^4)] \nonumber \\ 
&& -\: 12 \sbar^2 C_{u\sigma u}{}^{\cb }{}_{;}{}_{\sigma } [16 C_{u\sigma \sigma \cb } \rbar (\rbar^2 - 3 \sbar^2) + 13 C_{u\sigma u\cb } (\rbar^4 - 6 \rbar^2 \sbar^2 - 3 \sbar^4)] + 72 \sbar^4 C_{u}{}^{\cb }{}_{u}{}^{\db }{}_{;}{}_{u} [C_{\sigma \cb \sigma \db } \rbar (\rbar^2 + 3 \sbar^2) \nonumber \\ 
&& +\: 2 C_{u\cb \sigma \db } (\rbar^4 + 6 \rbar^2 \sbar^2 + \sbar^4) + C_{u\cb u\db } \rbar (\rbar^4 + 10 \rbar^2 \sbar^2 + 5 \sbar^4)]\} u_{\ab }.
\end{IEEEeqnarray}

\subsection{Gravitational singular field}
To $\mathcal{O}(\epsilon^4)$, the gravitational singular field is
\begin{IEEEeqnarray}{rCl} \label{eq:hS-approx}
\hb_{ab}^{\rm \sing} &=&
4 \mu g_{a}{}^\ab g_b{}^\bb \Big\{
    \frac{u_\ab u_\bb}{\sbar}
  + \frac{1}{6 \sbar^3} \Big[
      (\rbar^2 - \sbar^2) C_{u \sigma u \sigma} u_\ab u_\bb
    - 6 \rbar \sbar^2 C_{ u \sigma u (\ab} u_{\bb)}
    - 6 \sbar^4 C_{\ab u \bb u}
    \Big]
  + \frac{1}{24 \sbar^3} \Big[
      12 \sbar^4 (C_{\ab u \bb u ;\sigma} - \rbar C_{\ab u \bb u ; u})
\nonumber \\
&&  +\: 8 \sbar^2 \Big(u_{(\ab}  C_{\bb) uu \sigma; u} (\rbar^2 + \sbar^2) - \rbar u_{(\ab}  C_{\bb) u u \sigma ; \sigma}\Big)
    + u_\ab u_\bb \Big( \rbar (\rbar^2 - 3\sbar^2) C_{u \sigma u \sigma ; u}  - (\rbar^2-\sbar^2) C_{u \sigma u \sigma ; \sigma} \Big)
  \Big]
\nonumber \\
&&+ \frac{1}{1440 \sbar^5} \left[\hb_{\ab \bb}^{\rm \sing}\right]_{(3)}
  + \frac{1}{6480 \sbar^5} \left[\hb_{\ab \bb}^{\rm \sing}\right]_{(4)}
  + \mathcal{O}(\epsilon^5)
\Big\},
\end{IEEEeqnarray}
where
\begin{IEEEeqnarray}{rCl}
\Big[\hb_{\ab \bb}^{\rm \sing}&&\Big]_{(3)} =
-240 C_{\ab u\bb u}{}_{;}{}_{\sigma \sigma } \sbar^6 + 240 C_{\ab u\bb u}{}_{;}{}_{u\sigma } \rbar \sbar^6 + 480  C_{\ab }{}^{\cb }{}_{u\sigma } C_{\bb uu\cb } \rbar \sbar^6 - 120 C_{\ab u\sigma }{}^{\cb } C_{\bb u\sigma \cb } \sbar^6 + 960 C_{\ab u\bb }{}^{\cb } C_{u\sigma u\cb } \rbar  \sbar^6 \nonumber \\ 
&& + 40 C_{\ab u\bb u}{}^{;}{}^{\cb }{}_{\cb } \sbar^8 + 20 C_{\ab u}{}^{\cb \db } C_{\bb u\cb \db } \sbar^8 + 240 C_{\ab }{}^{\cb }{}_{\bb }{}^{\db } C_{u\cb u\db } \sbar^8 + 360 C_{\ab uu\sigma } C_{\bb uu\sigma } \sbar^4 (\rbar^2 + \sbar^2) \nonumber \\ 
&& + 240 C_{\ab u\bb u} C_{u\sigma u\sigma } \sbar^4 (\rbar^2 + \sbar^2) - 80 C_{\ab u\bb u}{}_{;}{}_{uu} \sbar^6 (3 \rbar^2  + \sbar^2) -40 C_{\ab uu}{}^{\cb } \sbar^6 [6  C_{\bb u\sigma \cb } \rbar + C_{\bb uu\cb } (3 \rbar^2 + \sbar^2)] \nonumber \\ 
&& + \{120 C_{\bb uu\sigma }{}_{;}{}_{\sigma \sigma } \rbar \sbar^4 + 240  C_{\bb uu\sigma } C_{u\sigma u\sigma } \rbar \sbar^2 (\rbar^2 - 3 \sbar^2) - 120 C_{\bb uu\sigma }{}_{;}{}_{u\sigma } \sbar^4 (\rbar^2 + \sbar^2) - 360 C_{\bb }{}^{\cb }{}_{u\sigma } C_{u\sigma u\cb } \sbar^4 (\rbar^2 + \sbar^2) \nonumber \\ 
&& + 120 C_{\bb uu\sigma }{}_{;}{}_{uu} \rbar \sbar^4 (\rbar^2 + 3 \sbar^2) + 120 C_{\bb }{}^{\cb }{}_{\sigma }{}^{\db } [ C_{u\cb u\db } \rbar +  C_{u\cb \sigma \db } ] \sbar^6 + 120 C_{\bb u\sigma }{}^{\cb } [ C_{u\sigma \sigma \cb }\rbar + 2 C_{u\sigma u\cb }  (\rbar^2 + \sbar^2)] \sbar^4 \nonumber \\ 
&& + 40 C_{\bb }{}^{\cb }{}_{u}{}^{\db } [3  C_{u\cb \sigma \db } \rbar  +  C_{u\cb u\db }  (3 \rbar^2 + \sbar^2)] \sbar^6+ 120 C_{\bb uu}{}^{\cb } [C_{u\sigma \sigma \cb }  (\rbar^2 + \sbar^2) +  C_{u\sigma u\cb } \rbar (\rbar^2 + 3 \sbar^2)]\sbar^4\} u_{\ab } \nonumber \\ 
&& + \{4 C_{\sigma \cb \sigma \db } C_{\sigma }{}^{\cb }{}_{\sigma }{}^{\db } \sbar^4 - 5 C_{\cb \db \eb \fb} C^{\cb \db \eb \fb} \sbar^8 + 12 C_{u\sigma u\sigma }{}_{;}{}_{\sigma \sigma } \sbar^2 (\rbar^2 - \sbar^2) 
+ 16 C_{u\sigma \sigma \cb } C_{u\sigma \sigma }{}^{\cb } \sbar^2  (\rbar^2 - \sbar^2)\nonumber \\ 
&&  - 12 C_{u\sigma u\sigma }{}_{;}{}_{u\sigma } \rbar \sbar^2 (\rbar^2 - 3 \sbar^2) + 64  C_{u\sigma u}{}^{\cb } C_{u\sigma \sigma \cb } \rbar\sbar^2 (\rbar^2 - 3 \sbar^2) + 60 C_{u\sigma u\sigma }^2 (\rbar^2 -  \sbar^2)^2 + 8 C_{u}{}^{\cb }{}_{u}{}^{\db } C_{\sigma \cb \sigma \db } \sbar^4 (\rbar^2 + \sbar^2) \nonumber \\ 
&&  + 12 C_{u\sigma u\sigma }{}_{;}{}_{uu} \sbar^2 (\rbar^4 - 6 \rbar^2 \sbar^2 - 3 \sbar^4)  + 16 C_{u\sigma u\cb } C_{u\sigma u}{}^{\cb } \sbar^2 (\rbar^4 - 6 \rbar^2 \sbar^2 - 3 \sbar^4) + 4 C_{u\cb u\db } C_{u}{}^{\cb }{}_{u}{}^{\db } \sbar^4 (\rbar^4 + 6 \rbar^2 \sbar^2 + \sbar^4) \nonumber \\ 
&& + 8 C_{u}{}^{\cb }{}_{\sigma }{}^{\db } [2  C_{\sigma \cb \sigma \db } \rbar+  C_{u\db \sigma \cb }  (\rbar^2 + \sbar^2)] \sbar^4+ 8 C_{u\cb \sigma \db } [C_{u}{}^{\cb }{}_{\sigma }{}^{\db }  (\rbar^2 + \sbar^2) + 2 C_{u}{}^{\cb }{}_{u}{}^{\db } \rbar  (\rbar^2 + 3 \sbar^2)]\sbar^4\} u_{\ab } u_{\bb }
\end{IEEEeqnarray}
and
\begin{IEEEeqnarray}{rCl}
\Big[\hb_{\ab \bb}^{\rm \sing}&&\Big]_{(4)} =
270 (C_{\ab u\bb u}{}_{;}{}_{\sigma \sigma \sigma } -  C_{\ab u\bb u}{}_{;}{}_{u\sigma \sigma } \rbar )\sbar^6 - 1080 (C_{u\sigma u}{}^{\cb }{}_{;}{}_{\sigma }  C_{\ab u\bb \cb } + C_{\ab }{}^{\cb }{}_{u\sigma }{}_{;}{}_{\sigma } C_{\bb uu\cb } ) \rbar\sbar^6 - 2700 C_{\ab u\bb }{}^{\cb }{}_{;}{}_{\sigma } \rbar C_{u\sigma u\cb } \sbar^6  \nonumber \\ 
&& - 90 C_{\ab u\bb u}{}^{;}{}^{\cb }{}_{\cb \sigma } \sbar^8+ 90 C_{\ab u\bb u}{}^{;}{}^{\cb }{}_{\cb u} \rbar \sbar^8 - 360 C_{u}{}^{\cb }{}_{u}{}^{\db }{}_{;}{}_{\sigma } C_{\ab \cb \bb \db } \sbar^8 + 720 C_{u}{}^{\cb }{}_{u}{}^{\db }{}_{;}{}_{u} \rbar C_{\ab \cb \bb \db } \sbar^8 + 180 C_{u}{}^{\cb }{}_{\sigma }{}^{\db }{}_{;}{}_{\db } C_{\ab u\bb \cb } \sbar^8 \nonumber \\ 
&&+ 180 C_{u}{}^{\cb }{}_{u}{}^{\db }{}_{;}{}_{\cb } \rbar C_{\ab u\bb \db } \sbar^8 - 120 C_{\ab u}{}^{\cb \db }{}_{;}{}_{\sigma } C_{\bb u\cb \db } \sbar^8  - 60 C_{\ab u\sigma }{}^{\cb }{}^{;}{}^{\db } C_{\bb u\cb \db } \sbar^8 + 60 C_{\ab u}{}^{\cb \db }{}_{;}{}_{u} \rbar C_{\bb u\cb \db } \sbar^8 
- 180 C_{\ab }{}^{\cb }{}_{\sigma }{}^{\db }{}_{;}{}_{\db } C_{\bb uu\cb } \sbar^8\nonumber \\ 
&& - 180 C_{\ab }{}^{\cb }{}_{u}{}^{\db }{}_{;}{}_{\db } \rbar C_{\bb uu\cb } \sbar^8 - 720 C_{\ab }{}^{\cb }{}_{\bb }{}^{\db }{}_{;}{}_{\sigma } C_{u\cb u\db } \sbar^8 + 360 C_{\ab }{}^{\cb }{}_{\bb }{}^{\db }{}_{;}{}_{u} \rbar C_{u\cb u\db } \sbar^8 - 270 C_{u\sigma u\sigma }{}_{;}{}_{\sigma } C_{\ab u\bb u} \sbar^4 (\rbar^2 + \sbar^2) \nonumber \\ 
&& - 1080 C_{\ab uu\sigma }{}_{;}{}_{\sigma } C_{\bb uu\sigma } \sbar^4 (\rbar^2 + \sbar^2) - 540 C_{\ab u\bb u}{}_{;}{}_{\sigma } C_{u\sigma u\sigma } \sbar^4 (\rbar^2 + \sbar^2)  - 270 C_{\ab u\bb u}{}_{;}{}_{uuu} \rbar \sbar^6 (\rbar^2 + \sbar^2)\nonumber \\ 
&& + 270 (C_{u\sigma u\sigma }{}_{;}{}_{u}  C_{\ab u\bb u}  + 4 C_{\ab uu\sigma }{}_{;}{}_{u}  C_{\bb uu\sigma }  + 2 C_{\ab u\bb u}{}_{;}{}_{u}  C_{u\sigma u\sigma } )\rbar \sbar^4 (\rbar^2 + 3 \sbar^2)  + 540 C_{\ab u\sigma }{}^{\cb }{}_{;}{}_{\sigma } ( C_{\bb uu\cb } \rbar  + C_{\bb u\sigma \cb } )\sbar^6 \nonumber \\ 
&&  + 90( C_{\ab u\bb u}{}_{;}{}_{uu\sigma }  + 6 C_{u\sigma u}{}^{\cb }{}_{;}{}_{u} C_{\ab u\bb \cb } 
+ 2 C_{\ab }{}^{\cb }{}_{u\sigma }{}_{;}{}_{u} C_{\bb uu\cb } + 6 C_{\ab u\bb }{}^{\cb }{}_{;}{}_{u} C_{u\sigma u\cb }) (3 \rbar^2  + \sbar^2) \sbar^6+ 60 C_{\ab uu}{}^{\cb }{}^{;}{}^{\db } (3  C_{\bb \cb u\db } \rbar\nonumber \\ 
&&    + 3 C_{\bb \cb \sigma \db }  -   C_{\bb u\cb \db } \rbar ) \sbar^8+ 180 C_{\ab u\bb }{}^{\cb }{}^{;}{}^{\db } ( C_{u\cb u\db } \rbar
+  C_{u\cb \sigma \db }) \sbar^8
-180 C_{\ab u\sigma }{}^{\cb }{}_{;}{}_{u} [ 3  C_{\bb u\sigma \cb }  \rbar+ C_{\bb uu\cb } (3 \rbar^2  + \sbar^2)] \sbar^6
\nonumber \\ 
&&     + 180 C_{\ab uu}{}^{\cb }{}_{;}{}_{\sigma } [-3 \rbar C_{\bb \cb u\sigma } + 3 \rbar C_{\bb u\sigma \cb } + C_{\bb uu\cb } (3 \rbar^2  + \sbar^2)]\sbar^6+ 90 C_{\ab u\bb u}{}^{;}{}^{\cb } [3  C_{u\sigma \sigma \cb } \rbar + 2 C_{u\sigma u\cb } (3 \rbar^2 + \sbar^2)]\sbar^6
\nonumber \\ 
&&  + 180 C_{\ab uu}{}^{\cb }{}_{;}{}_{u} [-3  C_{\bb uu\cb }  \rbar (\rbar^2 + \sbar^2)+ 3 C_{\bb \cb u\sigma } (3 \rbar^2 + \sbar^2) - C_{\bb u\sigma \cb } (3 \rbar^2 + \sbar^2)] \sbar^6\nonumber \\ 
&& + \big\{-108 C_{\bb uu\sigma }{}_{;}{}_{\sigma \sigma \sigma } \rbar \sbar^4 + 36 C_{u}{}^{\cb }{}_{\sigma }{}^{\db }{}^{;}{}^{\eb } C_{\bb \cb \db \eb } \sbar^8 + 36 C_{u}{}^{\cb }{}_{u}{}^{\db }{}^{;}{}^{\eb } \rbar C_{\bb \cb \db \eb } \sbar^8  + 12 C_{\bb }{}^{\cb \db \eb }{}_{;}{}_{\sigma } C_{u\cb \db \eb } \sbar^8 + 24 C_{\bb }{}^{\cb }{}_{\sigma }{}^{\db }{}^{;}{}^{\eb } C_{u\cb \db \eb } \sbar^8\nonumber \\ 
&& + 12 C_{\bb }{}^{\cb \db \eb }{}_{;}{}_{u}  C_{u\cb \db \eb } \rbar \sbar^8 + 24 C_{\bb }{}^{\cb }{}_{u}{}^{\db }{}^{;}{}^{\eb }  C_{u\cb \db \eb } \rbar \sbar^8 - 90(3 C_{u\sigma u\sigma }{}_{;}{}_{\sigma }  C_{\bb uu\sigma } +4 C_{\bb uu\sigma }{}_{;}{}_{\sigma } C_{u\sigma u\sigma }) \rbar\sbar^2 (\rbar^2 - 3 \sbar^2)\nonumber \\ 
&&  + 108 (C_{\bb uu\sigma }{}_{;}{}_{u\sigma \sigma }  + 7 C_{\bb }{}^{\cb }{}_{u\sigma }{}_{;}{}_{\sigma } C_{u\sigma u\cb }) \sbar^4 (\rbar^2 + \sbar^2) - 108 (C_{\bb uu\sigma }{}_{;}{}_{uu\sigma } +4  C_{\bb }{}^{\cb }{}_{u\sigma }{}_{;}{}_{u}  C_{u\sigma u\cb } )\rbar \sbar^4 (\rbar^2 + 3 \sbar^2)  \\ 
&&  + 90 (3 C_{u\sigma u\sigma }{}_{;}{}_{u} C_{\bb uu\sigma }  + 4 C_{\bb uu\sigma }{}_{;}{}_{u} C_{u\sigma u\sigma }) \sbar^2 (\rbar^4 - 6 \rbar^2 \sbar^2 - 3 \sbar^4) + 108 C_{\bb uu\sigma }{}_{;}{}_{uuu} \sbar^4 (\rbar^4 + 6 \rbar^2 \sbar^2  + \sbar^4)\nonumber \\ 
&& -216 C_{u}{}^{\cb }{}_{\sigma }{}^{\db }{}_{;}{}_{\sigma } ( C_{\bb \cb u\db } \rbar + C_{\bb \cb \sigma \db }) \sbar^6 -324 C_{\bb }{}^{\cb }{}_{\sigma }{}^{\db }{}_{;}{}_{\sigma } (C_{u\cb u\db }  \rbar  + C_{u\cb \sigma \db } )\sbar^6 + 8 C_{\bb u}{}^{\cb \db }{}^{;}{}^{\eb } ( C_{u\cb \db \eb } \rbar -   C_{u\eb \cb \db } \rbar   \nonumber \\ 
&& +  C_{\sigma \cb \db \eb }  -  C_{\sigma \eb \cb \db } ) \sbar^8-216 C_{u\sigma \sigma }{}^{\cb }{}_{;}{}_{\sigma } [  C_{\bb u\sigma \cb } \rbar+ C_{\bb uu\cb }  (\rbar^2 + \sbar^2)] \sbar^4
-252 C_{\bb u\sigma }{}^{\cb }{}_{;}{}_{\sigma } [ \rbar C_{u\sigma \sigma \cb }  +2 C_{u\sigma u\cb } (\rbar^2 + \sbar^2)]\sbar^4 \nonumber \\ 
&&+54  C_{u\sigma u}{}^{\cb }{}_{;}{}_{\sigma } [6 C_{\bb \cb u\sigma }  (\rbar^2 + \sbar^2) - 5 C_{\bb u\sigma \cb }  (\rbar^2 + \sbar^2) - 3 \rbar C_{\bb uu\cb }  (\rbar^2 + 3 \sbar^2)]\sbar^4   + 162 C_{u\sigma \sigma }{}^{\cb }{}_{;}{}_{u} [ C_{\bb u\sigma \cb }  (\rbar^2 + \sbar^2)   \nonumber \\ 
&&   + C_{\bb uu\cb } \rbar (\rbar^2  + 3 \sbar^2)] \sbar^4 +12 C_{\bb uu}{}^{\cb }{}_{;}{}_{\sigma } [-9 C_{u\sigma \sigma \cb }  (\rbar^2 + \sbar^2) - 17 \rbar C_{u\sigma u\cb }  (\rbar^2 + 3 \sbar^2)]\sbar^4 -48  C_{\bb uu\sigma }{}^{;}{}^{\cb } [3 C_{u\sigma \sigma \cb }  (\rbar^2+ \sbar^2) \nonumber \\ 
&&   +4 \rbar C_{u\sigma u\cb }  (\rbar^2 + 3 \sbar^2)] \sbar^4   + 48 C_{\bb u\sigma }{}^{\cb }{}_{;}{}_{u} [3 C_{u\sigma \sigma \cb } (\rbar^2 + \sbar^2) + 8  C_{u\sigma u\cb } \rbar (\rbar^2+ 3 \sbar^2)] \sbar^4   + 108 C_{u\sigma u}{}^{\cb }{}_{;}{}_{u} [  3 \rbar C_{\bb u\sigma \cb }  (\rbar^2 + 3 \sbar^2)  \nonumber \\ 
&& -6 C_{\bb \cb u\sigma } \rbar (\rbar^2 + 3 \sbar^2) + 2 C_{\bb uu\cb }  (\rbar^4+ 6 \rbar^2 \sbar^2 + \sbar^4)] \sbar^4  +36  C_{\bb uu}{}^{\cb }{}_{;}{}_{u} [8  C_{u\sigma \sigma \cb } \rbar (\rbar^2 + 3 \sbar^2) + 7 C_{u\sigma u\cb }  (\rbar^4 + 6 \rbar^2 \sbar^2+ \sbar^4)] \sbar^4 \nonumber \\ 
&&  -72 C_{u}{}^{\cb }{}_{u}{}^{\db }{}_{;}{}_{\sigma } [3 C_{\bb \cb \sigma \db } \rbar  + C_{\bb \cb u\db } (3 \rbar^2  + \sbar^2)] \sbar^6+ 108 C_{u}{}^{\cb }{}_{\sigma }{}^{\db }{}_{;}{}_{u} [3 C_{\bb \cb \sigma \db } \rbar + C_{\bb \cb u\db } (3 \rbar^2 + \sbar^2)] \sbar^6 \nonumber \\ 
&&  + 108 C_{u}{}^{\cb }{}_{u}{}^{\db }{}_{;}{}_{u} [3 C_{\bb \cb u\db } \rbar (\rbar^2 + \sbar^2) +  C_{\bb \cb \sigma \db } (3 \rbar^2 + \sbar^2)]\sbar^6  -108 C_{\bb }{}^{\cb }{}_{u}{}^{\db }{}_{;}{}_{\sigma } [3  C_{u\cb \sigma \db }  \rbar + C_{u\cb u\db } (3 \rbar^2  + \sbar^2)] \sbar^6 \nonumber \\ 
&& + 24 C_{\bb u\sigma }{}^{\cb }{}^{;}{}^{\db } [3  C_{u\cb \sigma \db } \rbar + 3  C_{u\db \sigma \cb }\rbar  + 3 C_{\sigma \cb \sigma \db }  +  C_{u\cb u\db } (3 \rbar^2  + \sbar^2)]\sbar^6+ 72 C_{\bb }{}^{\cb }{}_{\sigma }{}^{\db }{}_{;}{}_{u} [3 \rbar C_{u\cb \sigma \db }+ C_{u\cb u\db } (3 \rbar^2  + \sbar^2)] \sbar^6\nonumber \\ 
&&    + 72 C_{\bb }{}^{\cb }{}_{u}{}^{\db }{}_{;}{}_{u} [3  C_{u\cb u\db }  \rbar (\rbar^2 + \sbar^2) +  C_{u\cb \sigma \db } (3 \rbar^2  + \sbar^2)] \sbar^6 \nonumber \\ 
&&  + 24 C_{\bb uu}{}^{\cb }{}^{;}{}^{\db } [3  C_{\sigma \cb \sigma \db } \rbar + 3  C_{u\cb u\db } \rbar (\rbar^2 + \sbar^2) +  C_{u\cb \sigma \db } (3 \rbar^2 + \sbar^2) +  C_{u\db \sigma \cb } (3 \rbar^2 + \sbar^2)] \sbar^6\big\} u_{\ab } \nonumber \\ 
&&  + \{-18 (C_{\sigma }{}^{\cb }{}_{\sigma }{}^{\db }{}_{;}{}_{\sigma } -  C_{\sigma }{}^{\cb }{}_{\sigma }{}^{\db }{}_{;}{}_{u} \rbar) C_{\sigma \cb \sigma \db } \sbar^4 + 27 C^{\cb \db \eb \fb }{}_{\sigma } C_{\cb \db \eb \fb } \sbar^8 - 18 C^{\cb \db \eb \fb }{}_{u} \rbar C_{\cb \db \eb \fb } \sbar^8 + 9 C_{u\sigma u\sigma }{}_{;}{}_{u\sigma \sigma } \rbar \sbar^2 (\rbar^2 - 3 \sbar^2)\nonumber \\ 
&&  - 15 C_{u\sigma u\sigma }{}^{;}{}^{\cb } \rbar C_{u\sigma \sigma \cb } \sbar^2 (\rbar^2 - 3 \sbar^2) - 135 C_{u\sigma u\sigma }{}_{;}{}_{\sigma } C_{u\sigma u\sigma } (\rbar^2 -  \sbar^2)^2 + 9 C_{u\sigma u\sigma }{}_{;}{}_{uuu} \rbar \sbar^2 (\rbar^4 - 10 \rbar^2 \sbar^2 - 15 \sbar^4)  \nonumber \\ 
&&+ 9 C_{u\sigma u\sigma }{}_{;}{}_{uu\sigma } \sbar^2 (- \rbar^4 + 6 \rbar^2 \sbar^2 + 3 \sbar^4) + 15 C_{u\sigma u\sigma }{}^{;}{}^{\cb } C_{u\sigma u\cb } \sbar^2 (- \rbar^4 + 6 \rbar^2 \sbar^2 + 3 \sbar^4) + C_{u\sigma u\sigma }{}_{;}{}_{\sigma \sigma \sigma } (-9 \rbar^2 \sbar^2 + 9 \sbar^4) \nonumber \\ 
&& + 45 C_{u\sigma u\sigma }{}_{;}{}_{u} C_{u\sigma u\sigma } (3 \rbar^5 - 10 \rbar^3 \sbar^2 + 15 \rbar \sbar^4) - 54 C_{u\sigma \sigma }{}^{\cb }{}_{;}{}_{\sigma } [2  C_{u\sigma u\cb } \rbar  (\rbar^2 - 3 \sbar^2) +  C_{u\sigma \sigma \cb }  (\rbar^2 \nonumber - \sbar^2)]\sbar^2\\ 
&&  + 36 C_{u\cb \sigma \db } \sbar^4 [- C_{\sigma }{}^{\cb }{}_{\sigma }{}^{\db }{}_{;}{}_{\sigma } \rbar + C_{\sigma }{}^{\cb }{}_{\sigma }{}^{\db }{}_{;}{}_{u} (\rbar^2 + \sbar^2)] - 18 C_{u\cb u\db } \sbar^4 [C_{\sigma }{}^{\cb }{}_{\sigma }{}^{\db }{}_{;}{}_{\sigma } (\rbar^2 + \sbar^2)-  C_{\sigma }{}^{\cb }{}_{\sigma }{}^{\db }{}_{;}{}_{u} \rbar (\rbar^2 + 3 \sbar^2)]  \nonumber \\ 
&& -36 C_{u}{}^{\cb }{}_{\sigma }{}^{\db }{}_{;}{}_{\sigma } [ C_{\sigma \cb \sigma \db } \rbar + C_{u\cb \sigma \db }  (\rbar^2 + \sbar^2) + C_{u\db \sigma \cb }  (\rbar^2+ \sbar^2) +  C_{u\cb u\db } \rbar (\rbar^2 + 3 \sbar^2)] \sbar^4  + 15 C_{u\sigma \sigma }{}^{\cb }{}_{;}{}_{u} [2  C_{u\sigma \sigma \cb } \rbar (\rbar^2 - 3 \sbar^2)\nonumber \\ 
&& + 5 C_{u\sigma u\cb }  (\rbar^4 - 6 \rbar^2 \sbar^2 - 3 \sbar^4)] \sbar^2   + 18 C_{u\sigma u}{}^{\cb }{}_{;}{}_{u} [3 \rbar C_{u\sigma u\cb }  (\rbar^4 - 10 \rbar^2 \sbar^2 - 15 \sbar^4) + 5 C_{u\sigma \sigma \cb }  (\rbar^4  - 6 \rbar^2 \sbar^2 - 3 \sbar^4)]\sbar^2\nonumber \\ 
&& -18 C_{u}{}^{\cb }{}_{u}{}^{\db }{}_{;}{}_{\sigma } [ C_{\sigma \cb \sigma \db }  (\rbar^2 + \sbar^2) +2  C_{u\cb \sigma \db }  \rbar (\rbar^2 + 3 \sbar^2)  + 18 C_{u\cb u\db }  (\rbar^4 + 6 \rbar^2 \sbar^2 + \sbar^4)]\sbar^4 \nonumber \\ 
&& + 36 C_{u}{}^{\cb }{}_{\sigma }{}^{\db }{}_{;}{}_{u} [ C_{\sigma \cb \sigma \db }  (\rbar^2 + \sbar^2) +   C_{u\cb \sigma \db }\rbar (\rbar^2  + 3 \sbar^2) +   C_{u\db \sigma \cb } \rbar (\rbar^2 + 3 \sbar^2) +  C_{u\cb u\db } (\rbar^4 + 6 \rbar^2 \sbar^2 + \sbar^4)]  \sbar^4 \nonumber \\ 
&& + C_{u\sigma u}{}^{\cb }{}_{;}{}_{\sigma } [-48  C_{u\sigma \sigma \cb }  \rbar (\rbar^2 - 3 \sbar^2) + 39 C_{u\sigma u\cb }  (- \rbar^4 + 6 \rbar^2 \sbar^2 + 3 \sbar^4)] \sbar^2 + 18 C_{u}{}^{\cb }{}_{u}{}^{\db }{}_{;}{}_{u} [ C_{\sigma \cb \sigma \db } \rbar (\rbar^2 \nonumber + 3 \sbar^2) \\ 
&&  + 2 C_{u\cb \sigma \db }  (\rbar^4 + 6 \rbar^2 \sbar^2 + \sbar^4) +   C_{u\cb u\db } \rbar (\rbar^4 + 10 \rbar^2 \sbar^2  + 5 \sbar^4)]\sbar^4\big\} u_{\ab } u_{\bb }.
\end{IEEEeqnarray}

\section{Mode-sum regularization parameters} \label{sec:ModeSum}

The singular field expansions derived in the previous sections have several applications in
explicit self-force calculations. One of the most successful computational approaches to date
is the \emph{mode-sum} scheme of Barack and Ori
\cite{Barack:Ori:2000,Barack:Mino:Nakano:Ori:Sasaki:2001}; the majority of existing calculations
are based on it in one form or another
\cite{Barack:Burko:2000,Burko:2000b,Detweiler:Messaritaki:Whiting:2002,DiazRivera:2004,Haas:Poisson:2006,Haas:2007,Canizares:Sopuerta:2009,Canizares:Sopuerta:Jaramillo:2010,Barack:Sago:2007,Barack:Lousto:2002,Sago:Barack:Detweiler:2008,Detweiler:2008,Sago:2009,Barack:Sago:2010,Keidl:Shah:Friedman:Kim:Price:2010,Shah:Keidl:Friedman:Kim:Price:2010,Warburton:Barack:2009,Warburton:Barack:2010,Thornburg:2010,Haas:2011bt,Warburton:2011fk,Hopper:2010uv,Keidl:2010pm,Shah:2010bi}.
The basic idea is to decompose the singular retarded field
into spherical harmonic modes which are continuous and finite in general for the scalar case
and in Lorenz gauge for the electromagnetic and gravitational cases. A key component of the
calculation involves the subtraction of so-called \emph{regularization parameters} - analytically
derived expressions which render the formally divergent sum over spherical harmonic modes finite.
In this section, we derive these parameters from our singular field expressions and show how
they may be used to compute the self-force with extremely high accuracy.

\subsection{Rotated Coordinates}
In order to obtain expressions which are readily written as mode-sums, previous calculations
\cite{Barack:Ori:2000,Detweiler:Messaritaki:Whiting:2002,Haas:Poisson:2006} found it useful
to work in a rotated coordinate frame.  We found it most efficient to carry out this rotation
prior to doing any calculations.  To this end, we introduce Riemann normal coordinates on the 2-sphere at $\xb$ in the
form
\begin{equation}
w_{1} = 2 \sin\left(\frac{\alpha}{2}\right) \cos\beta, \quad\quad w_2 = 2 \sin\left(\frac{\alpha}{2}\right) \sin\beta,
\end{equation}
where $\alpha$ and $\beta$ are rotated angular coordinates given by
\begin{eqnarray}
\sin \theta \cos \phi &=& \cos \alpha, \\
\sin \theta \sin \phi &=& \sin \alpha \cos \beta, \\
\cos \theta &=& \sin \alpha \sin \beta.
\end{eqnarray}
In these coordinates, the Schwarzschild metric is given by the line element
\begin{align}
ds^2 =& -\left(\frac{r - 2m}{r}\right) dt^2 + \left(\frac{r}{r - 2m}\right) dr^2 +
r^2 \Bigg\{\left[\frac{16 - w_2^2 \left(8 - w_1^2 - w_2^2\right)}{4 \left(4 - w_1^2 - w_2^2 \right)}\right] dw_1^2 \nonumber \\ 
& \qquad + 2 dw_1 dw_2 \left[ \frac{w_1 w_2 \left( 8 - w_1^2 - w_2^2 \right)}{4 \left( 4 - w_1^2 - w_2^2\right)} \right]
+ \left[ \frac{16 - w_1^2 \left(8 - w_1^2 - w_2^2\right)}{4 \left(4 - w_1^2 - w_2^2 \right)}\right] dw_2^2\Bigg\}.
\end{align}
The algebraic form of the metric makes it very suitable for using with computer algebra programmes such as Mathematica. 
The apparent complexity of having a non-diagonal metric on $S^2$ is in fact minimal since the determinant of that
metric is simply $1$.

\subsection{Mode decomposition}
The method of regularization of the self force through $l$-mode decomposition is by now
standard, see, for example, \cite{Barack:Ori:2000}, \cite{Detweiler:Messaritaki:Whiting:2002}
and \cite{Haas:Poisson:2006}.  Having calculated the singular field, it is straightforward
to calculate the component of the self-force that arises from the singular field\footnote{In this section, for notational convenience
we drop the implied $\sing$ superscript denoting ``singular'' as we are always referring
to the singular component.}, $F_a$, for scalar, electromagnetic and gravitational cases using Eqs.~(\ref{eqn:SelfForceScalar}), (\ref{eqn:SelfForceEM}) and (\ref{eqn:SelfForceGravityBasic}) with the singular field substituted for the regular field.  We study the multipole decomposition of $F_a$ by writing
\begin{equation}\label{eqn:falpha}
F_a \left(r, t, \alpha, \beta \right) = \sum_{lm} F_a^{lm} \left(r, t \right) Y^{lm} \left( \alpha, \beta \right),
\end{equation}
where $Y^{lm} \left( \theta, \phi \right)$ are scalar spherical harmonics, and accordingly
\begin{equation} \label{eqn:flm}
F_a^{lm} \left(r, t\right) = \int F_a \left(r, t, \alpha, \beta \right) Y^{lm*} \left( \alpha, \beta \right) d \Omega.
\end{equation}
To calculate the $l$-mode contribution at $\xb = \left( t_0, r_0, \alpha_0, \beta_0 \right)$, we have
\begin{equation} \label{eqn:falphal}
F_a^l \left(r_0, t_0 \right) = \lim_{\Delta r \rightarrow 0} \sum_{m} F_a^{lm} \left(r_0+\Delta r, t_0 \right) Y^{lm} \left( \alpha_0, \beta_0 \right).
\end{equation}

In previous calculations, Eq. (\ref{eqn:falpha}) has naturally arisen in Schwarzschild coordinates with $\theta_0 = \frac{\pi}{2}$, and it was necessary to perform a rotation to move the coordinate location of the particle from the equatorial plane to a pole in the new coordinate system.  However, by choosing to work in an $S^2$ Riemann normal coordinate system from the start, our particle is already located on the pole.  This saves us from further transformation and expansions at this stage.  With the particle on the pole, $Y^{lm} \left( \alpha_0 = 0, \beta_0 \right) = 0$  for all $m \neq 0$.  This also allows us, without loss of generality, to take $\beta_0 = 0$.  Taking $\alpha_0$, $\beta_0$ and $m$ all to be equal to zero in Eq.~(\ref{eqn:falphal}) gives us
\begin{equation} \label{eqn:fla}
F_a^l \left(r_0, t_0\right) = \lim_{\Delta r \rightarrow 0} \sqrt{\frac{2 l + 1}{4 \pi}} F_a^{l,m=0} \left(r_0+\Delta r, t_0 \right)
= \frac{2 l + 1}{4 \pi} \lim_{\Delta r \rightarrow 0} \int F_a \left( r_0+\Delta r, t_0, \alpha, \beta \right) P_l \left( \cos \alpha \right) d \Omega .
\end{equation}

For each spin field, the singular self-force, $F_a \left( r, t, \alpha, \beta \right)$, has the form
\begin{equation}
\label{eqn:fasum}
F_a \left( r, t, \alpha, \beta \right) = \sum_{n=1}^{\infty} \frac{B_a^{(3 n -2)}}{ \zrho^{2 n + 1}} \epsilon^{n-3},
\end{equation}
where $B_a^{(k)} = b_{a_1 a_2 \dots a_k}(\bar{x}) \Delta x^{a_1} \Delta x^{a_2} \dots \Delta x^{a_k}$.  On identifying $\tau_1 = \rbar_{(1)} \pm \zrho$, this form can be easily seen to follow from the coordinate representation of the above expressions for the singular field. In using Eq.~(\ref{eqn:fasum}) to determine the regularization parameters, we only need to take the sum to the appropriate order: $n=1$ for $F_{a\lnpow{1}}$, $n=2$ for $F_{a\lpow{0}}$, etc.

Explicitly, in our coordinates 
$\zrho = \sqrt{(g_{\alphab \betab} u^{\alphab} \Delta x^b)^2 +g_{\alphab \betab} \Delta x^a \Delta x^b}$ takes the form
\begin{align}
\zrho \left(r, t, \alpha, \beta \right)^2 =& \frac{\left(E^2 \rb^3-L^2 (\rb-2M)\right)}{\rb (\rb-2 M)^2}\Delta r^2 +\left(L^2+\rb^2\right)\Delta w_1^2 -\left(\frac{2 E \rb \rbdot}{\rb-2M}\Delta r +2 E L\Delta w_1 \right)\Delta t +\frac{2  L \rb \rbdot}{\rb-2M} \Delta r \Delta w_1\nonumber \\
& +\: \left(E^2+\frac{2 M}{\rb}-1\right)\Delta t^2 +\rb^2\Delta w_2^2 ,
\end{align}
where the $\alpha$, $\beta$ dependence is contained exclusively in $\Delta w_1$ and $\Delta w_2$.  $E = -u_t$ and $L = u_{\phi}$ are the energy per unit mass and angular momentum along the axis of symmetry, respectively.  In particular, taking $t = t_0$ ($\Delta t = 0$) allows us to write
\begin{equation}
\zrho \left(r, t_0, \alpha, \beta \right)^2 = \frac{E^2  \rb^4}{\left(L^2+\rb^2\right) (\rb-2 M)^2}\Delta r^2+ \left(L^2+\rb^2\right)\left(\Delta w_1 + \frac{L \rb \rbdot}{\left( \rb -2M\right)\left( L^2 + \rb^2 \right)} \Delta r\right)^2+\rb^2\Delta w_2^2.
\end{equation}

We further define
\begin{equation}
\zrhoz\left(\alpha, \beta \right)^2 =  \zrho\left(r_0, t_0, \alpha, \beta \right)^2 =  \left(L^2+\rb^2\right)\Delta w_1 ^2+\rb^2\Delta w_2^2 
\end{equation}
 and this allows us to rewrite our $\Delta w$'s in the alternate forms
\begin{align}
\Delta w_1 ^2 &= 2 \left( 1-\cos{\alpha}\right) \cos^2 \beta = \frac{\zrhoz^2}{(\rb^2+L^2)\chi}\cos^2\beta= \frac{\zrhoz^2}{L^2\chi}\left(k-(1-\chi)\right), \\
 \Delta w_2^2 &= 2 \left( 1-\cos{\alpha}\right) \sin^2 \beta = \frac{\zrhoz^2}{(\rb^2+L^2)\chi}\sin^2\beta= \frac{\zrhoz^2}{L^2\chi} (1 - \chi),
\end{align}
where
\begin{equation}\label{eqn:chi k}
\chi(\beta) = 1 - k \sin^2 \beta, \quad \quad k = \frac{L^2}{\rb^2 + L^2}  .
\end{equation}

Suppose, for the moment, that we may take the limit in Eq.~(\ref{eqn:fla}) through the integral sign, then using our alternate forms we have
\begin{equation}
\lim_{\Delta r \to 0} \frac{B_a^{(3 n -2)}}{ \zrho^{2 n + 1}} \epsilon^{n-3} = \frac{b_{i_1 i_2 \dots i_{3n-2}}(\rb) \Delta w^{i_1} \Delta w^{i_2} \dots \Delta w^{i_{3n-2}}} { \zrhoz^{2 n + 1}} \epsilon^{n-3} = \zrhoz^{n -3 } \epsilon^{n-3} c_{a(n)}(\rb,\chi).
\end{equation}
In \cite{Barack:Ori:2002}, it was shown that the integral and limit in Eq.(\ref{eqn:fla}) are indeed interchangeable for all orders except the leading order, $n=1$ term, where the limiting $\zrhoz^{-3 }$ would not be integrable. Thus we
find the singular self-force now has the form
\begin{align} \label{eqn:fa}
F_a^l \left(r_0, t_0, \alpha, \beta \right) &= \frac{2 l + 1}{4 \pi}\left[\epsilon^{-2}  \lim_{\Delta r \rightarrow 0} \int\frac{B_a^{(1)}}{\zrho^3} P_l \left( \cos \alpha \right) d \Omega + \sum_{n=2} \epsilon^{n-3} \int \zrhoz^{n-3} c_{a (n)} \left(r_0, \chi \right) P_l \left( \cos \alpha \right) d \Omega \right]\nonumber\\
&\equiv F^l_{a\lnpow{1}}\left(r_0,t_0\right) \epsilon^{-2} + F^l_{a[0]}\left(r_0,t_0\right) \epsilon^{-1} + F^l_{a\lpow{2}}\left(r_0,t_0\right) \epsilon^1 + F^l_{a\lpow{4}}\left(r_0,t_0\right) \epsilon^3 + F^l_{a\lpow{6}}\left(r_0,t_0\right) \epsilon^5 +  \dots,
\end{align}
where the $\beta$ dependence in the $c_n$'s are hidden in $\chi$, while the $\alpha$ and $\beta$ dependence of $F_a \left(r, t_0, \alpha, \beta \right)$ is hidden in both the $\zrho$'s and $c_n$'s. Note here that we use the convention that
a subscript in square brackets denotes the term which will contribute at that order in $1/l$. Furthermore the integrand in the summation is odd or even
under $\Delta w_i \to - \Delta w_i$ according to whether $n$ (and so $3n-2$) is odd or even.  This means only the even terms are non-vanishing, while
$F^l_{a\lpow{1}}\left(r_0,t_0\right) = F^l_{a\lpow{3}}\left(r_0,t_0\right)= F^l_{a\lpow{5}}\left(r_0,t_0\right)=0$ etc. 

Some care is required in dealing with taking the limit in the first term. As this has been addressed previously 
\cite{Detweiler:Messaritaki:Whiting:2002,Barack:Ori:2002,Mino:Nakano:Sasaki:2002,Haas:Poisson:2006} and our main interest is in the higher orders we omit the details here and only discuss the calculation of the higher order terms.

In the higher order tems in Eq.~(\ref{eqn:fa}),
we may immediately work with  $\zrhoz^2= 2\chi(L^2+r_0^2)(1-\cos\alpha)$ so,
\begin{IEEEeqnarray}{rCl}
\zrhoz \left(r_0, t_0, \alpha, \beta \right)^n
= \left[2\chi\left(L^2+r_0^2\right) \left( 1 - \cos \alpha \right)\right]^{{n}/{2}} 
= \left[2\chi\left(L^2+r_0^2\right) \right]^{n/2} \sum_{l=0} \mathcal{A}_l^{{n}/{2}} (0) P_l \left( \cos \alpha \right),
\end{IEEEeqnarray}
where $\mathcal{A}_l^{-\frac{1}{2}} (0) = \sqrt{2}$ from the generating function of the Legendre polynomials
and, as derived  in Appendix~D of \cite{Detweiler:Messaritaki:Whiting:2002}, for $(n+1)/2\in\mathbb{N}$
\begin{align}
\mathcal{P}_{n/2} = &\left(-1\right)^{(n+1)/2} 2^{1 + n/2} \left( n!! \right)^2, \nonumber \\
\mathcal{A}^{n/2}_l \left(0 \right) =& 
		 \frac{\mathcal{P}_{n/2} \left( 2 l + 1\right) }{\left(2 l - n\right)\left(2 l - n +2\right) \dots \left(2 l + n\right) \left(2 l + n +2 \right)}  .
\end{align}
In this case the angular integrals involve
\begin{eqnarray} \label{eqn:zeta minus n}
\frac{1}{2 \pi} \int  \frac{d \beta}{\chi(\beta)^{n/2}} &=& 
\left<\chi^{-{n/2}}(\beta) \right> 
={}_2 F_1 \left(\frac{n}{2}, \frac{1}{2}, 1, k \right)
\end{eqnarray}
where $(n+1)/2\in\mathbb{N}\cup \{0\}$. The resulting equations can then be tidied up using the following special cases of hypergeometric functions
\begin{IEEEeqnarray}{lClClCl}
\left<\chi^{-\frac{1}{2}}\right> &=& \mathcal{F}_{\frac{1}{2}}(k) &=& {}_2F_1 \left(\frac{1}{2}, \frac{1}{2};1;k \right) &=& \frac{2}{\pi} \mathcal{K}, \\
\left<\chi^{\frac{1}{2}}\right> &=& \mathcal{F}_{-\frac{1}{2}}(k) &=& {}_2F_1 \left(-\frac{1}{2}, \frac{1}{2};1;k \right) &=& \frac{2}{\pi} \mathcal{E} ,
\end{IEEEeqnarray}
where
\begin{equation}
\mathcal{K} \equiv \int_0^{\pi/2} (1 - k \sin^2 \beta)^{-1/2} d\beta, \quad
\mathcal{E} \equiv \int_0^{\pi/2} (1 - k \sin^2 \beta)^{1/2} d\beta
\end{equation}
are complete elliptic integrals of the first and second kinds, respectively.  All other powers of $\chi$ can be integrated to hypergeometric functions that can then be manipulated to be one of the above by the use of the recurrence relation in Eq. (15.2.10) of \cite{Abramowitz:Stegun}.  That is
\begin{equation}
\mathcal{F}_{p+1} (k) = \frac{p-1}{p \left(k - 1\right)} \mathcal{F}_{p-1}(k) + \frac{1 - 2p + \left(p - \frac{1}{2}\right) k}{p \left(k - 1\right)} \mathcal{F}_p(k).
\end{equation}

In the next sections, we give the results of applying this calculation to each of scalar,
electromagnetic and gravitational cases in turn. In doing so, we omit the explicit dependence on $l$ which in each case is
\begin{gather}
F^l_{a\lnpow{1}} = (2l+1) F_{a\lnpow{1}}, \quad F^l_{a[0]} = F_{a[0]}, \quad F^l_{a\lpow{2}} = \frac{F_{a\lpow{2}}}{(2l-1)(2l+3)}, \quad
F^l_{a\lpow{4}} = \frac{F_{a\lpow{4}}}{(2l-3)(2l-1)(2l+3)(2l+5)}, \nonumber \\
F^l_{a\lpow{6}} = \frac{F_{a\lpow{6}}}{(2l-5)(2l-3)(2l-1)(2l+3)(2l+5)(2l+7)}.
\end{gather}
It is also also worth pointing out that there exists in the literature several different
notations for the regularization parameters. We have adopted a notation which is readily
extensible to other orders and which makes the dependence on $l$ explicit. To avoid
confusion, in Table \ref{table:rp} we give the relation between our notation and
other common notations.

\begin{table}[htb]
 \begin{tabular}{|c|c|c|}
\hline
  RP & BO & DMW \\
\hline
 $F_{a\lnpow{1}}$ & $A_a$ & $A_a$ \\
 $F_{a[0]}$ & $B_a$ & $B_a$ \\
 $F_{a\lpow{1}}$ & $C_a$ & $C_a$ \\
 $F_{a\lpow{2}}$ & --- & $D_a$ \\
 $F_{a\lpow{4}}$ & --- & $E^1_a$ \\
 $F_{a\lpow{6}}$ & --- & $E^2_a$ \\
\hline
 \end{tabular}
\caption{Relation between notational choices for the regularization parameters (RPs). The 
  most common choices are those of either Barack and Ori \cite{Barack:Ori:2000} or
  Detweiler, Messaritaki and Whiting \cite{Detweiler:Messaritaki:Whiting:2002}.}
\label{table:rp}
\end{table}

\subsection{Scalar case} \label{sec:scalar-regularization}
In the scalar case, the regularization parameters are given by
\begin{gather}
F_{t\lnpow{1}} = \frac{\rbdot \sgn(\Delta r)}{2 (L^2+\rb^2)}, \quad
F_{r\lnpow{1}} = -\frac{E\rb \sgn(\Delta r)}{2 (\rb-2M) (L^2+\rb^2)}, \quad
F_{\theta\lnpow{1}} = 0, \quad
F_{\phi\lnpow{1}} = 0,
\end{gather}
\begin{gather}
F_{t[0]} = - \frac{ E \rb \rbdot}{\pi (L^2+\rb^2)^{3/2}} (2\mathcal{E} - \mathcal{K}), \\
F_{r[0]} = \frac{1}{\pi \rb (\rb-2 M) (L^2+\rb^2)^{3/2}} \Big\{
    [2 E^2 \rb^3 - (\rb - 2 M) (L^2 + \rb^2)]  \mathcal{E}
  - [E^2 \rb^3 + (\rb - 2 M) (L^2 + \rb^2)] \mathcal{K} 
  \Big\},
\nonumber \\
F_{\theta[0]} = 0, \quad
F_{\phi[0]} = - \frac{ \rb \rbdot}{L \pi (L^2+\rb^2)^{1/2}} (\mathcal{E} - \mathcal{K}),
\end{gather}

\begin{equation}
F_{t\lpow{2}} = \frac{E \rbdot}{2 \pi \rb^4 (L^2+\rb^2)^{7/2}} (F^{\mathcal{E}}_{t\lpow{2}} \mathcal{E} + F^{\mathcal{K}}_{t\lpow{2}} \mathcal{K}),
\end{equation}
where
\begin{IEEEeqnarray*}{rCl}
F^{\mathcal{E}}_{t\lpow{2}} &=& 8 E^2 (L^2 - \rb^2 ) \rb^7
  -  (L^2 + \rb^2) (36 L^6 M + 104 L^4 M \rb^2 + 98 L^2 M \rb^4 + L^2 \rb^5 + 46 M \rb^6 - 7 \rb^7),
\nonumber \\
F^{\mathcal{K}}_{t\lpow{2}} &=& - E^2 \rb^7 (3 L^2 - 5 \rb^2)
+ 2 \rb^2 (L^2 + \rb^2) (9 L^4 M + 18 L^2 M \rb^2 + 13 M \rb^4 - 2 \rb^5),
\end{IEEEeqnarray*}

\begin{equation}
F_{r\lpow{2}} = \frac{1}{2 \pi \rb^6 (\rb - 2M) (L^2+\rb^2)^{7/2}} (F^{\mathcal{E}}_{r\lpow{2}} \mathcal{E} + F^{\mathcal{K}}_{r\lpow{2}} \mathcal{K}),
\end{equation}
where
\begin{IEEEeqnarray*}{rCl}
F^{\mathcal{E}}_{r\lpow{2}} &=& - 8 E^4 \rb^{10} (L^2 -\rb^2 )
  + 4 E^2 \rb^3 (L^2 + \rb^2) (9 L^6 M + 26 L^4 M \rb^2 + 23 L^2 M \rb^4 + L^2 \rb^5 + 14 M \rb^6 - 3 \rb^7)
\nonumber \\ &&
  -\: (\rb - 2 M) (L^2 + \rb^2)^2 (28 L^6 M + 82 L^4 M \rb^2 + 82 L^2 M \rb^4 -  L^2 \rb^5 + 32 M \rb^6 - 3 \rb^7),
\nonumber \\
F^{\mathcal{K}}_{r\lpow{2}} &=& E^4 \rb^{10} (3 L^2 - 5 \rb^2)
  -  E^2 \rb^5 (L^2 + \rb^2) (18 L^4 M + 34 L^2 M \rb^2 + L^2 \rb^3 + 32 M \rb^4 - 7 \rb^5)
\nonumber \\ &&
  +\:  (\rb - 2 M ) \rb^2 (L^2 + \rb^2)^2 (14 L^4 M + 28 L^2 M \rb^2 + 16 M \rb^4 -  \rb^5), 
\end{IEEEeqnarray*}

\begin{equation}
F_{\theta\lpow{2}} = 0,
\end{equation}

\begin{equation}
F_{\phi\lpow{2}} = \frac{\rbdot}{2 \pi L \rb^4 (L^2+\rb^2)^{5/2}} (F^{\mathcal{E}}_{\phi\lpow{2}} \mathcal{E} + F^{\mathcal{K}}_{\phi\lpow{2}} \mathcal{K}),
\end{equation}
where
\begin{IEEEeqnarray*}{rCl}
F^{\mathcal{E}}_{\phi\lpow{2}} &=& E^2 \rb^7 (7 L^2 - \rb^2)
  + (L^2 + \rb^2) (28 L^6 M + 58 L^4 M \rb^2 + 34 L^2 M \rb^4 -  L^2 \rb^5 + \rb^7),
\nonumber \\
F^{\mathcal{K}}_{\phi\lpow{2}} &=& - E^2 \rb^7 (3 L^2 - \rb^2)
  - \rb^2 (L^2 + \rb^2) (14 L^4 M + 16 L^2 M \rb^2 + \rb^5),
\end{IEEEeqnarray*}

\begin{equation}
F_{t\lpow{4}} = \frac{3 E \rbdot}{40 \pi \rb^{11} (L^2+\rb^2)^{11/2}} (F^{\mathcal{E}}_{t\lpow{4}} \mathcal{E} + F^{\mathcal{K}}_{t\lpow{4}} \mathcal{K}),
\end{equation}
where
\begin{IEEEeqnarray*}{rCl}
F^{\mathcal{E}}_{t\lpow{4}} &=& -30 E^4 \rb^{16} (23 L^4 - 82 L^2 \rb^2 + 23 \rb^4)
\nonumber \\ &&
  +\: 2 E^2 \rb^5 (L^2 + \rb^2) (44800 L^{12} M + 219136 L^{10} M \rb^2 + 428252 L^8 M \rb^4 + 418776 L^6 M \rb^6
\nonumber \\ &&
  \quad +\: 206374 L^4 M \rb^8 + 45 L^4 \rb^9 + 45188 L^2 M \rb^{10} - 1230 L^2 \rb^{11} - 166 M \rb^{12} + 645 \rb^{13})
\nonumber \\ &&
  -\: 2 (L^2 + \rb^2)^2 (20480 L^{14} M^2 - 97280 L^{12} M^2 \rb^2 + 85120 L^{12} M \rb^3 - 700832 L^{10} M^2 \rb^4
\nonumber \\ &&
  \quad +\: 388480 L^{10} M \rb^5 - 1426472 L^8 M^2 \rb^6 + 704552 L^8 M \rb^7 - 1358276 L^6 M^2 \rb^8
\nonumber \\ &&
  \quad +\: 635226 L^6 M \rb^9 - 635180 L^4 M^2 \rb^{10} + 286498 L^4 M \rb^{11} - 15 L^4 \rb^{12} - 124540 L^2 M^2 \rb^{12}
\nonumber \\ &&
  \quad +\: 54086 L^2 M \rb^{13} - 90 L^2 \rb^{14} - 2796 M^2 \rb^{14} + 182 M \rb^{15} + 285 \rb^{16}),
\nonumber \\
F^{\mathcal{K}}_{t\lpow{4}} &=& 15 E^4 \rb^{16} (15 L^4 - 82 L^2 \rb^2 + 31 \rb^4)
\nonumber \\ &&
  -\: 4 E^2 \rb^7 (L^2 + \rb^2) (11200 L^{10} M + 44984 L^8 M \rb^2 + 68227 L^6 M \rb^4
\nonumber \\ &&
  \quad +\: 46849 L^4 M \rb^6 + 13493 L^2 M \rb^8 - 270 L^2 \rb^9 + 127 M \rb^{10} + 210 \rb^{11})
\nonumber \\ &&
  +\: \rb^2 (L^2 + \rb^2)^2 (20480 L^{12} M^2 - 115200 L^{10} M^2 \rb^2 + 85120 L^{10} M \rb^3 - 599072 L^8 M^2 \rb^4
\nonumber \\ &&
  \quad +\: 314000 L^8 M \rb^5 - 908104 L^6 M^2 \rb^6 + 433792 L^6 M \rb^7 - 589164 L^4 M^2 \rb^8 + 268648 L^4 M \rb^9
\nonumber \\ &&
  \quad -\: 151484 L^2 M^2 \rb^{10} + 66380 L^2 M \rb^{11} - 15 L^2 \rb^{12} - 5592 M^2 \rb^{12} + 1204 M \rb^{13} + 345 \rb^{14}),
\end{IEEEeqnarray*}

\begin{equation}
F_{r\lpow{4}} = \frac{3}{40 \pi \rb^{13} (\rb - 2M) (L^2+\rb^2)^{11/2}} (F^{\mathcal{E}}_{r\lpow{4}} \mathcal{E} + F^{\mathcal{K}}_{r\lpow{4}} \mathcal{K}),
\end{equation}
where
\begin{IEEEeqnarray*}{rCl}
F^{\mathcal{E}}_{r\lpow{4}} &=& 30 E^6 \rb^{19} (23 L^4 - 82 L^2 \rb^2 + 23 \rb^4)
\nonumber \\ &&
  -\: E^4 \rb^8 (L^2 + \rb^2) (89600 L^{12} M + 438272 L^{10} M \rb^2 + 856504 L^8 M \rb^4 + 837552 L^6 M \rb^6
\nonumber \\ &&
  \quad +\: 411938 L^4 M \rb^8 + 495 L^4 \rb^9 + 92836 L^2 M \rb^{10} - 3690 L^2 \rb^{11} - 902 M \rb^{12} + 1575 \rb^{13})
\nonumber \\ &&
  +\: 8 E^2 \rb^3 (L^2 + \rb^2)^2 (5120 L^{14} M^2 - 35200 L^{12} M^2 \rb^2 + 26720 L^{12} M \rb^3 - 227368 L^{10} M^2 \rb^4 + 123200 L^{10} M \rb^5\nonumber \\ &&
  \quad -\: 456300 L^8 M^2 \rb^6 + 225979 L^8 M \rb^7 - 434510 L^6 M^2 \rb^8 + 206277 L^6 M \rb^9 - 203983 L^4 M^2 \rb^{10}\nonumber \\ &&
  \quad +\: 94211 L^4 M \rb^{11} - 40376 L^2 M^2 \rb^{12} + 18367 L^2 M \rb^{13} - 135 L^2 \rb^{14} - 571 M^2 \rb^{14} - 146 M \rb^{15} + 135 \rb^{16})
\nonumber \\ &&
  -\: (\rb - 2 M ) (L^2 + \rb^2)^3 (40960 L^{14} M^2 - 86016 L^{12} M^2 \rb^2 + 116480 L^{12} M \rb^3 - 860224 L^{10} M^2 \rb^4
\nonumber \\ &&
  \quad +\: 510080 L^{10} M \rb^5 - 1780112 L^8 M^2 \rb^6 + 882400 L^8 M \rb^7 - 1657392 L^6 M^2 \rb^8 + 752340 L^6 M \rb^9
\nonumber \\ &&
  \quad -\: 743164 L^4 M^2 \rb^{10} + 316100 L^4 M \rb^{11} + 30 L^4 \rb^{12} - 136236 L^2 M^2 \rb^{12} + 53200 L^2 M \rb^{13} + 75 L^2 \rb^{14}
\nonumber \\ &&
  \quad -\: 3120 M^2 \rb^{14} + 160 M \rb^{15} + 165 \rb^{16}),
\nonumber \\
F^{\mathcal{K}}_{r\lpow{4}} &=&-15 E^6 \rb^{19} (15 L^4 - 82 L^2 \rb^2 + 31 \rb^4)
\nonumber \\ &&
  +\: E^4 \rb^{10} (L^2 + \rb^2) (44800 L^{10} M + 179936 L^8 M \rb^2 + 272908 L^6 M \rb^4
\nonumber \\ &&
  \quad + 187126 L^4 M \rb^6 + 135 L^4 \rb^7 + 55232 L^2 M \rb^8 - 1710 L^2 \rb^9 + 118 M \rb^{10} + 1035 \rb^{11})
\nonumber \\ &&
  -\:  E^2 \rb^5 (L^2 + \rb^2)^2 (20480 L^{12} M^2 - 158720 L^{10} M^2 \rb^2 + 106880 L^{10} M \rb^3 - 769632 L^8 M^2 \rb^4
\nonumber \\ &&
  \quad +\: 399280 L^8 M \rb^5 - 1159632 L^6 M^2 \rb^6 + 559556 L^6 M \rb^7 - 755876 L^4 M^2 \rb^8 + 352004 L^4 M \rb^9
\nonumber \\ &&
  \quad -\: 196524 L^2 M^2 \rb^{10} + 89680 L^2 M \rb^{11} - 405 L^2 \rb^{12} - 5528 M^2 \rb^{12} + 512 M \rb^{13} + 675 \rb^{14})
\nonumber \\ &&
  +\:  (\rb - 2 M) \rb^2 (L^2 + \rb^2)^3 (20480 L^{12} M^2 - 60928 L^{10} M^2 \rb^2 + 58240 L^{10} M \rb^3 - 375840 L^8 M^2 \rb^4
\nonumber \\ &&
  \quad +\: 204080 L^8 M \rb^5 - 564472 L^6 M^2 \rb^6 + 265360 L^6 M \rb^7 - 350956 L^4 M^2 \rb^8 + 152380 L^4 M \rb^9
\nonumber \\ &&
  \quad -\: 84276 L^2 M^2 \rb^{10} + 33740 L^2 M \rb^{11} + 15 L^2 \rb^{12} - 3120 M^2 \rb^{12} + 640 M \rb^{13} + 75 \rb^{14}),
\end{IEEEeqnarray*}

\begin{equation}
F_{\theta\lpow{4}} = 0,
\end{equation}

\begin{equation}
F_{\phi\lpow{4}} = \frac{3 \rbdot}{40 \pi L \rb^{11} (L^2+\rb^2)^{9/2}}  (F^{\mathcal{E}}_{\phi\lpow{4}} \mathcal{E} + F^{\mathcal{K}}_{\phi\lpow{4}} \mathcal{K}),
\end{equation}
where
\begin{IEEEeqnarray*}{rCl}
F^{\mathcal{E}}_{\! \phi\lpow{4}} &=& -15 E^4 \rb^{16} (43 L^4 - 82 L^2 \rb^2 + 3 \rb^4)
\nonumber \\ &&
  -\: 10 E^2 \rb^5 (L^2 + \rb^2) (4352 L^{12} M + 16512 L^{10} M \rb^2 + 22948 L^8 M \rb^4
\nonumber \\ &&
  \quad +\: 13346 L^6 M \rb^6 + 2136 L^4 M \rb^8 - 9 L^4 \rb^9 - 710 L^2 M \rb^{10} + 126 L^2 \rb^{11} - 9 \rb^{13})
\nonumber \\ &&
  +\: (L^2 + \rb^2)^2 (40960 L^{14} M^2 - 96256 L^{12} M^2 \rb^2 + 116480 L^{12} M \rb^3 - 704064 L^{10} M^2 \rb^4
\nonumber \\ &&
  \quad +\: 429440 L^{10} M \rb^5 - 1134992 L^8 M^2 \rb^6 + 595040 L^8 M \rb^7 - 755632 L^6 M^2 \rb^8
\nonumber \\ &&
  \quad +\: 372500 L^6 M \rb^9 - 194724 L^4 M^2 \rb^{10} + 94940 L^4 M \rb^{11} + 30 L^4 \rb^{12} - 6276 L^2 M^2 \rb^{12}
\nonumber \\ &&
  \quad +\: 4040 L^2 M \rb^{13} + 105 L^2 \rb^{14} + 480 M^2 \rb^{14} - 45 \rb^{16}),
\nonumber \\
F^{\mathcal{K}}_{\! \phi\lpow{4}} &=& 15 E^4 \rb^{16} (L^2 - 3 \rb^2) (15 L^2 -  \rb^2)
\nonumber \\ &&
  +\: 10 E^2 \rb^7 (L^2 + \rb^2) (2176 L^{10} M + 6352 L^8 M \rb^2 + 6018 L^6 M \rb^4
\nonumber \\ &&
  \quad +\: 1666 L^4 M \rb^6 - 320 L^2 M \rb^8 + 63 L^2 \rb^9 - 9 \rb^{11})
\nonumber \\ &&
  -\:  \rb^2 (L^2 + \rb^2)^2 (20480 L^{12} M^2 - 66048 L^{10} M^2 \rb^2 + 58240 L^{10} M \rb^3 - 293280 L^8 M^2 \rb^4
\nonumber \\ &&
  \quad +\: 163760 L^8 M \rb^5 - 314392 L^6 M^2 \rb^6 + 156960 L^6 M \rb^7 - 114876 L^4 M^2 \rb^8
\nonumber \\ &&
  \quad +\: 55420 L^4 M \rb^9 - 6516 L^2 M^2 \rb^{10} + 3740 L^2 M \rb^{11} + 15 L^2 \rb^{12} + 480 M^2 \rb^{12} - 45 \rb^{14}),
\end{IEEEeqnarray*}

\begin{equation}
F_{t\lpow{6}} = \frac{-3 E \rbdot}{560 \pi  \rb^{16} \left(L^2+\rb^2\right)^{15/2}} \left(F^{\mathcal{E}}_{t\lpow{6}} \mathcal{E} + F^{\mathcal{K}}_{t\lpow{6}} \mathcal{K} \right),
\end{equation}
where
\begin{IEEEeqnarray*}{rCl}
F^{\mathcal{E}}_{t\lpow{6}} &=& 28000 E^6 \rb^{23} (\rb-L) (L+\rb) \left(11 L^4-74 L^2 \rb^2+11 \rb^4\right) \nonumber \\
&& -\: 25 E^4 \rb^8 \left(L^2+\rb^2\right) \big(-16056320 L^{18} M-107151360 L^{16} M \rb^2-302586880 L^{14} M \rb^4-464979968 L^{12} M \rb^6 \nonumber \\
&& \quad -\: 412568652 L^{10} M \rb^8-201055024 L^8 M \rb^{10}-39268410 L^6 M \rb^{12}-1575 L^6 \rb^{13}+5226426 L^4 M \rb^{14} \nonumber \\
&& \quad +\: 99435 L^4 \rb^{15}+3185118 L^2 M \rb^{16}-186165 L^2 \rb^{17}+19662 M \rb^{18}+35385 \rb^{19}\big) \nonumber \\
&& +\: 2 E^2 \rb^3 \left(L^2+\rb^2\right)^2 \big(-1007616000 L^{20} M^2-2885324800 L^{18} M^2 \rb^2-1548288000 L^{18} M \rb^3 \nonumber \\
&& \quad +\: 5271990272 L^{16} M^2 \rb^4 - 9940582400 L^{16} M \rb^5+35832487264 L^{14} M^2 \rb^6-27145052800 L^{14} M \rb^7 \nonumber \\
&& \quad +\: 69571689904 L^{12} M^2 \rb^8 - 40793731200 L^{12} M \rb^9+69887626312 L^{10} M^2 \rb^{10}-36329433800 L^{10} M \rb^{11} \nonumber \\
&& \quad +\: 39015325900 L^8 M^2 \rb^{12}-19063343950 L^8 M \rb^{13}+11166709052 L^6 M^2 \rb^{14}-5373108900 L^6 M \rb^{15} \nonumber \\
&& \quad +\: 7875 L^6 \rb^{16}+1046817944 L^4 M^2 \rb^{16}-575985300 L^4 M \rb^{17}+94500 L^4 \rb^{18}-118281276 L^2 M^2 \rb^{18} \nonumber \\
&& \quad +\: 29440500 L^2 M \rb^{19}-1178625 L^2 \rb^{20}-8271468 M^2 \rb^{20}+479850 M \rb^{21}+414750 \rb^{22}\big) \nonumber \\
&& +\: \left(L^2+\rb^2\right)^3 \big(-5775360000 L^{20} M^3+2580480000 L^{20} M^2 \rb-18980904960 L^{18} M^3 \rb^2+750796800 L^{18} M^2 \rb^3 \nonumber \\
&& \quad +\: 3429888000 L^{18} M \rb^4+10876463104 L^{16} M^3 \rb^4-53063915520 L^{16} M^2 \rb^5+21396480000 L^{16} M \rb^6 \nonumber \\
&& \quad +\: 143196789568 L^{14} M^3 \rb^6-191859546624 L^{14} M^2 \rb^7+56793312800 L^{14} M \rb^8+292841560608 L^{12} M^3 \rb^8 \nonumber \\
&& \quad -\: 317782413664 L^{12} M^2 \rb^9+83096000800 L^{12} M \rb^{10}+297880915104 L^{10} M^3 \rb^{10}-295661821784 L^{10} M^2 \rb^{11} \nonumber \\
&& \quad +\: 72364880400 L^{10} M \rb^{12}+168534399040 L^8 M^3 \rb^{12}-159472848000 L^8 M^2 \rb^{13}+37560515600 L^8 M \rb^{14} \nonumber \\
&& \quad +\: 50707761864 L^6 M^3 \rb^{14}-46826640820 L^6 M^2 \rb^{15}+10839698800 L^6 M \rb^{16}+7000 L^6 \rb^{17}\nonumber \\
&& \quad +\: 6272875728 L^4 M^3 \rb^{16} - 5878415984 L^4 M^2 \rb^{17}+1398754050 L^4 M \rb^{18}+41125 L^4 \rb^{19}-86931352 L^2 M^3 \rb^{18}\nonumber \\
&& \quad +\: 212436 L^2 M^2 \rb^{19}+21357300 L^2 M \rb^{20}+124250 L^2 \rb^{21}-29620256 M^3 \rb^{20}+16081176 M^2 \rb^{21}\nonumber \\
&& \quad -\: 597150 M \rb^{22}-245875 \rb^{23}\big), 
\end{IEEEeqnarray*}
\begin{IEEEeqnarray*}{rCl}
F^{\mathcal{K}}_{t\lpow{6}} &=& -875 E^6 \rb^{23} \left(-105 L^6+1189 L^4 \rb^2-1531 L^2 \rb^4+247 \rb^6\right) \nonumber \\
&& +\: 50 E^4 \rb^{10} \left(L^2+\rb^2\right) \big(-4014080 L^{16} M-23275520 L^{14} M \rb^2-55468800 L^{12} M \rb^4-68718512 L^{10} M \rb^6\nonumber \\
&& \quad -\: 45183275 L^8 M \rb^8-13052460 L^6 M \rb^{10}+362358 L^4 M \rb^{12}+18900 L^4 \rb^{13}+919476 L^2 M \rb^{14}-49560 L^2 \rb^{15}\nonumber \\
&& \quad +15501 M \rb^{16}+12180 \rb^{17}\big)\nonumber \\
&& -\: E^2 \rb^5 \left(L^2+\rb^2\right)^2 \big(-1007616000 L^{18} M^2-2003660800 L^{16} M^2 \rb^2-1548288000 L^{16} M \rb^3\nonumber \\
&& \quad +\: 6977961472 L^{14} M^2 \rb^4-8585830400 L^{14} M \rb^5+29653513376 L^{12} M^2 \rb^6-19705027200 L^{12} M \rb^7 \nonumber \\
&& \quad +\: 43981240344 L^{10} M^2 \rb^8-23922541200 L^{10} M \rb^9+32635169200 L^8 M^2 \rb^{10}-16162950800 L^8 M \rb^{11}\nonumber \\
&& \quad +\: 12004692860 L^6 M^2 \rb^{12}-5726827800 L^6 M \rb^{13}+1578230992 L^4 M^2 \rb^{14}-803087700 L^4 M \rb^{15}+7875 L^4 \rb^{16} \nonumber \\
&& \quad -\: 113069964 L^2 M^2 \rb^{16}+24092400 L^2 M \rb^{17}-1118250 L^2 \rb^{18}-13324056 M^2 \rb^{18}+1526700 M \rb^{19}+553875 \rb^{20}\big)\nonumber \\
&& +\: 2 \rb^2 \left(L^2+\rb^2\right)^3 \big(1443840000 L^{18} M^3-645120000 L^{18} M^2 \rb+3481866240 L^{16} M^3 \rb^2+376780800 L^{16} M^2 \rb^3\nonumber \\
&& \quad -\: 857472000 L^{16} M \rb^4-5698068736 L^{14} M^3 \rb^4+12906055680 L^{14} M^2 \rb^5-4598832000 L^{14} M \rb^6\nonumber \\
&& \quad -\: 30679784768 L^{12} M^3 \rb^6+36702979536 L^{12} M^2 \rb^7-10214544200 L^{12} M \rb^8-46687160912 L^{10} M^3 \rb^8 \nonumber \\
&& \quad +\: 47920180732 L^{10} M^2 \rb^9-12034259400 L^{10} M \rb^{10}-34910457500 L^8 M^3 \rb^{10}+33450421620 L^8 M^2 \rb^{11} \nonumber \\
&& \quad -\: 7955786400 L^8 M \rb^{12}-13231827540 L^6 M^3 \rb^{12}+12237529680 L^6 M^2 \rb^{13}-2830778675 L^6 M \rb^{14}\nonumber \\
&& \quad -\: 2081061396 L^4 M^3 \rb^{14}+1920152080 L^4 M^2 \rb^{15}-448289925 L^4 M \rb^{16}-1750 L^4 \rb^{17}-4481148 L^2 M^3 \rb^{16}\nonumber \\
&& \quad +\: 24248796 L^2 M^2 \rb^{17}-11278125 L^2 M \rb^{18}-8750 L^2 \rb^{19}+11067088 M^3 \rb^{18}-6431148 M^2 \rb^{19} \nonumber \\
&& \quad +\: 440325 M \rb^{20}+77000 \rb^{21}\big),
\end{IEEEeqnarray*}

\begin{equation}
F_{r\lpow{6}} = \frac{-3}{560 \pi  \rb^{18} \left(L^2+\rb^2\right)^{15/2} (\rb-2 M)} \left(F^{\mathcal{E}}_{r\lpow{6}} \mathcal{E} + F^{\mathcal{K}}_{r\lpow{6}} \mathcal{K} \right),
\end{equation}
where
\begin{IEEEeqnarray*}{rCl}
F^{\mathcal{E}}_{r\lpow{6}} &=& -28000 E^8 (\rb-L) (L+\rb) \left(11 L^4-74 \rb^2 L^2+11 \rb^4\right) \rb^{26} \nonumber \\
&& +\: 50 E^6 \left(L^2+\rb^2\right) \big(19565 \rb^{19}+6086 M \rb^{18}-113785 L^2 \rb^{17}+1633964 L^2 M \rb^{16}+76615 L^4 \rb^{15}+2559418 L^4 M \rb^{14} \nonumber \\
&& \quad -\: 5075 L^6 \rb^{13}-19625630 L^6 M \rb^{12}-100527512 L^8 M \rb^{10}-206284326 L^{10} M \rb^8-232489984 L^{12} M \rb^6 \nonumber \\
&& \quad -\: 151293440 L^{14} M \rb^4-53575680 L^{16} M \rb^2-8028160 L^{18} M\big) \rb^{11} \nonumber \\
&& -\: E^4 \left(L^2+\rb^2\right)^2 \big(1094625 \rb^{22}+545450 M \rb^{21}-4202625 L^2 \rb^{20}-16774936 M^2 \rb^{20}+110101000 L^2 M \rb^{19} \nonumber \\
&& \quad +\: 1414875 L^4 \rb^{18}-331621052 L^2 M^2 \rb^{18}-822447000 L^4 M \rb^{17}-7875 L^6 \rb^{16}+1429685188 L^4 M^2 \rb^{16} \nonumber \\
&& \quad -\: 9480504900 L^6 M \rb^{15}+19802086804 L^6 M^2 \rb^{14}-35145688050 L^8 M \rb^{13}+72068652100 L^8 M^2 \rb^{12} \nonumber \\
&& \quad -\: 68030273700 L^{10} M \rb^{11}+130518064824 L^{10} M^2 \rb^{10}-76873222400 L^{12} M \rb^9+129714899808 L^{12} M^2 \rb^8 \nonumber \\
&& \quad -\: 51274604800 L^{14} M \rb^7+65633972928 L^{14} M^2 \rb^6-18785894400 L^{16} M \rb^5+8353439744 L^{16} M^2 \rb^4 \nonumber \\
&& \quad -\: 2924544000 L^{18} M \rb^3-6114713600 L^{18} M^2 \rb^2-2015232000 L^{20} M^2\big) \rb^6 \nonumber \\
&& +\: 2 E^2 \left(L^2+\rb^2\right)^3 \big(242375 \rb^{23}+138500 M \rb^{22}-419125 L^2 \rb^{21}-10992820 M^2 \rb^{21}+20399392 M^3 \rb^{20} \nonumber \\
&& \quad +\: 6959650 L^2 M \rb^{20}+9625 L^4 \rb^{19}-81665756 L^2 M^2 \rb^{19}+138887552 L^2 M^3 \rb^{18}-822884550 L^4 M \rb^{18} \nonumber \\
&& \quad -\: 875 L^6 \rb^{17}+3518901164 L^4 M^2 \rb^{17}-3802035608 L^4 M^3 \rb^{16}-7067551050 L^6 M \rb^{16}+30648837404 L^6 M^2 \rb^{15} \nonumber \\
&& \quad -\: 33234129320 L^6 M^3 \rb^{14}-25352340750 L^8 M \rb^{14}+106617027800 L^8 M^2 \rb^{13}-111740075320 L^8 M^3 \rb^{12} \nonumber \\
&& \quad -\: 49660036200 L^{10} M \rb^{12}+197830827680 L^{10} M^2 \rb^{11}-195029907128 L^{10} M^3 \rb^{10}-57553509200 L^{12} M \rb^{10} \nonumber \\
&& \quad +\: 209550665904 L^{12} M^2 \rb^9-183717663248 L^{12} M^3 \rb^8-39556896400 L^{14} M \rb^8+121346652064 L^{14} M^2 \rb^7 \nonumber \\
&& \quad -\: 77791192288 L^{14} M^3 \rb^6-14956032000 L^{16} M \rb^6+28755231744 L^{16} M^2 \rb^5+7146388480 L^{16} M^3 \rb^4 \nonumber \\
&& \quad -\: 2403072000 L^{18} M \rb^4-3619737600 L^{18} M^2 \rb^3+18731642880 L^{18} M^3 \rb^2-2297856000 L^{20} M^2 \rb \nonumber \\
&& \quad +\: 4902912000 L^{20} M^3\big) \rb^3 \nonumber \\
&& +\: (2 M-\rb) \left(L^2+\rb^2\right)^4 \big(53375 \rb^{23}+173600 M \rb^{22}+29750 L^2 \rb^{21}-5212944 M^2 \rb^{21}+10317184 M^3 \rb^{20} \nonumber \\
&& \quad -\: 14183750 L^2 M \rb^{20}+25375 L^4 \rb^{19}+62499436 L^2 M^2 \rb^{19}-76153328 L^2 M^3 \rb^{18}-679641900 L^4 M \rb^{18} \nonumber \\
&& \quad +\: 7000 L^6 \rb^{17}+3345821696 L^4 M^2 \rb^{17}-4111469784 L^4 M^3 \rb^{16}-5518261350 L^6 M \rb^{16}+25823273900 L^6 M^2 \rb^{15} \nonumber \\
&& \quad -\: 30032966752 L^6 M^3 \rb^{14}-20150264400 L^8 M \rb^{14}+88797134680 L^8 M^2 \rb^{13}-96877958296 L^8 M^3 \rb^{12} \nonumber \\
&& \quad -\: 40738068000 L^{10} M \rb^{12}+166165277616 L^{10} M^2 \rb^{11}-165473393280 L^{10} M^3 \rb^{10}-48857572400 L^{12} M \rb^{10} \nonumber \\
&& \quad +\: 177451524416 L^{12} M^2 \rb^9-149256330784 L^{12} M^3 \rb^8-34733020000 L^{14} M \rb^8+101711045376 L^{14} M^2 \rb^7 \nonumber \\
&& \quad -\: 51263318208 L^{14} M^3 \rb^6-13563648000 L^{16} M \rb^6+21232814080 L^{16} M^2 \rb^5+21391316992 L^{16} M^3 \rb^4 \nonumber \\
&& \quad -\: 2247168000 L^{18} M \rb^4-5520998400 L^{18} M^2 \rb^3+23777402880 L^{18} M^3 \rb^2-2580480000 L^{20} M^2 \rb \nonumber \\
&& \quad +\: 5775360000 L^{20} M^3\big),
\end{IEEEeqnarray*}
\begin{IEEEeqnarray*}{rCl}
F^{\mathcal{K}}_{r\lpow{6}} &=& 875 E^8 \left(-105 L^6+1189 \rb^2 L^4-1531 \rb^4 L^2+247 \rb^6\right) \rb^{26} \nonumber \\
&& -\: 25 E^6 \left(L^2+\rb^2\right) \big(27055 \rb^{17}+25612 M \rb^{16}-123235 L^2 \rb^{15}+1887182 L^2 M \rb^{14}+62125 L^4 \rb^{13}+676066 L^4 M \rb^{12} \nonumber \\
&& \quad -\: 2625 L^6 \rb^{11}-26099670 L^6 M \rb^{10}-90366550 L^8 M \rb^8-137437024 L^{10} M \rb^6-110937600 L^{12} M \rb^4 \nonumber \\
&& \quad -\: 46551040 L^{14} M \rb^2-8028160 L^{16} M\big) \rb^{13} \nonumber \\
&& +\: 2 E^4 \left(L^2+\rb^2\right)^2 \big(370125 \rb^{20}+697975 M \rb^{19}-1065750 L^2 \rb^{18}-6904028 M^2 \rb^{18}+27977700 L^2 M \rb^{17} \nonumber \\
&& \quad +\: 244125 L^4 \rb^{16}-86371482 L^2 M^2 \rb^{16}-315079050 L^4 M \rb^{15}+615225146 L^4 M^2 \rb^{14}-2582807100 L^6 M \rb^{13} \nonumber \\
&& \quad +\: 5441132830 L^6 M^2 \rb^{12}-7522573725 L^8 M \rb^{11}+15199781250 L^8 M^2 \rb^{10}-11253096600 L^{10} M \rb^9 \nonumber \\
&& \quad +\: 20574272172 L^{10} M^2 \rb^8-9303284800 L^{12} M \rb^7+13728299088 L^{12} M^2 \rb^6-4056729600 L^{14} M \rb^5 \nonumber \\
&& \quad +\: 3016609536 L^{14} M^2 \rb^4-731136000 L^{16} M \rb^3-1087846400 L^{16} M^2 \rb^2-503808000 L^{18} M^2\big) \rb^8 \nonumber \\
&& -\: E^2 \left(L^2+\rb^2\right)^3 \big(314125 \rb^{21}+970000 M \rb^{20}-358750 L^2 \rb^{19}-18222440 M^2 \rb^{19}+31216064 M^3 \rb^{18} \nonumber \\
&& \quad -\: 3780450 L^2 M \rb^{18}-875 L^4 \rb^{17}-36108732 L^2 M^2 \rb^{17}+87877152 L^2 M^3 \rb^{16}-1092352500 L^4 M \rb^{16} \nonumber \\
&& \quad +\: 4756636984 L^4 M^2 \rb^{15}-5211718840 L^4 M^3 \rb^{14}-7494234450 L^6 M \rb^{14}+32418083300 L^6 M^2 \rb^{13} \nonumber \\
&& \quad -\: 35018994560 L^6 M^3 \rb^{12}-21674793600 L^8 M \rb^{12}+89821855320 L^8 M^2 \rb^{11}-92610055960 L^8 M^3 \rb^{10} \nonumber \\
&& \quad -\: 33236721600 L^{10} M \rb^{10}+127823271040 L^{10} M^2 \rb^9-120667329776 L^{10} M^3 \rb^8-28422864400 L^{12} M \rb^8 \nonumber \\
&& \quad +\: 95019791488 L^{12} M^2 \rb^7-72612130368 L^{12} M^3 \rb^6-12853344000 L^{14} M \rb^6+30055494144 L^{14} M^2 \rb^5 \nonumber \\
&& \quad -\: 5260183040 L^{14} M^3 \rb^4-2403072000 L^{16} M \rb^4-1609113600 L^{16} M^2 \rb^3+14441594880 L^{16} M^3 \rb^2 \nonumber \\
&& \quad -\: 2297856000 L^{18} M^2 \rb+4902912000 L^{18} M^3\big) \rb^5 \nonumber \\
&& +\: (\rb-2 M) \left(L^2+\rb^2\right)^4 \big(27125 \rb^{21}+299600 M \rb^{20}+9625 L^2 \rb^{19}-4164624 M^2 \rb^{19}+7521664 M^3 \rb^{18} \nonumber \\
&& \quad -\: 12312300 L^2 M \rb^{18}+3500 L^4 \rb^{17}+60136468 L^2 M^2 \rb^{17}-77649520 L^2 M^3 \rb^{16}-435265950 L^4 M \rb^{16} \nonumber \\
&& \quad +\: 2136130980 L^4 M^2 \rb^{15}-2610250432 L^4 M^3 \rb^{14}-2918350050 L^6 M \rb^{14}+13485739400 L^6 M^2 \rb^{13} \nonumber \\
&& \quad -\: 15471992072 L^6 M^3 \rb^{12}-8700997200 L^8 M \rb^{12}+37444962000 L^8 M^2 \rb^{11}-39767055768 L^8 M^3 \rb^{10} \nonumber \\
&& \quad -\: 13875397200 L^{10} M \rb^{10}+54025898376 L^{10} M^2 \rb^9-50546368608 L^{10} M^3 \rb^8-12345326000 L^{12} M \rb^8 \nonumber \\
&& \quad +\: 40319920928 L^{12} M^2 \rb^7-27561410368 L^{12} M^3 \rb^6-5798688000 L^{14} M \rb^6+11983523840 L^{14} M^2 \rb^5 \nonumber \\
&& \quad +\: 2639284736 L^{14} M^3 \rb^4-1123584000 L^{16} M \rb^4-1631539200 L^{16} M^2 \rb^3+9361981440 L^{16} M^3 \rb^2 \nonumber \\
&& \quad -\: 1290240000 L^{18} M^2 \rb+2887680000 L^{18} M^3\big) \rb^2,
\end{IEEEeqnarray*}

\begin{equation}
F_{\theta\lpow{6}} = 0,
\end{equation}

\begin{equation}
F_{\phi\lpow{6}} = \frac{-3\rbdot}{560 \pi  L \rb^{16} \left(L^2+\rb^2\right)^{13/2}} \left(F^{\mathcal{E}}_{\phi\lpow{6}} \mathcal{E} + F^{\mathcal{K}}_{\phi\lpow{6}} \mathcal{K} \right),
\end{equation}
where
\begin{IEEEeqnarray*}{rCl}
F^{\mathcal{E}}_{\phi\lpow{6}} &=& 875 E^6 \rb^{23} \left(1773 L^4 \rb^2 -337 L^6-947 L^2 \rb^4+15 \rb^6\right) \nonumber \\
&& +\: 175 E^4 \rb^8 \left(L^2+\rb^2\right) \big(983040 L^{18} M + 7045120 L^{16} M \rb^2 + 21570560 L^{14} M \rb^4 + 36582912 L^{12} M \rb^6  \nonumber \\
&& \quad +\: 37139572 L^{10} M \rb^8 + 22617566 L^8 M \rb^{10} + 7708830 L^6 M \rb^{12} + 225 L^6 \rb^{13} + 1199730 L^4 M \rb^{14} - 9375 L^4 \rb^{15}  \nonumber \\
&& \quad +\: 4926 L^2 M \rb^{16} + 9375 L^2 \rb^{17} - 225 \rb^{19}\big) \nonumber \\
&& +\: E^2 \rb^3 \left(L^2+\rb^2\right)^2 \big(2015232000 L^{20} M^2+8400691200  L^{18} M^2 \rb^2+1376256000 L^{18} M \rb^3+11063078912 L^{16} M^2 \rb^4 \nonumber \\
&& \quad +\: 7276953600 L^{16} M \rb^5+82626752 L^{14} M^2 \rb^6+15631481600 L^{14} M \rb^7-12481032128 L^{12} M^2 \rb^8 \nonumber \\
&& \quad +\: 17141443200 L^{12} M \rb^9-10632497080 L^{10} M^2 \rb^{10}+9613116800 L^{10} M \rb^{11}-2120762600 L^8 M^2 \rb^{12} \nonumber \\
&& \quad +\: 2046335900 L^8 M \rb^{13}+1035946292 L^6 M^2 \rb^{14}-316454600 L^6 M \rb^{15}+15750 L^6 \rb^{16}+431941952 L^4 M^2 \rb^{16} \nonumber \\
&& \quad -\: 166655300 L^4 M \rb^{17}+133875 L^4 \rb^{18}+17346492 L^2 M^2 \rb^{18}-2290400 L^2 M \rb^{19}-850500 L^2 \rb^{20} \nonumber \\
&& \quad -\: 312480 M^2 \rb^{20}+39375 \rb^{22}\big) \nonumber \\
&& -\: \left(L^2+\rb^2\right)^3 \big(2580480000 L^{20} M^2 \rb -5775360000 L^{20} M^3-21381242880 L^{18} M^3 \rb^2+4448665600 L^{18} M^2 \rb^3 \nonumber \\
&& \quad +\: 2247168000 L^{18} M \rb^4-16436058112 L^{16} M^3 \rb^4-20228515840 L^{16} M^2 \rb^5+12144384000 L^{16} M \rb^6 \nonumber \\
&& \quad +\: 37967372736 L^{14} M^3 \rb^6-78865174016 L^{14} M^2 \rb^7+27288335200 L^{14} M \rb^8+91117248928 L^{12} M^3 \rb^8 \nonumber \\
&& \quad -\: 114866020480 L^{12} M^2 \rb^9+32738062000 L^{12} M \rb^{10}+80974789248 L^{10} M^3 \rb^{10}-86565871136  L^{10} M^2 \rb^{11} \nonumber \\
&& \quad +\: 22304895200 L^{10} M \rb^{12}+35112838392  L^8 M^3 \rb^{12}-34540540744 L^8 M^2 \rb^{13}+8392988800 L^8 M \rb^{14} \nonumber \\
&& \quad +\: 6757023136 L^6 M^3 \rb^{14}-6373891596  L^6 M^2 \rb^{15}+1516716950 L^6 M \rb^{16}-7000 L^6 \rb^{17}+273916728 L^4 M^3 \rb^{16} \nonumber \\
&& \quad -\: 286552800 L^4 M^2 \rb^{17}+80573500 L^4 M \rb^{18}-28875 L^4 \rb^{19}-24711184 L^2 M^3 \rb^{18}+14372004 L^2 M^2 \rb^{19} \nonumber \\
&& \quad -\: 855050 L^2 M \rb^{20}-50750 L^2 \rb^{21}+309120 M^3 \rb^{20}-245280 M^2 \rb^{21}+13125 \rb^{23}\big),
\end{IEEEeqnarray*}
\begin{IEEEeqnarray*}{rCl}
F^{\mathcal{K}}_{\phi\lpow{6}} &=& -875 E^6 \rb^{23} \left(-105 L^6+829 L^4 \rb^2-587 L^2 \rb^4+15 \rb^6\right) \nonumber \\
&& +\: 175 E^4 \rb^{10} \left(L^2+\rb^2\right) \big(-491520 L^{16} M-3092480 L^{14} M \rb^2-8102400 L^{12} M \rb^4-11336736 L^{10} M \rb^6\nonumber \\
&& \quad -\: 8972282 L^8 M \rb^8-3855372 L^6 M \rb^{10}-750282 L^4 M \rb^{12}+3825 L^4 \rb^{13}-12696 L^2 M \rb^{14}\nonumber \\
&& \quad -\: 5550 L^2 \rb^{15}+225 \rb^{17}\big) \nonumber \\
&& -\: E^2 \rb^5 \left(L^2+\rb^2\right)^2 \big(1007616000 L^{18} M^2+3318681600 L^{16} M^2 \rb^2+688128000 L^{16} M \rb^3+2674925056 L^{14} M^2 \rb^4\nonumber \\
&& \quad +\: 3036364800 L^{14} M \rb^5-2164346848 L^{12} M^2 \rb^6+5191177600 L^{12} M \rb^7-4277576360 L^{10} M^2 \rb^8\nonumber \\
&& \quad +\: 4156658800 L^{10} M \rb^9-1698491920 L^8 M^2 \rb^{10}+1358613200 L^8 M \rb^{11}+281096500 L^6 M^2 \rb^{12}\nonumber \\
&& \quad -\: 46924500 L^6 M \rb^{13}+248366216 L^4 M^2 \rb^{14}-97522600 L^4 M \rb^{15}+7875 L^4 \rb^{16}+15819372 L^2 M^2 \rb^{16}\nonumber \\
&& \quad -\: 3686900 L^2 M \rb^{17}-456750 L^2 \rb^{18}-312480 M^2 \rb^{18}+39375 \rb^{20}\big) \nonumber \\
&& +\: \rb^2 \left(L^2+\rb^2\right)^3 \big(-2887680000 L^{18} M^3+1290240000 L^{18} M^2 \rb-8163901440 L^{16} M^3 \rb^2+1095372800 L^{16} M^2 \rb^3\nonumber \\
&& \quad +\: 1123584000 L^{16} M \rb^4-1209975296 L^{14} M^3 \rb^4-11012229120 L^{14} M^2 \rb^5+5089056000 L^{14} M \rb^6\nonumber \\
&& \quad +\: 19718951872 L^{12} M^3 \rb^6-29772000928  L^{12} M^2 \rb^7+9243911600 L^{12} M \rb^8+28381407744 L^{10} M^3 \rb^8\nonumber \\
&& \quad -\: 31906051368 L^{10} M^2 \rb^9+8496115600 L^{10} M \rb^{10}+16529207256 L^8 M^3 \rb^{10}-16531893472 L^8 M^2 \rb^{11} \nonumber \\
&& \quad +\: 4060456400 L^8 M \rb^{12}+4051974744 L^6 M^3 \rb^{12}-3820351368 L^6 M^2 \rb^{13}+903318850 L^6 M \rb^{14} \nonumber \\
&& \quad +\: 233659760 L^4 M^3 \rb^{14}-228732252 L^4 M^2 \rb^{15}+59520650 L^4 M \rb^{16}-3500 L^4 \rb^{17}-17332624 L^2 M^3 \rb^{16}\nonumber \\
&& \quad +\: 10640724 L^2 M^2 \rb^{17}-891800 L^2 M \rb^{18}-11375 L^2 \rb^{19}+309120 M^3 \rb^{18}-245280 M^2 \rb^{19}+13125 \rb^{21}\big).
\end{IEEEeqnarray*}

\subsection{Electromagnetic case}
In the electromagnetic case, an ambiguity arises in the definition of $u^a$ in the angular
directions \emph{away} from the world-line. In \eqref{eqn:SelfForceEM} one is free to define
$u^a(x)$ as they wish provided $\lim_{x->\xb} u^a(x) = u^\ab$. A natural covariant choice
would be to define this through parallel transport, $u^a(x) = g^a{}_\bb u^\bb$. However,
in reality it is more practical in numerical calculations to define $u^a$ such that its
components in Schwarzschild
coordinates are equal to the components of $u^\ab$ in Schwarzschild coordinates. Doing so,
the regularization parameters are given by
\begin{gather}
F_{t\lnpow{1}} = -\frac{\rbdot \sgn(\Delta r)}{2 (L^2+\rb^2)}, \quad
F_{r\lnpow{1}} = \frac{E\rb \sgn(\Delta r)}{2 (\rb-2M) (L^2+\rb^2)}, \quad
F_{\theta\lnpow{1}} =0, \quad
F_{\phi\lnpow{1}} = 0,
\end{gather}

\begin{equation}
F_{t[0]} = -\frac{E \rbdot}{\pi  \rb \left(\rb^2+L^2\right)^{3/2}} \left( \rb^2 \mathcal{K} +2 L^2 \mathcal{E}\right),
\end{equation}

\begin{equation}
F_{r[0]} = \frac{1}{\pi  \rb^3 \left(\rb^2+L^2\right)^{3/2} \left(\rb-2 M\right)} \left(F^{\mathcal{E}}_{r[0]} \mathcal{E} + F^{\mathcal{K}}_{r[0]} \mathcal{K} \right),
\end{equation}
where
\begin{IEEEeqnarray*}{rCl}
F^{\mathcal{E}}_{r[0]} &=& 2 E^2 L^2 \rb^3 + \left(L^2+\rb^2\right) \left(2 L^2+\rb^2\right) (2 M-\rb), \nonumber \\
F^{\mathcal{K}}_{r[0]} &=& E^2 \rb^5 + \rb^2 \left(L^2+\rb^2\right) (\rb-2 M),
\end{IEEEeqnarray*}

\begin{equation}
F_{\theta[0]} = 0,
\end{equation}

\begin{equation} 
F_{\phi[0]} = \frac{ \rbdot}{\pi  L \rb \sqrt{L^2+\rb^2}} \left[\mathcal{E} \left(2 L^2+\rb^2\right)-\mathcal{K} \rb^2 \right],
\end{equation}

\begin{equation}
F_{t\lpow{2}} = -\frac{E \rbdot}{2 \pi  \rb^4 \left(L^2+\rb^2\right)^{7/2}} \left( F^{\mathcal{E}}_{t\lpow{2}} \mathcal{E} + F^{\mathcal{K}}_{t\lpow{2}} \mathcal{K} \right),
\end{equation}
where
\begin{IEEEeqnarray*}{rCl}
F^{\mathcal{E}}_{t\lpow{2}} &=& 2 E^2 \rb^5 \left(-L^4+10 L^2 \rb^2+3 \rb^4\right)+\left(L^2+\rb^2\right) \left(60 L^6 M+168 L^4 M \rb^2+182 L^2 M \rb^4-13 L^2 \rb^5+58 M \rb^6-5 \rb^7\right), \nonumber \\
F^{\mathcal{K}}_{t\lpow{2}} &=& 2 \rb^2 \left(L^2+\rb^2\right) \left(-21 L^4 M-48 L^2 M \rb^2+3 L^2 \rb^3-23 M \rb^4+\rb^5\right)-E^2 \rb^7 \left(11 L^2+3\rb^2\right),
\end{IEEEeqnarray*}

\begin{equation}
F_{r\lpow{2}} = -\frac{1}{2 \pi  \rb^6 \left(L^2+\rb^2\right)^{7/2} (\rb-2 M)} \left(F^{\mathcal{E}}_{r\lpow{2}} \mathcal{E} + F^{\mathcal{K}}_{r\lpow{2}} \mathcal{K} \right),
\end{equation}
where
\begin{IEEEeqnarray*}{rCl}
F^{\mathcal{E}}_{r\lpow{2}} &=& \: -2 E^4 \rb^8 \left(-L^4+10 L^2 \rb^2+3 \rb^4\right)-\left(L^2+\rb^2\right)^2 (2 M-\rb) \left(44 L^6 M+94 L^4 M \rb^2+54 L^2 M \rb^4+L^2 \rb^5+3 \rb^7\right) \nonumber \\
&& +\: 2 E^2 \rb^3 \left(L^2+\rb^2\right) \left(-30 L^6 M-86 L^4 M \rb^2+L^4 \rb^3-98 L^2 M \rb^4+10 L^2 \rb^5-26 M \rb^6+\rb^7\right), \nonumber \\
F^{\mathcal{K}}_{r\lpow{2}} &=& \: E^4 \rb^{10} \left(11 L^2+3 \rb^2\right)-\rb^2 \left(L^2+\rb^2\right)^2 (\rb-2 M) \left(22 L^4 M+24 L^2 M \rb^2+2 L^2 \rb^3+3 \rb^5\right) \nonumber \\
&& +\: E^2 \rb^5 \left(L^2+\rb^2\right) \left(42 L^4 M+98 L^2 M \rb^2-7 L^2 \rb^3+40 M \rb^4+\rb^5\right),
\end{IEEEeqnarray*}

\begin{equation}
F_{\theta\lpow{2}} = 0,
\end{equation}

\begin{equation}
F_{\phi\lpow{2}} = -\frac{\rbdot}{2 \pi  L \rb^4 \left(L^2+\rb^2\right)^{5/2}} \left(F^{\mathcal{E}}_{\phi\lpow{2}} \mathcal{E} + F^{\mathcal{K}}_{\phi\lpow{2}} \mathcal{K} \right),
\end{equation}
where
\begin{IEEEeqnarray*}{rCl}
F^{\mathcal{E}}_{\phi\lpow{2}} &=& E^2 \rb^5 \left(-2 L^4-7 L^2 \rb^2+3 \rb^4\right)-\left(L^2+\rb^2\right) \left(44 L^6 M+94 L^4 M \rb^2+54 L^2 M \rb^4+L^2 \rb^5+3 \rb^7\right), \nonumber \\
F^{\mathcal{K}}_{\phi\lpow{2}} &=& \rb^2 \left(L^2+\rb^2\right) \left(22 L^4 M+24 L^2 M \rb^2+2 L^2 \rb^3+3 \rb^5\right)-E^2 \rb^7 \left(3 \rb^2-L^2\right),
\end{IEEEeqnarray*}

\begin{equation}
F_{t\lpow{4}} = \frac{3 E \rbdot}{40 \pi  \rb^{11} \left(L^2+\rb^2\right)^{11/2}} \left(F^{\mathcal{E}}_{t\lpow{4}} \mathcal{E} + F^{\mathcal{K}}_{t\lpow{4}} \mathcal{K} \right),
\end{equation}
where
\begin{IEEEeqnarray*}{rCl}
F^{\mathcal{E}}_{t\lpow{4}} &=& -30 E^4 \rb^{14} \left(3 L^6-102 L^4 \rb^2+43 L^2 \rb^4+20 \rb^6\right) \nonumber \\
&& +\: 2 E^2 \rb^5 \left(L^2+\rb^2\right) \big(34560 L^{12} M+169728 L^{10} M \rb^2 + 333564 L^8 M \rb^4 + 328912 L^6 M \rb^6+167074 L^4 M \rb^8\nonumber \\
&& \quad -\: 1245 L^4 \rb^9+32948 L^2 M \rb^{10}+1230 L^2 \rb^{11}+230 M \rb^{12} + 555 \rb^{13}\big) \nonumber \\
&& -\: 4 \left(L^2+\rb^2\right)^2 \big(11520 L^{14} M^2-18240 L^{12} M^2 \rb^2+33600 L^{12} M \rb^3-226624 L^{10} M^2\rb^4 + 153960 L^{10} M \rb^5 \nonumber \\
&& \quad -\: 496164 L^8 M^2 \rb^6+280764 L^8 M \rb^7-485652 L^6 M^2 \rb^8+255197 L^6 M \rb^9-230930 L^4 M^2 \rb^{10} + 116771 L^4 M \rb^{11}\nonumber \\
&& \quad -\: 30 L^4 \rb^{12}-45200 L^2 M^2 \rb^{12}+21487 L^2 M \rb^{13}+270 L^2 \rb^{14}-1342 M^2 \rb^{14}+229 M \rb^{15}+120 \rb^{16}\big), \nonumber \\
F^{\mathcal{K}}_{t\lpow{4}} &=& 15 E^4 \rb^{16} \left(-87 L^4+66 L^2 \rb^2+25 \rb^4\right) \nonumber \\
&& -\: 4 E^2 \rb^7 \left(L^2+\rb^2\right) \big(8640 L^{10} M+34872 L^8 M \rb^2+53283 L^6 M \rb^4 + 37555 L^4 M \rb^6-225 L^4 \rb^7+9809 L^2 M \rb^8 \nonumber \\
&& \quad +\: 420 L^2 \rb^9+265 M \rb^{10}+165 \rb^{11}\big) \nonumber \\
&& +\: \rb^2 \left(L^2+\rb^2\right)^2 \big(23040 L^{12} M^2 - 56640 L^{10} M^2 \rb^2+67200 L^{10} M \rb^3-402608 L^8 M^2 \rb^4+249120 L^8 M \rb^5 \nonumber \\
&& \quad -\: 643736 L^6 M^2 \rb^6+346608 L^6 M \rb^7 - 427796 L^4 M^2 \rb^8+217192 L^4 M \rb^9-110916 L^2 M^2 \rb^{10} \nonumber \\
&& \quad +\: 52580 L^2 M \rb^{11}+615 L^2 \rb^{12}-5368 M^2 \rb^{12} + 1516 M \rb^{13}+255 \rb^{14}\big),
\end{IEEEeqnarray*}

\begin{equation}
F_{r\lpow{4}} = \frac{3}{40 \pi  \rb^{13} \left(L^2+\rb^2\right)^{11/2} (\rb-2 M)} \left(F^{\mathcal{E}}_{r\lpow{4}} \mathcal{E} + F^{\mathcal{K}}_{r\lpow{4}} \mathcal{K} \right),
\end{equation}
where
\begin{IEEEeqnarray*}{rCl}
F^{\mathcal{E}}_{r\lpow{4}} &=& 30 E^6 \rb^{17} \left(3 L^6-102 L^4 \rb^2+43 L^2 \rb^4+20 \rb^6\right) \nonumber \\
&& -\: E^4 \rb^8 \left(L^2+\rb^2\right) \big(69120 L^{12} M+339456 L^{10} M \rb^2+667128 L^8 M \rb^4 + 657884 L^6 M \rb^6-30 L^6 \rb^7+334658 L^4 M \rb^8 \nonumber \\
&& \quad -\: 2745 L^4 \rb^9+62656 L^2 M \rb^{10}+4080 L^2 \rb^{11}+610 M \rb^{12}+1035 \rb^{13}\big) \nonumber \\
&& +\: \left(L^2+\rb^2\right)^3 (2 M-\rb) \big(46080 L^{14} M^2+89856 L^{12} M^2 \rb^2+53760 L^{12} M \rb^3-86336 L^{10} M^2 \rb^4+211520 L^{10} M \rb^5 \nonumber \\
&& \quad -\: 344128 L^8 M^2 \rb^6+317600 L^8 M \rb^7-306808 L^6 M^2 \rb^8+221140 L^6 M \rb^9-98676 L^4 M^2 \rb^{10}+66220 L^4 M \rb^{11}  \nonumber \\
&& \quad +\: 60 L^4 \rb^{12}-5244 L^2 M^2 \rb^{12}+4440 L^2 M \rb^{13}+105 L^2 \rb^{14}+160 M^2 \rb^{14}-75 \rb^{16}\big)  \nonumber \\
&&+\: 2 E^2 \rb^3 \left(L^2+\rb^2\right)^2 \big(23040 L^{14} M^2 - 36480 L^{12} M^2 \rb^2+67200 L^{12} M \rb^3-445056 L^{10} M^2 \rb^4+303824 L^{10} M \rb^5 \nonumber \\
&& \quad -\: 959952 L^8 M^2 \rb^6+545340 L^8 M \rb^7 - 922996 L^6 M^2 \rb^8+486240 L^6 M \rb^9-428644 L^4 M^2 \rb^{10}+216604 L^4 M \rb^{11} \nonumber \\
&& \quad +\: 105 L^4 \rb^{12}-78428 L^2 M^2 \rb^{12} + 35368 L^2 M \rb^{13}+1350 L^2 \rb^{14}-2684 M^2 \rb^{14}+608 M \rb^{15}+165 \rb^{16}\big), \nonumber \\
F^{\mathcal{K}}_{r\lpow{4}} &=& -15 E^6 \rb^{19} \left(-87 L^4+66 L^2 \rb^2+25 \rb^4\right) \nonumber \\
&& +\: E^4 \rb^{10} \left(L^2+\rb^2\right) \big(34560 L^{10} M+139488 L^8 M \rb^2+213132 L^6 M \rb^4 + 150250 L^4 M \rb^6-915 L^4 \rb^7+37496 L^2 M \rb^8 \nonumber \\
&& \quad +\: 2550 L^2 \rb^9+1210 M \rb^{10}+585 \rb^{11}\big) \nonumber \\
&& -\: \rb^2 \left(L^2+\rb^2\right)^3 \left(\rb-2 M \right) \big(63760 L^8 M^2 \rb^4 -23040 L^{12} M^2-24768 L^{10} M^2 \rb^2-26880 L^{10} M \rb^3-82240 L^8 M \rb^5 \nonumber \\
&& \quad +\: 115848 L^6 M^2 \rb^6-87920 L^6 M \rb^7+56084 L^4 M^2 \rb^8-36340 L^4 M \rb^9+5324 L^2 M^2 \rb^{10}-3540 L^2 M \rb^{11} \nonumber \\
&& \quad +\: 15 L^2 \rb^{12} - 160 M^2 \rb^{12}+75 \rb^{14}\big) \nonumber \\
&& -\: E^2 \rb^5 \left(L^2+\rb^2\right)^2 \big(23040 L^{12} M^2-56640 L^{10} M^2 \rb^2+67200 L^{10} M \rb^3-394416 L^8 M^2 \rb^4 + 245024 L^8 M \rb^5 \nonumber \\
&& \quad -\: 618528 L^6 M^2 \rb^6+334004 L^6 M \rb^7-400636 L^4 M^2 \rb^8+203252 L^4 M \rb^9+180 L^4 \rb^{10} - 97892 L^2 M^2 \rb^{10} \nonumber \\
&& \quad +\: 44568 L^2 M \rb^{11}+1365 L^2 \rb^{12}-5368 M^2 \rb^{12}+1816 M \rb^{13}+105 \rb^{14}\big),
\end{IEEEeqnarray*}

\begin{equation}
F_{\theta\lpow{4}} = 0,
\end{equation}

\begin{equation}
F_{\phi\lpow{4}} = \frac{3\rbdot}{40 \pi  L \rb^{11} \left(L^2+\rb^2\right)^{9/2}} \left(F^{\mathcal{E}}_{\phi\lpow{4}} \mathcal{E} + F^{\mathcal{K}}_{\phi\lpow{4}} \mathcal{K} \right),
\end{equation}
where
\begin{IEEEeqnarray*}{rCl}
F^{\mathcal{E}}_{\phi\lpow{4}} &=& -15 E^4 \rb^{14} \left(2 L^6+17 L^4 \rb^2-108 L^2 \rb^4+5 \rb^6\right) \nonumber \\
&& +\: 2 E^2 \rb^7 \left(L^2+\rb^2\right) \big(4096 L^{10} M+16188 L^8 M \rb^2+24154 L^6 M \rb^4 + 16608 L^4 M \rb^6-165 L^4 \rb^7+5986 L^2 M \rb^8 \nonumber \\
&& \quad -\: 810 L^2 \rb^9+75 \rb^{11}\big) \nonumber \\
&& +\left(L^2+\rb^2\right)^2 \big(46080 L^{14} M^2+89856 L^{12} M^2 \rb^2 + 53760 L^{12} M \rb^3-86336 L^{10} M^2 \rb^4+211520 L^{10} M \rb^5 \nonumber \\
&& \quad -\: 344128 L^8 M^2 \rb^6+317600 L^8 M \rb^7-306808 L^6 M^2 \rb^8 + 221140 L^6 M \rb^9-98676 L^4 M^2 \rb^{10} \nonumber \\
&& \quad +\: 66220 L^4 M \rb^{11}+60 L^4 \rb^{12}-5244 L^2 M^2 \rb^{12}+4440 L^2 M \rb^{13}+105 L^2 \rb^{14} + 160 M^2 \rb^{14}-75 \rb^{16}\big), \nonumber \\
F^{\mathcal{K}}_{\phi\lpow{4}} &=& 15 E^4 \rb^{16} \left(L^4-58 L^2 \rb^2+5 \rb^4\right) \nonumber \\
&& -\: 2 E^2 \rb^9 \left(L^2+\rb^2\right) \big(2048 L^8 M+6302 L^6 M \rb^2+6790 L^4 M \rb^4-90 L^4 \rb^5 + 3256 L^2 M \rb^6-375 L^2 \rb^7+75 \rb^9\big) \nonumber \\
&& +\rb^2 \left(L^2+\rb^2\right)^2 \big(-23040 L^{12} M^2-24768 L^{10} M^2 \rb^2-26880 L^{10} M \rb^3 + 63760 L^8 M^2 \rb^4-82240 L^8 M \rb^5 \nonumber \\
&& \quad +\: 115848 L^6 M^2 \rb^6-87920 L^6 M \rb^7+56084 L^4 M^2 \rb^8-36340 L^4 M \rb^9 + 5324 L^2 M^2 \rb^{10}-3540 L^2 M \rb^{11} \nonumber \\
&& \quad +\: 15 L^2 \rb^{12}-160 M^2 \rb^{12}+75 \rb^{14}\big).
\end{IEEEeqnarray*}

\subsection{Gravitational case}
\subsubsection{Self-force regularization}
The self force on a gravitational particle is given by
\begin{equation}
F^a = k^{abcd} \hb^{\rm R}_{bc;d},
\end{equation}
where
\begin{equation}
k^{abcd} \equiv \frac12 g^{ad} u^b u^c - g^{ab} u^c u^d - 
  \frac12 u^a u^b u^c u^d + \frac14 u^a g^{b c} u^d + 
 \frac14 g^{a d} g^{b c}.
\end{equation}
Note that, as in the electromagnetic case, an ambiguity arises here due to the presense of
terms involving the four-velocity at $x$. One is free to arbitrarily choose how to define this
provided $\lim_{x\to\bar{x}} u^a = u^{\ab}$. Following Barack and Sago~\cite{Barack:Sago:2010},
we choose to take the Schwarzschild
components of the four velocity at $x$ to be exactly those at $\xb$.  The regularisation parameters in the gravitational case are given by

\begin{gather}
F^t_{\lnpow{1}} = \mp\frac{\rb \rbdot}{2 (L^2+\rb^2) (\rb-2M)}, \quad
F^r_{\lnpow{1}} = \mp\frac{E}{2 (L^2+\rb^2)}, \quad
F^\theta_{\lnpow{1}} = 0, \quad
F^\phi_{\lnpow{1}} = 0,
\end{gather}
\begin{gather}
F^t_{[0]} = -\frac{\rbdot E}{\pi (\rb-2 M) (L^2+\rb^2)^{3/2}} (2 L^2 \mathcal{E}+ \rb^2 \mathcal{K}), \\
F^r_{[0]} = \frac{1}{\pi \rb^4 (L^2+\rb^2)^{3/2}} \Bigg\{
    \big[(\rb-2 M) (L^2 + \rb^2) (2 L^2 + \rb^2)-2 E^2 L^2 \rb^3\big]\mathcal{E}
  - \rb^2 \big[( \rb - 2 M ) (L^2 + \rb^2)+E^2 \rb^3\big] \mathcal{K}
  \Big\},
\nonumber \\
F^\theta_{[0]} = 0, \quad
F^\phi_{[0]} = - \frac{\rbdot }{\pi L \rb^3 (L^2+\rb^2)^{1/2}}[(2L^2+\rb^2) \mathcal{E}-\rb^2 \mathcal{K}],
\end{gather}

\begin{equation}
F^t_{\lpow{2}} = \frac{E \rbdot}{2 \pi \rb^3 (\rb-2 M) (L^2+r^2)^{7/2}}  (F_{\mathcal{E}\lpow{2}}^t \mathcal{E} + D_{\mathcal{K}\lpow{2}}^t \mathcal{K}),
\end{equation}
where
\begin{IEEEeqnarray*}{rCl}
F_{\mathcal{E}\lpow{2}}^t &=& - 2 E^2 \rb^5 (11 L^4 + 34 L^2 \rb^2 + 15 \rb^4) \nonumber \\
 && - (L^2 + \rb^2) (276 L^6 M + 768 L^4 M \rb^2 + 782 L^2 M \rb^4 - 37 L^2 \rb^5 + 274 M \rb^6 - 29 \rb^7), \\
F_{\mathcal{K}\lpow{2}}^t &=& E^2 \rb^5 (12 L^4 + 35 L^2 \rb^2 + 15 \rb^4) + 2 \rb^2 (L^2 + \rb^2) (93 L^4 M + 204 L^2 M \rb^2 - 9 L^2 \rb^3 + 107 M \rb^4 - 7 \rb^5),
\end{IEEEeqnarray*}

\begin{equation}
F^r_{\lpow{2}} = \frac{1}{2 \pi \rb^7 (L^2+\rb^2)^{7/2}}(F_{\mathcal{E}\lpow{2}}^r \mathcal{E} + F_{\mathcal{K}\lpow{2}}^r \mathcal{K}),
\end{equation}
where
\begin{IEEEeqnarray*}{rCl}
F_{\mathcal{E}\lpow{2}}^r &=& (\rb-2M) (L^2 + \rb^2)^2 (188 L^6 M + 406 L^4 M \rb^2 + 222 L^2 M \rb^4 + 13 L^2 \rb^5 + 15 \rb^7) - 2 E^4 \rb^8 (11 L^4 + 34 L^2 \rb^2 + 15 \rb^4)\nonumber \\
&& -\: 2 E^2 \rb^3 (L^2 + \rb^2) (138 L^6 M + 422 L^4 M \rb^2 - 19 L^4 \rb^3 + 422 L^2 M \rb^4 - 34 L^2 \rb^5 + 122 M \rb^6 - 7 \rb^7), \nonumber\\
F_{\mathcal{K}\lpow{2}}^r &=& - \rb^2 (\rb-2 M) (L^2 + \rb^2)^2 (94 L^4 M + 96 L^2 M \rb^2 + 14 L^2 \rb^3 + 15 \rb^5) + E^4 \rb^8 (12 L^4 + 35 L^2 \rb^2 + 15 \rb^4)\nonumber \\
&& + E^2 \rb^5 (L^2 + \rb^2) (210 L^4 M - 12 L^4 \rb + 410 L^2 M \rb^2 - 19 L^2 \rb^3 + 184 M \rb^4 + \rb^5)],
\end{IEEEeqnarray*}

\begin{equation}
F^\theta_{\lpow{2}} = 0,
\end{equation}

\begin{equation}
F^\phi_{\lpow{2}} = \frac{\rbdot}{2 \pi  L^3 \rb^6 (L^2+\rb^2)^{5/2}}(F_{\mathcal{E}\lpow{2}}^\phi \mathcal{E} + F_{\mathcal{K}\lpow{2}}^\phi \mathcal{K}),
\end{equation}
where
\begin{IEEEeqnarray*}{rCl}
F_{\mathcal{E}\lpow{2}}^\phi &=&  - (L^2 + \rb^2) (188 L^8 M + 406 L^6 M \rb^2 - 64 L^6 \rb^3 + 222 L^4 M \rb^4 - 163 L^4 \rb^5 - 145 L^2 \rb^7 - 48 \rb^9) \nonumber \\
&& -\: E^2 L^2 \rb^5 (38 L^4 + 31 L^2 \rb^2 - 15 \rb^4), \\
F_{\mathcal{K}\lpow{2}}^\phi &=& E^2 L^2 \rb^5 (12 L^4 + L^2 \rb^2 - 15 \rb^4) + \rb^2 (L^2 + \rb^2) (94 L^6 M + 96 L^4 M \rb^2 - 74 L^4 \rb^3 - 121 L^2 \rb^5 - 48 \rb^7),
\end{IEEEeqnarray*}

\begin{equation}
F^t_{\lpow{4}} = \frac{3 E \rbdot }{40 \pi \rb^{10} (\rb-2 M) (L^2+r^2)^{11/2}}(F_{\mathcal{E}\lpow{4}}^t \mathcal{E} + F_{\mathcal{K}\lpow{4}}^t \mathcal{K}),
\end{equation}
where
\begin{IEEEeqnarray*}{rCl}
F_{\mathcal{E}\lpow{4}}^t &=& 30 E^4 \rb^{10} (64 L^{10} + 384 L^8 \rb^2 + 989 L^6 \rb^4 + 1222 L^4 \rb^6 + 437 L^2 \rb^8 + 12 \rb^{10})
\nonumber \\ &&
  +\: 2 E^2 \rb^5 (L^2 + \rb^2) (92160 L^{12} M + 445008 L^{10} M \rb^2 + 859044 L^8 M \rb^4 - 1920 L^8 \rb^5 + 838312 L^6 M \rb^6 - 9780 L^6 \rb^7
\nonumber \\ &&
  \quad +\: 433114 L^4 M \rb^8 - 19065 L^4 \rb^9 + 102188 L^2 M \rb^{10} - 9870 L^2 \rb^{11} + 5030 M \rb^{12} - 585 \rb^{13})
\nonumber \\ &&
  +\: 4 (L^2 + \rb^2)^2 (46080 L^{14} M^2 + 403200 L^{12} M^2 \rb^2 - 48000 L^{12} M \rb^3 + 1231984 L^{10} M^2 \rb^4 - 219840 L^{10} M \rb^5
\nonumber \\ &&
  \quad +\: 1841004 L^8 M^2 \rb^6 - 411324 L^8 M \rb^7 + 1490772 L^6 M^2 \rb^8 - 406397 L^6 M \rb^9 + 480 L^6 \rb^{10} + 668810 L^4 M^2 \rb^{10}
\nonumber \\ &&
  \quad - 232331 L^4 M \rb^{11} + 2040 L^4 \rb^{12} + 161840 L^2 M^2 \rb^{12} - 75367 L^2 M \rb^{13} + 1590 L^2 \rb^{14} + 18382 M^2 \rb^{14}
\nonumber \\ &&
  \quad -\: 10669 M \rb^{15} + 210 \rb^{16}), \\
F_{\mathcal{K}\lpow{4}}^t &=& -15 E^4 \rb^{12} (64 L^8 + 328 L^6 \rb^2 + 495 L^4 \rb^4 + 22 L^2 \rb^6 - 81 \rb^8)
\nonumber \\ &&
  -\: 4 E^2 \rb^7 (L^2 + \rb^2) (25920 L^{10} M + 102372 L^8 M \rb^2 + 152523 L^6 M \rb^4 - 480 L^6 \rb^5 + 103375 L^4 M \rb^6 - 1575 L^4 \rb^7
\nonumber \\ &&
  \quad +\: 25889 L^2 M \rb^8 - 120 L^2 \rb^9 - 455 M \rb^{10} + 495 \rb^{11})
\nonumber \\ &&
  -\:  \rb^2 (L^2 + \rb^2)^2 (92160 L^{12} M^2 + 766080 L^{10} M^2 \rb^2 - 130560 L^{10} M \rb^3 + 2014208 L^8 M^2 \rb^4 - 497760 L^8 M \rb^5
\nonumber \\ &&
  \quad +\: 2443016 L^6 M^2 \rb^6 - 743088 L^6 M \rb^7 + 1527236 L^4 M^2 \rb^8 - 552712 L^4 M \rb^9 + 960 L^4 \rb^{10} + 496596 L^2 M^2 \rb^{10}
\nonumber \\ &&
  \quad -\: 206180 L^2 M \rb^{11} - 135 L^2 \rb^{12} + 73528 M^2 \rb^{12} - 30796 M \rb^{13} - 735 \rb^{14}),
\end{IEEEeqnarray*}

\begin{equation}
F^r_{\lpow{4}} = \frac{3 }{40 \pi \rb^{14} (L^2+\rb^2)^{11/2}}(F_{\mathcal{E}\lpow{4}}^r \mathcal{E} + F_{\mathcal{K}\lpow{4}}^r \mathcal{K}),
\end{equation}
where
\begin{IEEEeqnarray*}{rCl}
F_{\mathcal{E}\lpow{4}}^r &=& 30 E^6 \rb^{13} (64 L^{10} + 384 L^8 \rb^2 + 989 L^6 \rb^4 + 1222 L^4 \rb^6 + 437 L^2 \rb^8 + 12 \rb^{10})
\nonumber \\ &&
  +\: E^4 \rb^8 (L^2 + \rb^2) (184320 L^{12} M + 893856 L^{10} M \rb^2 - 1920 L^{10} \rb^3 + 1746888 L^8 M \rb^4 - 18240 L^8 \rb^5
\nonumber \\ &&
  \quad  + 1744604 L^6 M \rb^6 - 53550 L^6 \rb^7 + 907298 L^4 M \rb^8 - 58665 L^4 \rb^9 + 197536 L^2 M \rb^{10} - 16320 L^2 \rb^{11} + 9010 M \rb^{12}
\nonumber \\ &&
  \quad  - 645 \rb^{13})
\nonumber \\ &&
  +\: 2 E^2 \rb^3 (L^2 + \rb^2)^2 (92160 L^{14} M^2 + 1036800 L^{12} M^2 \rb^2 - 211200 L^{12} M \rb^3 + 3363456 L^{10} M^2 \rb^4 - 889424 L^{10} M \rb^5
\nonumber \\ &&
  \quad +\: 5003472 L^8 M^2 \rb^6 - 1487220 L^8 M \rb^7 + 1920 L^8 \rb^8 + 3866356 L^6 M^2 \rb^8 - 1272840 L^6 M \rb^9 + 9780 L^6 \rb^{10}
\nonumber \\ &&
  \quad +\: 1570564 L^4 M^2 \rb^{10} - 594724 L^4 M \rb^{11} + 10875 L^4 \rb^{12} + 321308 L^2 M^2 \rb^{12} - 146848 L^2 M \rb^{13} + 1830 L^2 \rb^{14}
\nonumber \\ &&
  \quad +\: 36764 M^2 \rb^{14} - 20288 M \rb^{15} - 105 \rb^{16})
\nonumber \\ &&
  - \:(\rb - 2 M) (L^2 + \rb^2)^3 (184320 L^{14} M^2 + 1711104 L^{12} M^2 \rb^2 - 245760 L^{12} M \rb^3 + 4872896 L^{10} M^2 \rb^4
\nonumber \\ &&
  \quad -\: 884480 L^{10} M \rb^5 + 6311728 L^8 M^2 \rb^6 - 1185120 L^8 M \rb^7 + 4083688 L^6 M^2 \rb^8 - 721620 L^6 M \rb^9 + 1920 L^6 \rb^{10}
\nonumber \\ &&
  \quad +\: 1299396 L^4 M^2 \rb^{10} - 198700 L^4 M \rb^{11} + 2460 L^4 \rb^{12} + 209484 L^2 M^2 \rb^{12} - 28120 L^2 M \rb^{13} - 105 L^2 \rb^{14}
\nonumber \\ &&
  \quad +\: 28640 M^2 \rb^{14} - 5120 M \rb^{15} - 525 \rb^{16}), \\
F_{\mathcal{K}\lpow{4}}^r &=& -15 E^6 \rb^{15} (64 L^8 + 328 L^6 \rb^2 + 495 L^4 \rb^4 + 22 L^2 \rb^6 - 81 \rb^8)
\nonumber \\ &&
  -\: E^4 \rb^{10} (L^2 + \rb^2) (103680 L^{10} M + 415248 L^8 M \rb^2 - 2880 L^8 \rb^3 + 627612 L^6 M \rb^4 - 10680 L^6 \rb^5 + 418570 L^4 M \rb^6
\nonumber \\ &&
  \quad -\: 8835 L^4 \rb^7 + 93896 L^2 M \rb^8 + 4350 L^2 \rb^9 - 2870 M \rb^{10} + 2505 \rb^{11})
\nonumber \\ &&
  -\: E^2 \rb^5 (L^2 + \rb^2)^2 (92160 L^{12} M^2 + 1019520 L^{10} M^2 \rb^2 - 257280 L^{10} M \rb^3 + 2806176 L^8 M^2 \rb^4 - 893744 L^8 M \rb^5
\nonumber \\ &&
  \quad +\: 3309168 L^6 M^2 \rb^6 - 1180004 L^6 M \rb^7 + 1920 L^6 \rb^8 + 1878796 L^4 M^2 \rb^8 - 724772 L^4 M \rb^9 - 900 L^4 \rb^{10}
\nonumber \\ &&
  \quad +\: 517652 L^2 M^2 \rb^{10} - 205608 L^2 M \rb^{11} - 5685 L^2 \rb^{12} + 73528 M^2 \rb^{12} - 28696 M \rb^{13} - 1785 \rb^{14})
\nonumber \\ &&
  +\: \rb^2 (\rb - 2 M) (L^2 + \rb^2)^3 (92160 L^{12} M^2 + 815232 L^{10} M^2 \rb^2 - 157440 L^{10} M \rb^3 + 1939840 L^8 M^2 \rb^4
\nonumber \\ &&
  \quad -\: 468160 L^8 M \rb^5 + 1942488 L^6 M^2 \rb^6 - 494560 L^6 M \rb^7 + 892484 L^4 M^2 \rb^8 - 217780 L^4 M \rb^9 - 960 L^4 \rb^{10}
\nonumber \\ &&
  \quad +\: 195164 L^2 M^2 \rb^{10} - 38820 L^2 M \rb^{11} - 1545 L^2 \rb^{12} + 28640 M^2 \rb^{12} - 5120 M \rb^{13} - 525 \rb^{14}),
\end{IEEEeqnarray*}

\begin{equation}
F^\theta_{\lpow{4}} = 0,
\end{equation}

\begin{equation}
F^\phi_{\lpow{4}} = \frac{3 \rbdot }{40 \pi \rb^{13} L^5 (L^2+r^2)^{9/2}} (F_{\mathcal{E}\lpow{4}}^\phi \mathcal{E} + F_{\mathcal{K}\lpow{4}}^\phi  \mathcal{K}) ,
\end{equation}
where
\begin{IEEEeqnarray*}{rCl}
F_{\mathcal{E}\lpow{4}}^\phi  &=& 15 E^4 L^4 \rb^{10} (128 L^{10} + 960 L^8 \rb^2 + 2266 L^6 \rb^4 + 1369 L^4 \rb^6 - 228 L^2 \rb^8 - 35 \rb^{10})
\nonumber \\ &&
  +\: 2 E^2 L^2 \rb^5 (L^2 + \rb^2) (115200 L^{14} M + 449744 L^{12} M \rb^2 + 660732 L^{10} M \rb^4 - 5760 L^{10} \rb^5 + 442406 L^8 M \rb^6
\nonumber \\ &&
  \quad -\: 21300 L^8 \rb^7 + 116472 L^6 M \rb^8 - 18475 L^6 \rb^9 - 1186 L^4 M \rb^{10} + 4310 L^4 \rb^{11} + 10765 L^2 \rb^{13} + 4240 \rb^{15})
\nonumber \\ &&
  +\: (L^2 + \rb^2)^2 (184320 L^{18} M^2 + 1711104 L^{16} M^2 \rb^2 - 245760 L^{16} M \rb^3 + 4872896 L^{14} M^2 \rb^4 - 884480 L^{14} M \rb^5
\nonumber \\ &&
  \quad +\: 6311728 L^{12} M^2 \rb^6 - 1090720 L^{12} M \rb^7 + 4083688 L^{10} M^2 \rb^8 - 436180 L^{10} M \rb^9 - 5760 L^{10} \rb^{10}
\nonumber \\ &&
  \quad +\: 1299396 L^8 M^2 \rb^{10} + 73140 L^8 M \rb^{11} - 83700 L^8 \rb^{12} + 209484 L^6 M^2 \rb^{12} + 15720 L^6 M \rb^{13} - 236585 L^6 \rb^{14}
\nonumber \\ &&
  \quad +\: 28640 L^4 M^2 \rb^{14} - 63200 L^4 M \rb^{15} - 271325 L^4 \rb^{16} - 21120 L^2 M \rb^{17} - 138400 L^2 \rb^{18} - 25600 \rb^{20}), \\
F_{\mathcal{K}\lpow{4}}^\phi  &=& -15 E^4 L^4 \rb^{12} (192 L^8 + 584 L^6 \rb^2 + 169 L^4 \rb^4 - 322 L^2 \rb^6 - 35 \rb^8)
\nonumber \\ &&
  -\: 2 E^2 L^2 \rb^7 (L^2 + \rb^2) (63360 L^{12} M + 197992 L^{10} M \rb^2 + 216538 L^8 M \rb^4 - 4800 L^8 \rb^5 + 87890 L^6 M \rb^6 - 7110 L^6 \rb^7
\nonumber \\ &&
  \quad +\: 5264 L^4 M \rb^8 + 2455 L^4 \rb^9 + 8645 L^2 \rb^{11} + 4240 \rb^{13})
\nonumber \\ &&
  -\: \rb^2 (L^2 + \rb^2)^2 (92160 L^{16} M^2 + 815232 L^{14} M^2 \rb^2 - 157440 L^{14} M \rb^3 + 1939840 L^{12} M^2 \rb^4 - 422080 L^{12} M \rb^5
\nonumber \\ &&
  \quad +\: 1942488 L^{10} M^2 \rb^6 - 309120 L^{10} M \rb^7 + 892484 L^8 M^2 \rb^8 + 9580 L^8 M \rb^9 - 39360 L^8 \rb^{10} + 195164 L^6 M^2 \rb^{10}
\nonumber \\ &&
  \quad +\: 22780 L^6 M \rb^{11} - 152745 L^6 \rb^{12} + 28640 L^4 M^2 \rb^{12} - 52640 L^4 M \rb^{13} - 213325 L^4 \rb^{14} - 21120 L^2 M \rb^{15}
\nonumber \\ &&
  \quad -\: 125600 L^2 \rb^{16} - 25600 \rb^{18}).
\end{IEEEeqnarray*}

\subsubsection{$huu$ regularization}
The quantity
\begin{equation}
H^{\reg} = \frac12 h^\reg_{ab} u^a u^b
\end{equation}
was first proposed by Detweiler \cite{Detweiler:2005} as a tool for constructing gauge invariant
measurements from self-force calculations. It has since proven invaluable in extracting gauge invariant
results from gauge dependent self-force calculations \cite{Detweiler:2008,Barack:2011ed}.

Much the same as with self-force calculations, the calculation of $H^{\reg}$ requires
the subtraction of the appropriate singular piece, $H^{\sing} = \frac12 h^{\sing}_{ab} u^a u^b$
from the full retarded field. In this section, we give this subtraction in the form of
mode-sum regularization parameters.
In doing so, we keep with our convention that the term proportional to
$l+\tfrac12$ is denoted by $H_{\lnpow{1}}$ ($=0$ in this case), the constant term is denoted by $H_{[0]}$,
and so on.

Note that, as in the self-force case, an ambiguity arises here due to the presense of
terms involving the four-velocity at $x$. One is free to arbitrarily choose how to define this
provided $\lim_{x\to\bar{x}} u^a = u^\ab$. As before, we choose this in such a way that the Schwarzschild
components of the four velocity at $x$ are exactly those at $\xb$.  The regularisation parameters are then given by

\begin{equation}
H_{[0]} = \frac{2 \mathcal{K}}{\pi \sqrt{L^2+\rb^2}},
\end{equation}
\begin{equation}
H_{\lpow{2}} = \frac{H^{\mathcal{E}}_{\lpow{2}} \mathcal{E} + H^{\mathcal{K}}_{\lpow{2}} \mathcal{K}}{\pi \rb^3 (L^2+\rb^2)^{3/2}},
\end{equation}
where
\begin{IEEEeqnarray*}{rCl}
H^{\mathcal{E}}_{\lpow{2}} &=& 2 E^2 \rb^5 + (L^2 + \rb^2) (36 L^2 M - 8 L^2 \rb + 38 M \rb^2 - 9 \rb^3), \\
H^{\mathcal{K}}_{\lpow{2}} &=& - E^2 \rb^3 (16 L^2 + 17 \rb^2) - 2 (L^2 + \rb^2) (16 L^2 M - 4 L^2 \rb + 33 M \rb^2 - 12 \rb^3),
\end{IEEEeqnarray*}
\begin{equation}
H_{\lpow{4}} = \frac{3(H^{\mathcal{E}}_{\lpow{4}} \mathcal{E} + H^{\mathcal{K}}_{\lpow{4}} \mathcal{K})}{20 \pi  r^{10} (L^2+r^2)^{7/2}},
\end{equation}
where
\begin{IEEEeqnarray*}{rCl}
H^{\mathcal{E}}_{\lpow{4}} &=& -120 E^4 \rb^{12} (8 L^4 + 17 L^2 \rb^2 + 7 \rb^4)
  + 2 E^2 \rb^5 (L^2 + \rb^2) (3584 L^8 M + 12712 L^6 M \rb^2 + 15516 L^4 M \rb^4
\nonumber \\ &&
  \quad +\: 120 L^4 \rb^5 + 6182 L^2 M \rb^6 + 735 L^2 \rb^7 + 34 M \rb^8 + 495 \rb^9)
\nonumber \\ &&
  +\: 2 (L^2 + \rb^2)^2 (1536 L^{10} M^2 + 13888 L^8 M^2 \rb^2 - 1600 L^8 M \rb^3 + 40584 L^6 M^2 \rb^4 - 9440 L^6 M \rb^5 + 46888 L^4 M^2 \rb^6
\nonumber \\ &&
  \quad -\: 14100 L^4 M \rb^7 + 120 L^4 \rb^8 + 18936 L^2 M^2 \rb^8 - 5350 L^2 M \rb^9 + 15 L^2 \rb^{10} + 340 M^2 \rb^{10} + 850 M \rb^{11} - 90 \rb^{12}), \\
H^{\mathcal{K}}_{\lpow{4}} &=& 15 E^4 \rb^{10} (64 L^6 + 224 L^4 \rb^2 + 259 L^2 \rb^4 + 91 \rb^6)
\nonumber \\ &&
  -\: 4 E^2 \rb^7 (L^2 + \rb^2) (1376 L^6 M + 3174 L^4 M \rb^2 + 420 L^4 \rb^3 + 1965 L^2 M \rb^4 + 960 L^2 \rb^5 + 227 M \rb^6 + 510 \rb^7)
\nonumber \\ &&
  -\: \rb^2 (L^2 + \rb^2)^2 (1536 L^8 M^2 + 15904 L^6 M^2 \rb^2 - 7360 L^6 M \rb^3 + 36160 L^4 M^2 \rb^4 - 19320 L^4 M \rb^5 + 22412 L^2 M^2 \rb^6
\nonumber \\ &&
  \quad -\: 11040 L^2 M \rb^7 - 720 L^2 \rb^8 + 680 M^2 \rb^8 + 860 M \rb^9 - 705 \rb^{10}).
\end{IEEEeqnarray*}

\subsection{Example}
As an example application of our high order regularization parameters, we consider the case
of a scalar particle on a circular geodesic of the Schwarzschild spacetime. In this case, the
retarded field may be computed using the frequency domain method
described in \cite{Detweiler:Messaritaki:Whiting:2002}, along with improved asymptotics for the
boundary conditions (by expanding inside the exponential rather than outside) and with the use of
the arbitrary precision differential equation solving support in
\emph{Mathematica} \cite{Mathematica}. These improvements allowed us to substantially increase the
accuracy of the computed retarded field. We found this to be necessary to get the full benefit from
the the higher order regularization parameters.

\begin{figure}
\includegraphics[width=8cm]{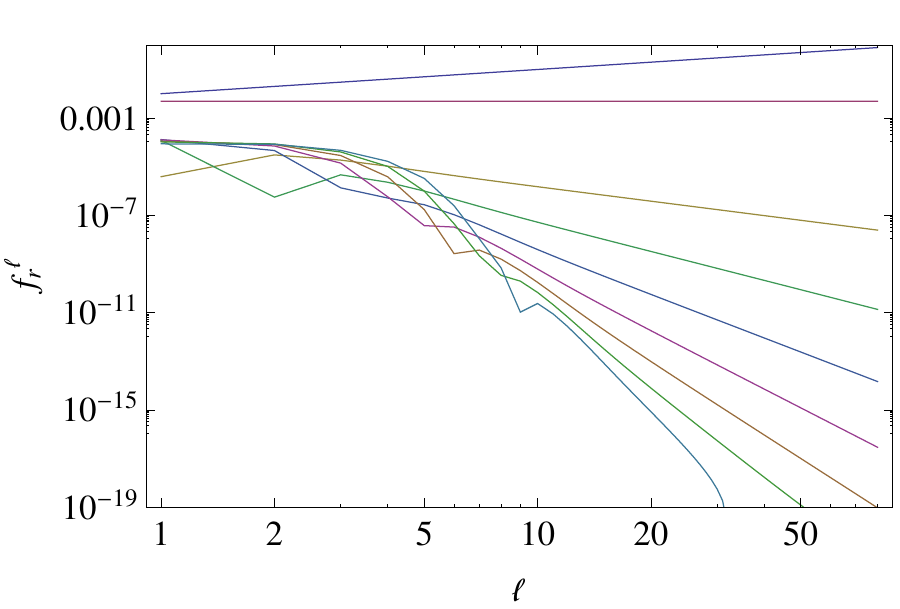}
\caption{Regularization of the radial component of the self-force for the case of
a scalar particle on a circular geodesic of radius $r_0 = 10M$ in Schwarzschild spacetime.  In decreasing slope the above lines represent the unregularised self-force, self-force regularised by subtracting from it in turn the cumulative sum of $F^l_{r[-1]}$, $F^l_{r[0]}$, $F^l_{r\lpow{2}}$, $F^l_{r\lpow{4}}$, $F^l_{r\lpow{6}}$, $F^l_{r\lpow{8}}$, $F^l_{r\lpow{10}}$, $F^l_{r\lpow{12}}$.}
\label{fig:scalar-circular-reg}
\end{figure}

In Fig.~\ref{fig:scalar-circular-reg}, we show the effect of subtracting in turn the cumulative sums of the regularization
parameters from the full retarded field.
Shown in the figure in order from top to bottom are $F_r^{\rm ret}$ and the result of subtracting
from it in turn the cumulative sum of the regularization terms $F^l_{r\lnpow{1}}$, $F^l_{r[0]}$, $F^l_{r\lpow{2}}$, $F^l_{r\lpow{4}}$, $F^l_{r\lpow{6}}$, $F^l_{r\lpow{8}}$, $F^l_{r\lpow{10}}$ and $F^l_{r\lpow{12}}$.
The parameters $F_{r\lnpow{1}}$, $F_{r[0]}$, $F_{r\lpow{2}}$, $F_{r\lpow{4}}$ and $F_{r\lpow{6}}$ are the analytically derived ones given in
Sec.~\ref{sec:scalar-regularization}. The  parameters $F_{r\lpow{8}}$, $F_{r\lpow{10}}$ and $F_{r\lpow{12}}$ were determined
through a numerical fit to the data. The resulting rapid convergence with $l$ enables the
calculation of an extremely accurate value for the self-force. Summing over $l$, we find
$F_r = 0.000013784482575667959(3)$, where the uncertainty in the last digit is estimated by
assuming that the error comes purely from the fact that the sum is only done up to a finite
$l_{\rm max} = 80$.

In addition to providing a highly accurate benchmark, this example may be used to assess
the benefits which can be obtained from the use of higher-order regularization parameters.
The most obvious benefit is that with fixed computational resources (i.e. fixed number of
spherical harmonic modes) one can obtain a much more accurate value for the self-force.
This is highlighted by comparison of our value for $F_r$ with that of the previous
benchmark given in Ref.~\cite{Detweiler:Messaritaki:Whiting:2002}. Both calculations
consider the same case of a scalar charge in a circular orbit of radius $10M$ around a
Schwarzschild black hole. Using $40$ $l$-modes and regularization parameters up
to $F^l_{r[2]}$, Ref.~\cite{Detweiler:Messaritaki:Whiting:2002}
obtained a value for the self-force with a fractional accuracy of $10^{-9}$. The inclusion
of the next two regularization parameters improves this to a fractional error of $10^{-12}$
which increases to a fractional error of $10^{-17}$ when $80$ modes are used.

This example represents a somewhat extreme case: it uses highly accurate frequency domain
methods combined with high-precision numerical integration and a relatively large number
of spherical harmonic modes. In more typical time-domain calculations, numerical
data up to $l \sim 15$ is used and it is common that the dominant source of error comes
from the tail fit. While it may seem that one merely needs to compute more modes to reduce
this error, this is not a realistic solution. In a mode-sum
calculation, the number of spherical harmonic modes required for each $l$ scales as $l^2$,
meaning that simply running simulations for larger and larger $l$ rapidly becomes
prohibitively expensive in terms of computational cost.  Additionally, the improvement
with each additional mode falls off as an inverse power in $l$, meaning that many more
$l$ modes are required for an increasingly small benefit. In this case, the inclusion of
higher order regularization parameters essentially eliminates this problem: without them
the tail fit is the dominant source of error, with them sufficiently accurate results
may be obtained without even fitting for a tail.

It should be pointed out that there is one caveat to our conclusions. The use of high
order regularization parameters requires the subtraction of increasingly
(relatively) large numbers to obtain a small regularized remainder. It is therefore
essential that any numerically provided data for the retarded field must be of sufficient
accuracy for the subtraction to yield meaningful results. As a result, calculations which were
previously deemed sufficient would not necessarily gain an immediate benefit from higher order
regularization parameters.

\section{Effective Source} \label{sec:EffectiveSource}

As another application of our high order expansions of the Detweiler-Whiting singular field,
we consider its use in the effective source approach to calculating the self-force. The effective
source approach -- independently proposed by Barack and Golbourn
\cite{Barack:Golbourn:2007} and by Vega and Detweiler \cite{Vega:Detweiler:2008} -- relies on
knowledge of the singular field to derive an equation for a regularized field that gives
the self-force without any need for post-processed regularization. If the singular field is known
exactly, then the regularized field is totally regular and is a solution of the homogeneous wave
equation. In reality, exact expressions for the singular field can only be obtained for very simple
spacetimes. More generally, the best one can do is an approximation such as
that given in Sec.~\ref{sec:SingularFieldExpansion}. Splitting the retarded field
into approximate singular and regularized parts,
\begin{equation}
\varphi^A_{\rm \ret} = \tilde{\varphi}^A_{\rm \sing} + \tilde{\varphi}^A_{\rm \reg},
\end{equation}
substituting into the wave equation, Eq.~\eqref{eq:Wave}, and rearranging, we obtain an equation
for the regularized field,
\begin{equation}
\mathcal{D}^{A}{}_B \tilde{\varphi}^{B}_{\rm \reg} = S_{\rm eff}^A,
\end{equation}
with an \emph{effective source},
\begin{equation}
S_{\rm eff}^{A} = -\mathcal{D}^{A}{}_B \tilde{\varphi}^{B}_{\rm \sing} - 4\pi \mathcal{Q} \int u^{A} \delta_4 \left( x,z(\tau') \right) d\tau'.
\end{equation}
For sufficiently good approximations to the singular field, $\tilde{\varphi}^A_{\rm \reg}$ and
$S^A$ are finite everywhere, in particular on the world-line. As a result, one never encounters
problematic singularities or $\delta$-functions, making the approach particularly suitable for use in
time domain numerical simulations. A detailed review of this approach can be found in
\cite{Vega:2011wf,Wardell:2011gb}. 

\begin{figure}
\includegraphics[width=4cm]{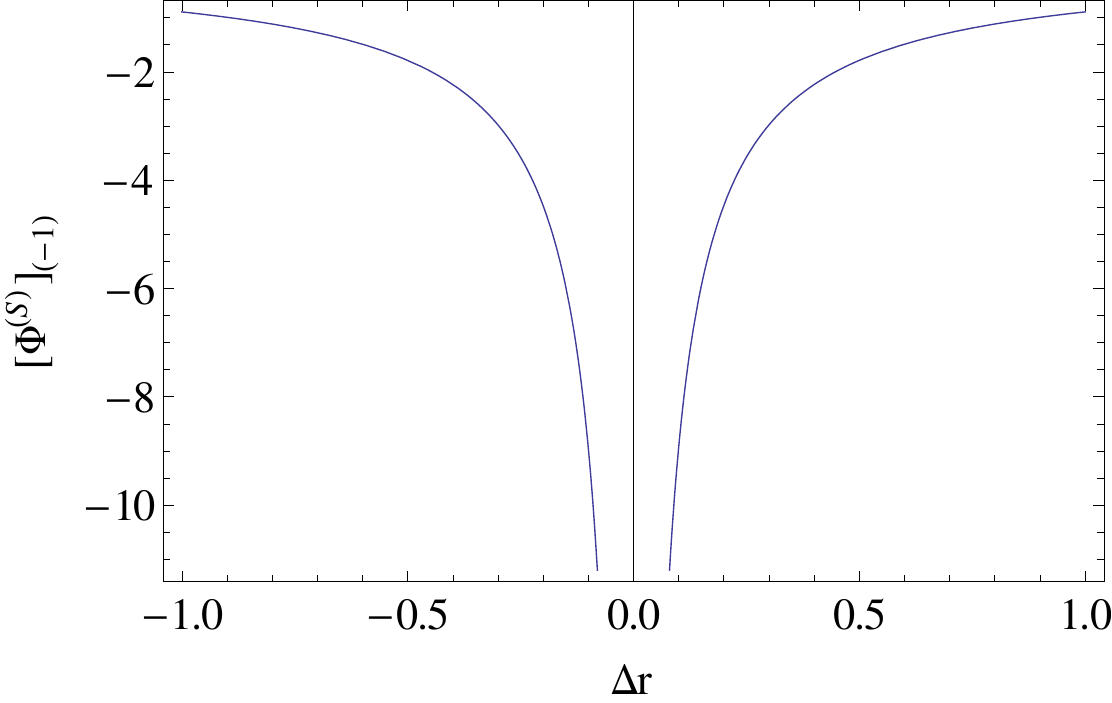}
\includegraphics[width=4cm]{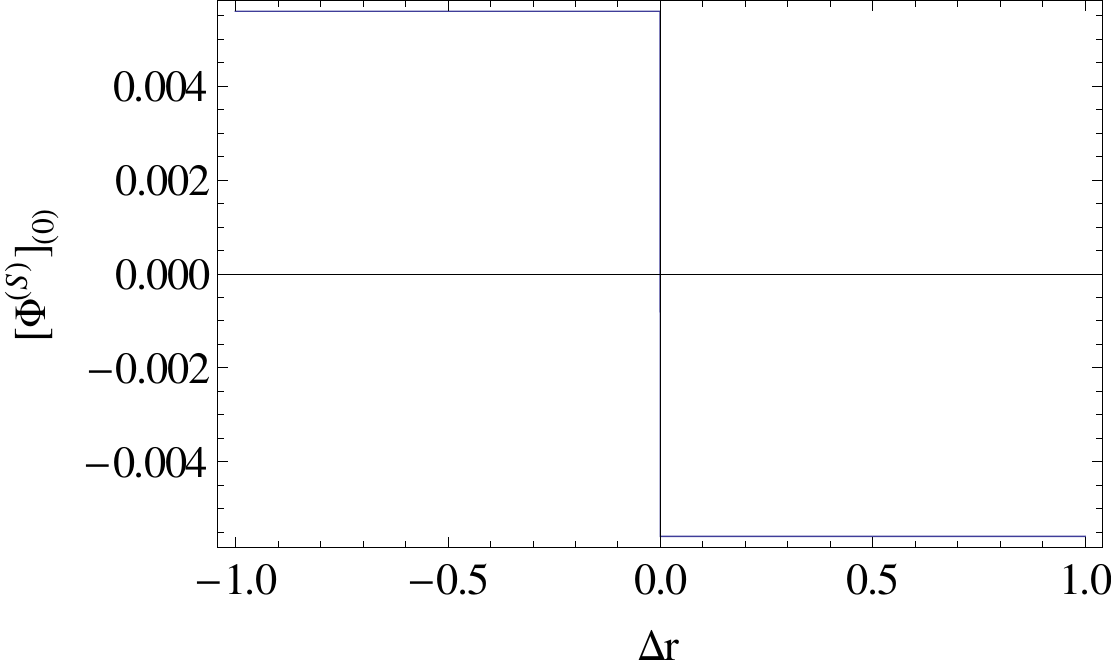}
\includegraphics[width=4cm]{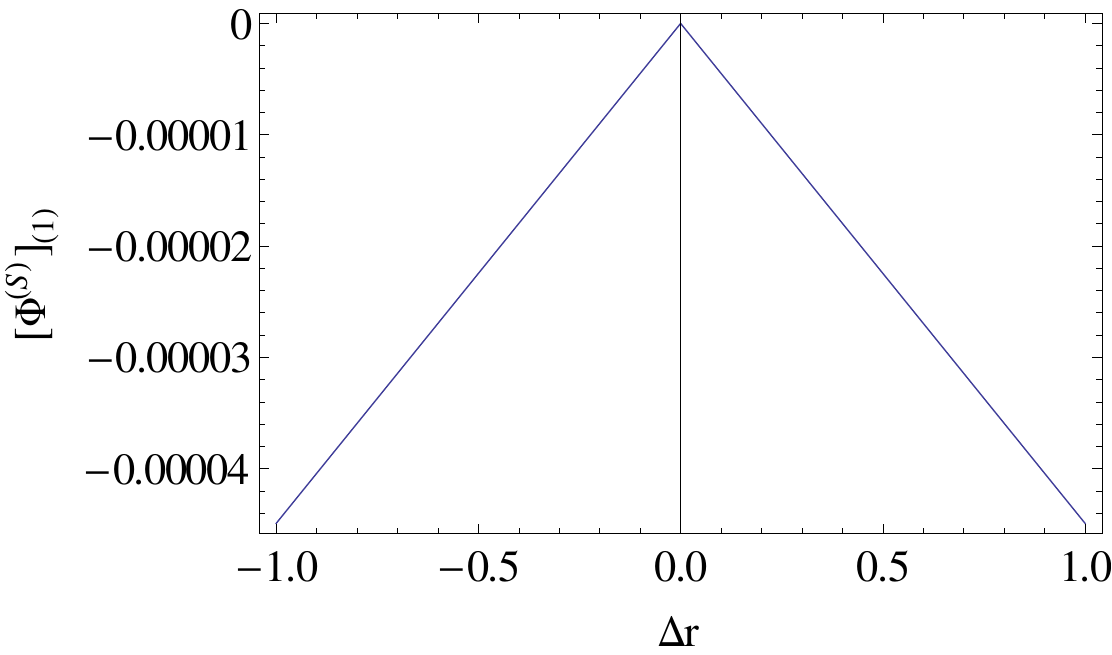}
\includegraphics[width=4cm]{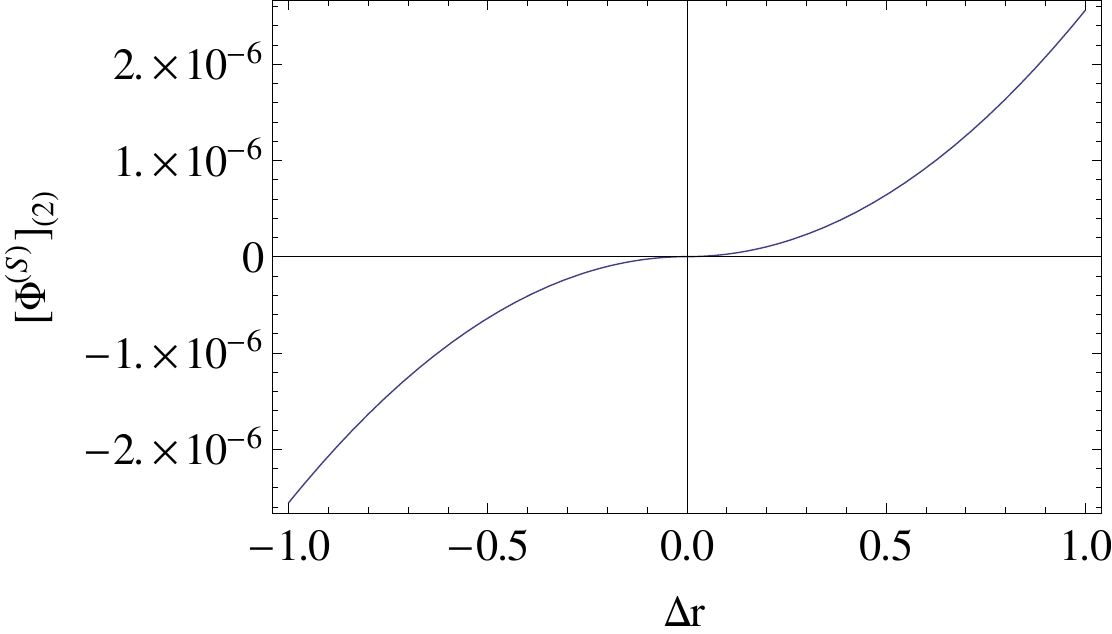}
\includegraphics[width=4cm]{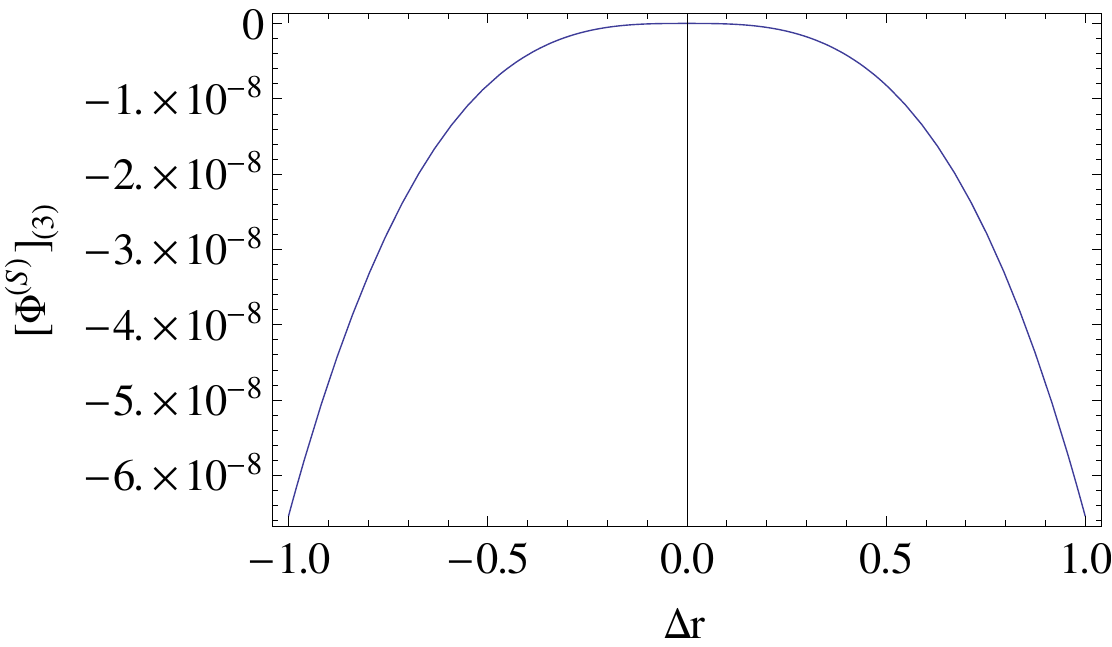}
\includegraphics[width=4cm]{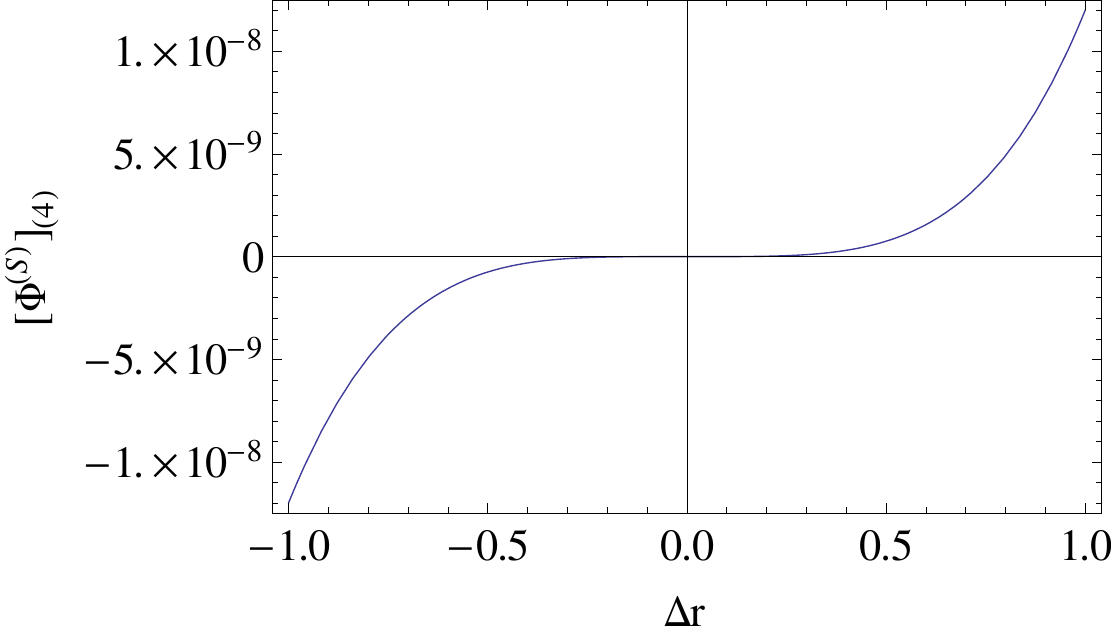}
\includegraphics[width=4cm]{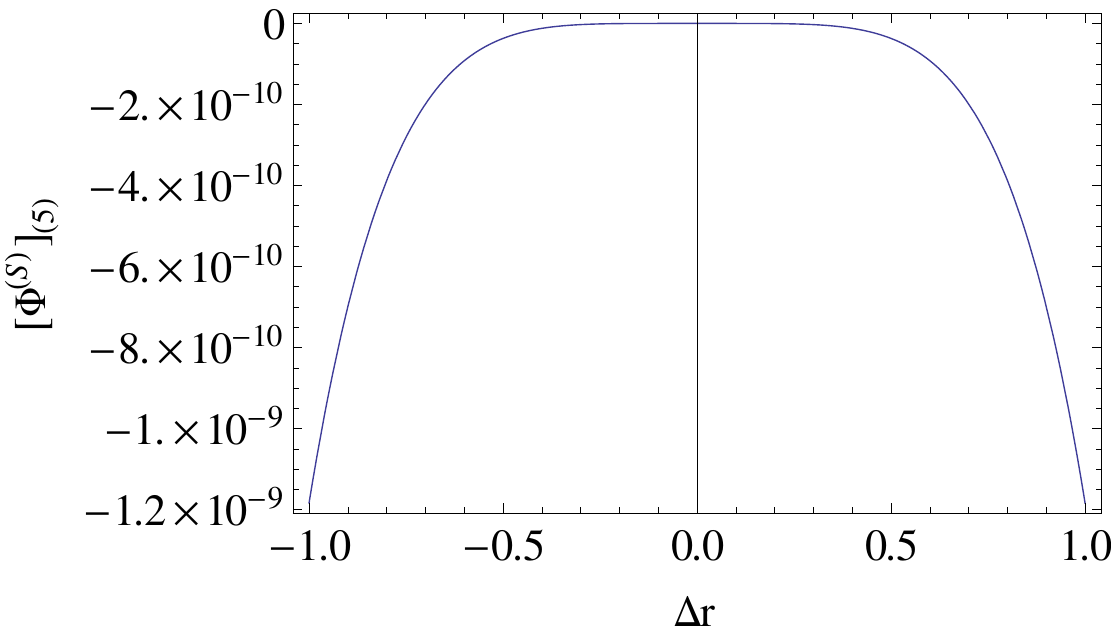}
\includegraphics[width=4cm]{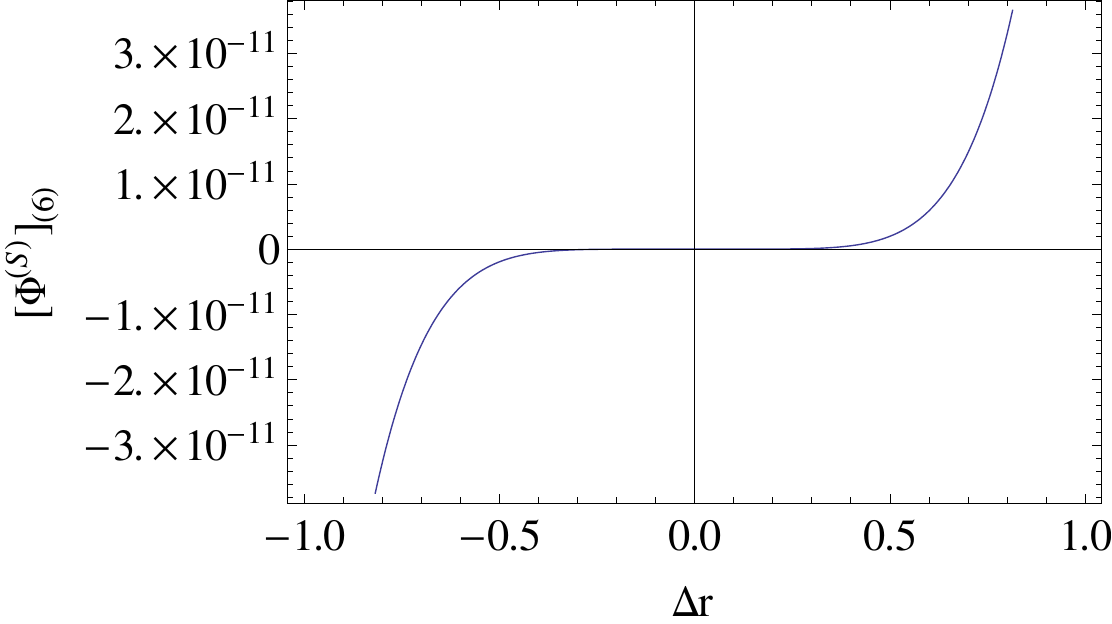}
\caption{Terms in the coordinate expansion of the singular field
for $\mathcal{O}(\epsilon^{-1})$ (top left) to $\mathcal{O}(\epsilon^{6})$ (bottom right).
Shown is the case of a circular geodesic of radius $r_0 = 10M$ in Schwarzschild spacetime.}
\label{fig:SingularField1D}
\end{figure}

\begin{figure}
\includegraphics[width=4cm]{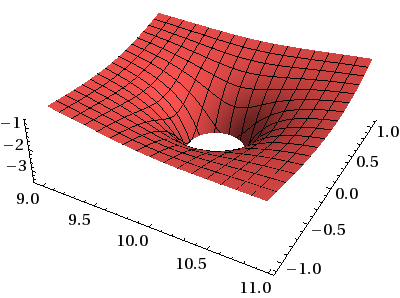}
\includegraphics[width=4cm]{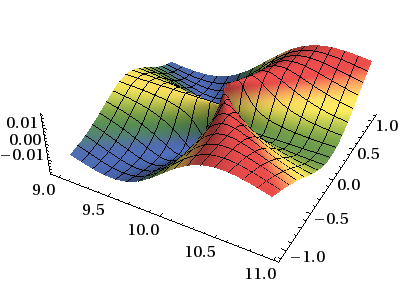}
\includegraphics[width=4cm]{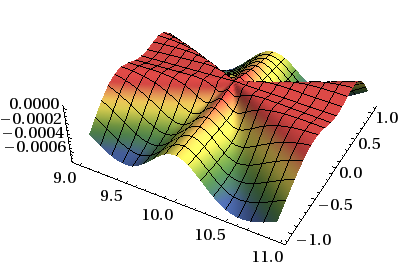}
\includegraphics[width=4cm]{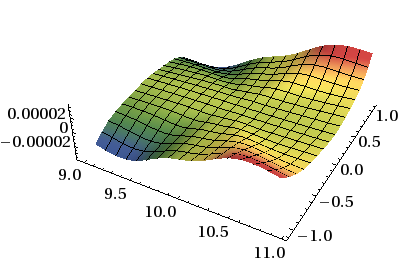}
\includegraphics[width=4cm]{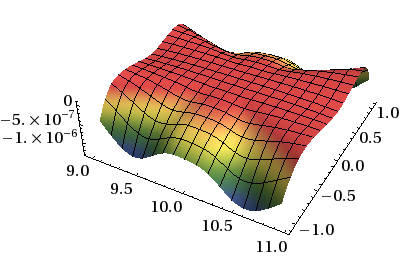}
\includegraphics[width=4cm]{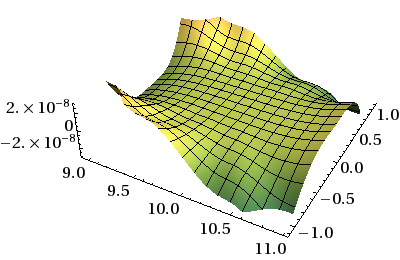}
\includegraphics[width=4cm]{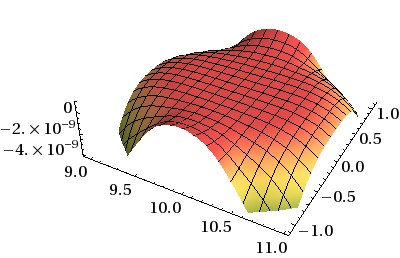}
\includegraphics[width=4cm]{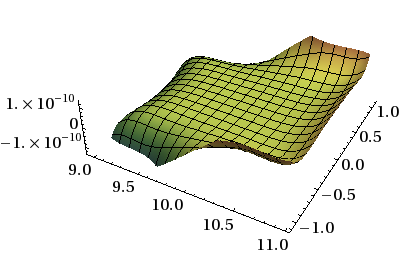}
\caption{Terms in the coordinate expansion of the singular field, $[\Phi^{\sing}]_{(n)}$, in the region of the particle
for $\mathcal{O}(\epsilon^{-1})$ (top left) to $\mathcal{O}(\epsilon^{6})$ (bottom right).
Shown is the case of a circular geodesic of radius $r_0 = 10M$ in Schwarzschild spacetime.}
\label{fig:SingularField2D}
\end{figure}

\begin{figure}
\includegraphics[width=4cm]{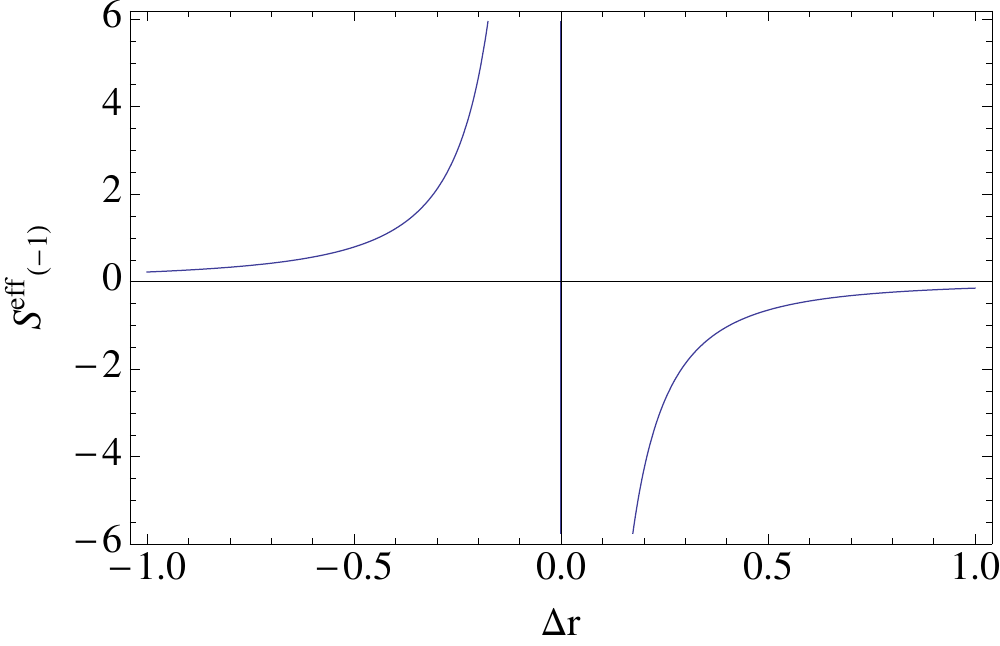}
\includegraphics[width=4cm]{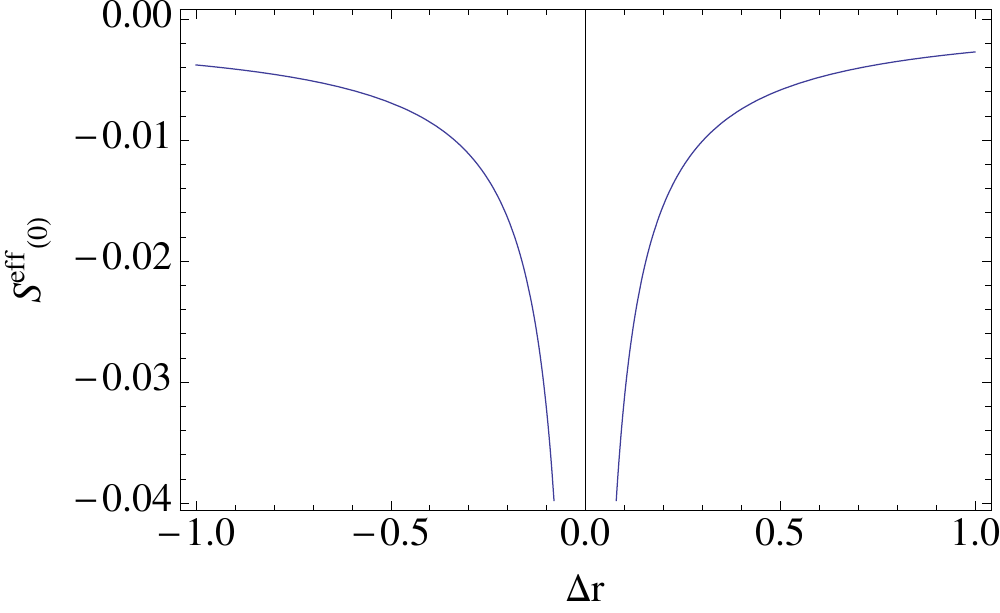}
\includegraphics[width=4cm]{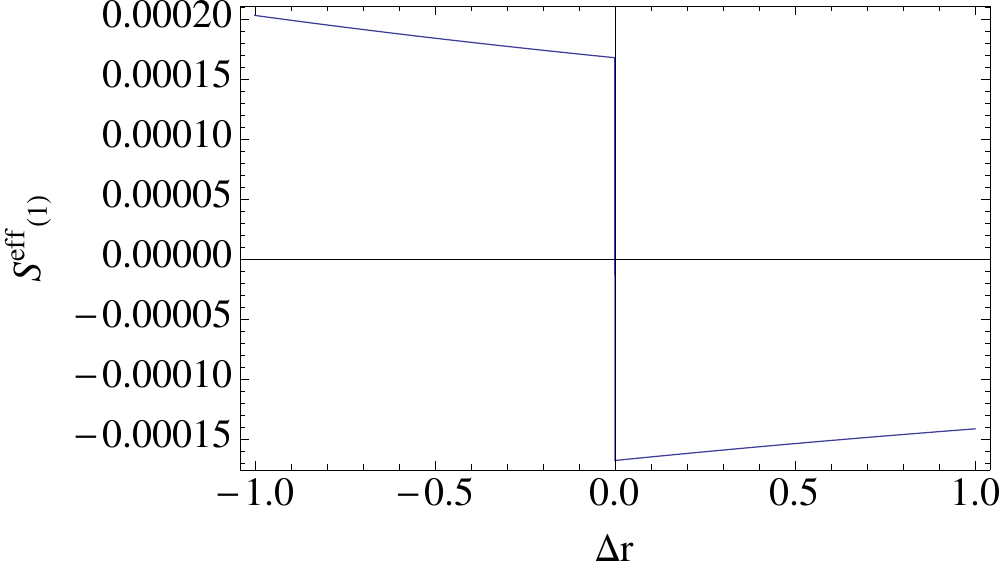}
\includegraphics[width=4cm]{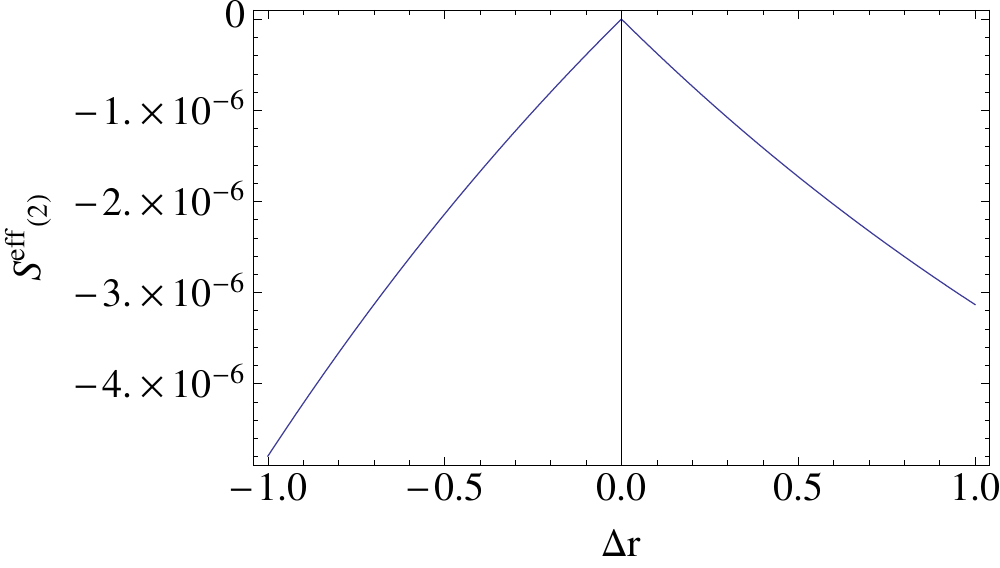}
\includegraphics[width=4cm]{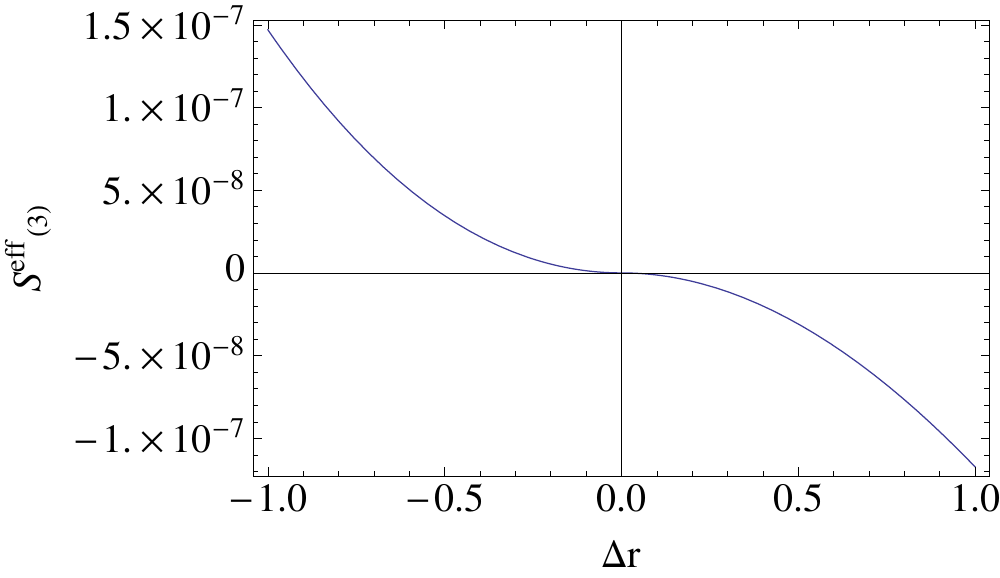}
\includegraphics[width=4cm]{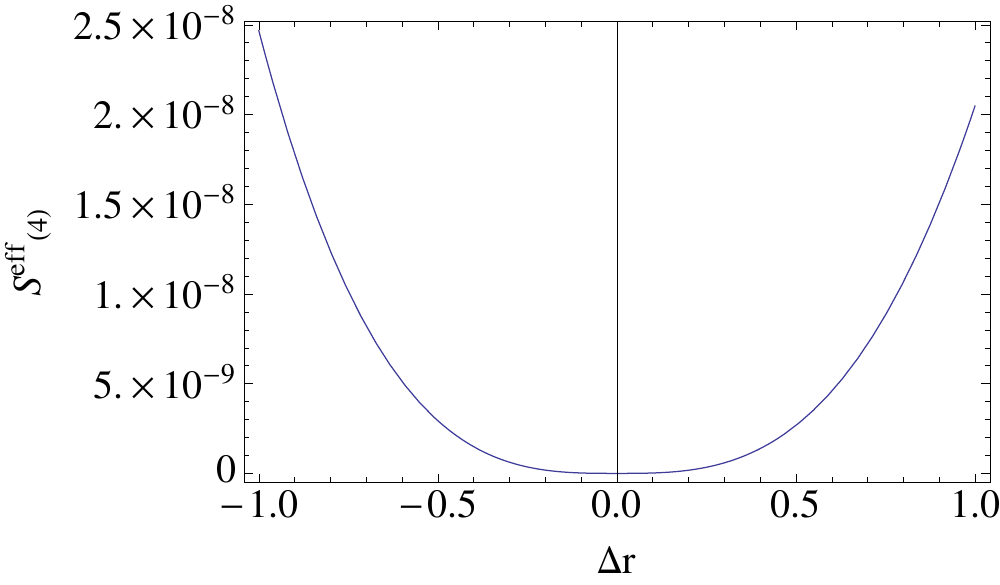}
\includegraphics[width=4cm]{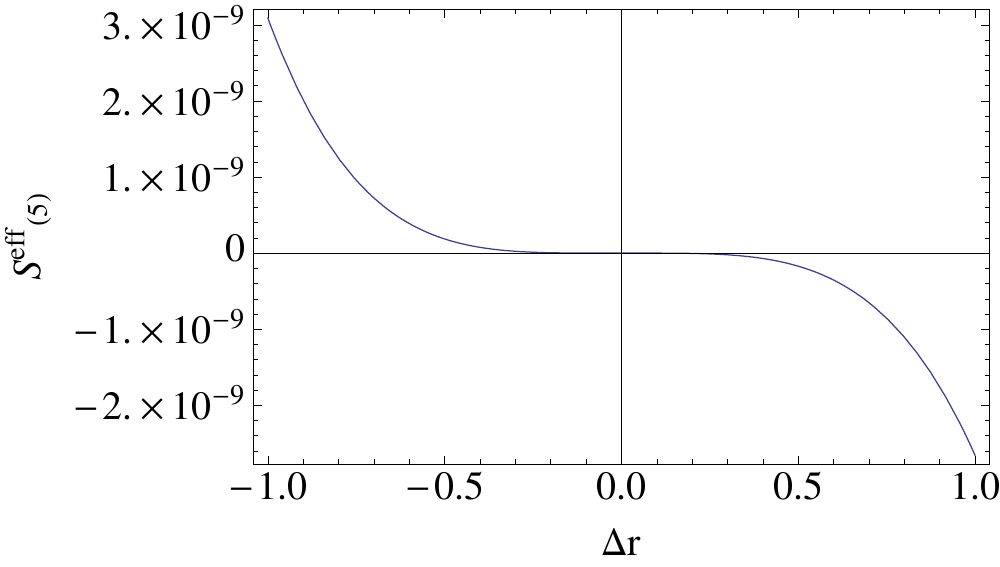}
\includegraphics[width=4cm]{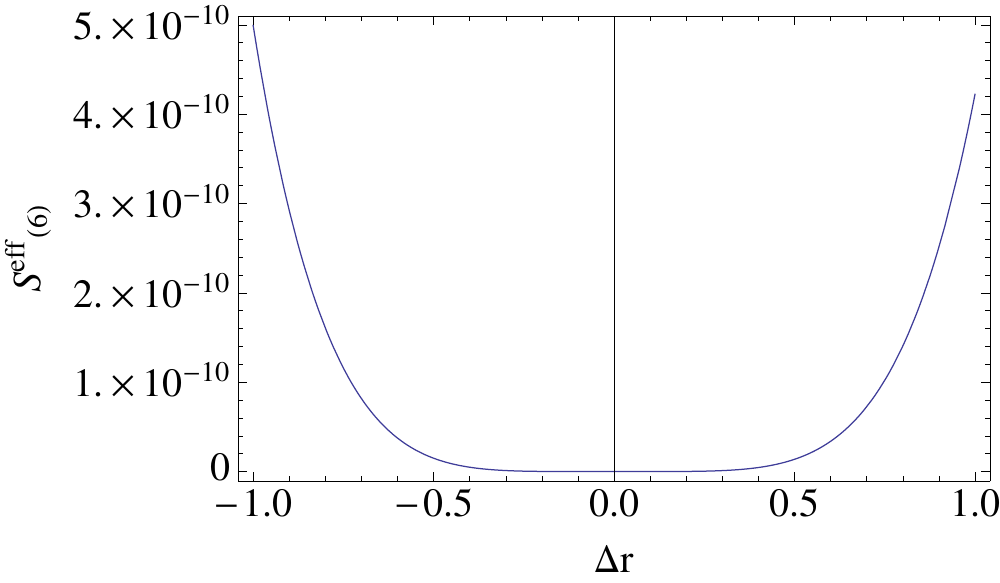}
\caption{Effective source for the approximate singular field of order
$\mathcal{O}(\epsilon^{-1})$ (top left) to $\mathcal{O}(\epsilon^{6})$ (bottom right).
Shown is the case of a circular geodesic of radius $r_0 = 10M$ in Schwarzschild spacetime.}
\label{fig:EffectiveSource1D}
\end{figure}

\begin{figure}
\includegraphics[width=4cm]{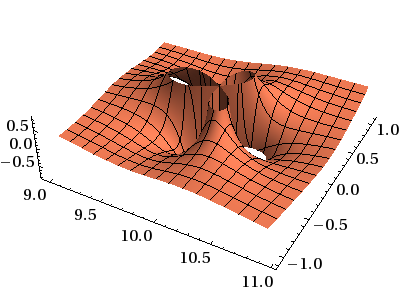}
\includegraphics[width=4cm]{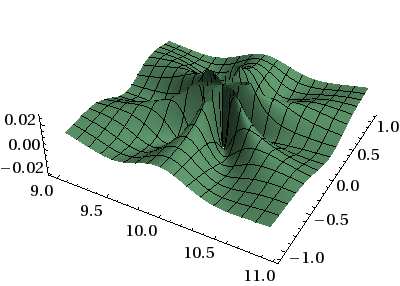}
\includegraphics[width=4cm]{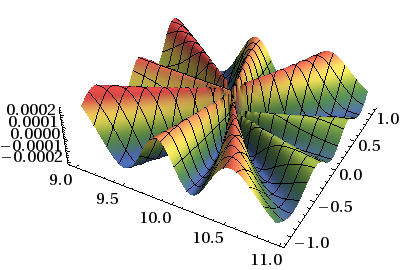}
\includegraphics[width=4cm]{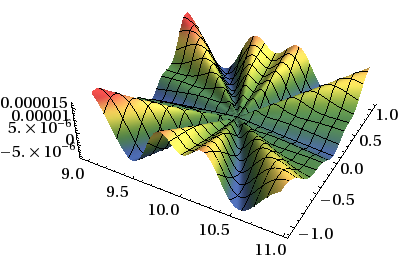}
\includegraphics[width=4cm]{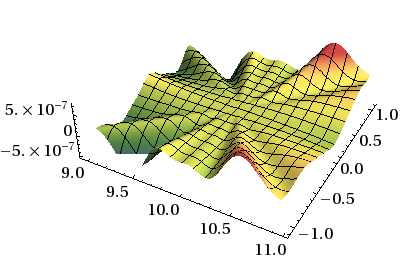}
\includegraphics[width=4cm]{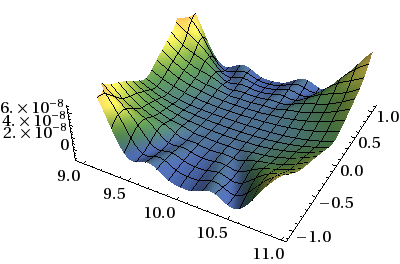}
\includegraphics[width=4cm]{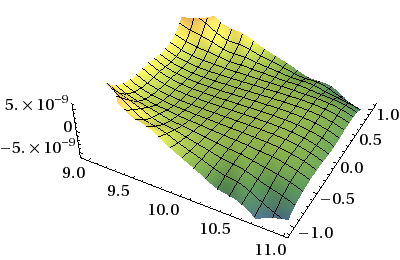}
\includegraphics[width=4cm]{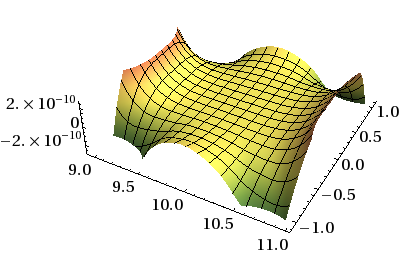}
\caption{Effective source for the approximate singular field, $S^{\rm eff}{}_{(n)}$, in the region of the particle of order
$\mathcal{O}(\epsilon^{-1})$ (top left) to $\mathcal{O}(\epsilon^{6})$ (bottom right).
Shown is the case of a circular geodesic of radius $r_0 = 10M$ in Schwarzschild spacetime.}
\label{fig:EffectiveSource2D}
\end{figure}

In Figs.~\ref{fig:SingularField1D},\ref{fig:SingularField2D}, \ref{fig:EffectiveSource1D}
and \ref{fig:EffectiveSource2D} we show the result of applying our expansions
to the case of a scalar particle on a circular geodesic of radius $r_0 = 10M$ in
Schwarzschild spacetime. Similar plots can be obtained for the electromagnetic case, gravitational case
and for more generic motion. However, the general structure does not change and is best illustrated
by this simple example.

\section{Discussion} \label{sec:Discussion}

In this paper we have developed high order expansions of the Detweiler-Whiting singular field
of a point scalar or electromagnetic particle and of a point mass. Many of our expressions
are very general, not necessarily being restricted to any particular choice of spacetime.
In our explicit coordinate calculations, however, for simplicity we chose to limit ourselves
to the Schwarzschild geometry. A logical extension of this work would be to consider, instead,
the Kerr spacetime. This has already been done to $\mathcal{O}(\epsilon^2)$ for the case of a
scalar charge in a circular geodesic orbit in Kerr spacetime and spherical-harmonic regularization
parameters have been computed \cite{AnnaCapra}. We will present a more general version of this
calculation in forthcoming work \cite{Heffernan:InPreparation}. However, the use of spherical
harmonics is not well suited to the Kerr spacetime. A more appropriate choice of basis functions
are the spheroidal harmonics; it may be more sensible to compute regularization parameters in this spheroidal harmonic basis.
Nevertheless, the full singular field is the same, independently of the choice of basis functions,
and the majority of our calculations can easily be carried over to the Kerr case with little
modification.

In the cases of electromagnetic and gravitational fields, there is an arbitraryness in the
choice of gauge. For this work, we have focused only on the case of Lorenz gauge, $A^a{}_{;a}=0$
and $h_{ab}{}^{;b} = 0$, where the singular field is best understood.
However, other gauges such as Regge-Wheeler and Radiation gauge are better suited to time
domain numerical calculations. Indeed, there are many outstanding problems in the time domain
evolution of metric perturbations in Lorenz gauge \cite{dolan:11}. Given recent
developments in understanding the singular behavior of the retarded field in other gauges
\cite{Keidl:2010pm,Shah:2010bi,Gralla:2011ke,Gralla:2011zr}, it may soon be possible to
implement calculations similar to what we have presented here for other gauges. This could be
achieved, for example, by either taking our Lorenz gauge expressions and performing the gauge
transformation to a different gauge or by directly computing the singular field in the other gauge.
It should be noted, however, that there are many non-trivial issues which can arise when considering
other gauges; for example, the gauge tranformation itself may be singular or the singular field may
be even more singular\footnote{A specific example of this is the case of metric perturbations
in Regge-Wheeler gauge, where there are additional delta-function divergences in the singular
field which are not smoothed out by a mode decomposition \cite{Hopper:2010uv}.} than in Lorenz
gauge. Any calculations will require considerable care to correctly deal with such issues.

In Sec.~\ref{sec:EffectiveSource} we gave plots of an effective source which may be used in
a numerical evolution. We have not yet, however, fed this effective source into an actual
numerical code. Existing results from effective source calculations have used a source obtained
by taking the first four terms in the expansion of the singular field (to
$\mathcal{O}(\epsilon^2)$) which yields a source which is $C^0$. This is sufficient to obtain
correct results, but limits the convergence of any numerical scheme with resolution or number of
modes required. By including the next four terms in the expansion of the singular field (to
$\mathcal{O}(\epsilon^6)$) the highest order effective source shown in
Sec.~\ref{sec:EffectiveSource} is now $C^4$, allowing for the use of high-order finite
differencing to potentially obtain a significant increase in numerical accuracy.

We have limited the focus of the present work to linear order. However, for calculations
with sufficient accuracy to match the requirements of gravitational wave detection, it will
be necessary to go beyond first order and to compute a second order self-force. Rapid progress
is being made in addressing the difficult problem of deriving the
second order equations of motion \cite{Detweiler:2011tt,Pound:2012nt,Gralla:2012db} and the computation of a
second order effective source. The next step is to implement this second order prescription
in a numerical calculation. As this involves an effective source which contains terms involving
the first order singular field, it seems likely that many of the calculations done here may be of
use for second order.

\section*{Acknowledgements}
We are extremely grateful to Niels Warburton, Sarp Akcay and Leor Barack  for making available their data for the spherical harmonic modes of the retarded field in scalar and gravitational cases,
and to Roland Hass for making the equivalent data available for the electromagnetic case.
We thank Sam Dolan, Marc Casals, Jos\'e Luis Jaramillo, Michael Jasiulek, Abraham Harte and
particularly Ian Vega for many insightful discussions during the progress of this work.  
Finally, we thank participants of the 2010 and 2011 Capra meetings (in Waterloo and Southampton,
respectively) for many illuminating conversations.

AH has been supported by the Irish Research Council for Science, Engineering and Technology, funded by the National Development Plan and the Institute of Phyics C. R. Barber Trust Fund. BW and ACO gratefully acknowledge support from the Science Foundation Ireland under Grant No. 10/RFP/PHY2847.

\appendix

\section{Coordinate Expansions in Schwarschild Space-Time} \label{sec:CoordinateExpansions}
In this Appendix we give coordinate expansions of the key quantities appearing in the singular field,
Eqs.~\eqref{eq:PhiS}, \eqref{eq:AS} and \eqref{eq:hS}.

\subsection*{Synge World function}

Letting $\cos \gamma = \cos \theta \cos \theta' - \sin \theta \sin \theta' \cos(\phi - \phi')$
so that $2(1-\cos\gamma) = \Delta w_1^2 + \Delta w_2^2$, the expansion of the
world function to the order required in this paper is
\begin{align} \label{eqn: sigma expansion}
\sigma(x&,x') = \sum_{i,j,k=0}^{i+j+2 k\le9} \sigma_{ijk} (t'-t)^i (r'-r)^j (1-\cos \gamma)^k
 +\mathcal{O}(\epsilon^{10}),
\end{align}
where the non-zero coefficients are
\begin{gather}
\sigma_{001}=r^2, \quad
\sigma_{002}=\frac{M r}{3},\quad
\sigma_{003}=\frac{1}{90} M (9 r-2 M),\quad
\sigma_{004}=\frac{M \left(14 M^2-19 M r+30 r^2\right)}{840 r},\quad
\sigma_{011}=r,\quad
\sigma_{012}=\frac{M}{6},\nonumber\\
\sigma_{013}=\frac{M}{20},\quad
\sigma_{014}=\frac{1}{840} M \left(15-\frac{7 M^2}{r^2}\right),\quad
\sigma_{020}=-\frac{r}{4 M-2 r},\quad
\sigma_{021}=\frac{M}{12 M-6 r},\quad
\sigma_{022}=\frac{M (r-M)}{60 r (2 M-r)},\nonumber\\
\sigma_{023}=\frac{M \left(-14 M^2+6 M r+3 r^2\right)}{840 r^2 (2 M-r)},\quad
\sigma_{030}=-\frac{M}{2 (r-2 M)^2},\quad
\sigma_{031}=\frac{M}{12 (r-2 M)^2},\quad
\sigma_{032}=\frac{M \left(2 M^2-2 M r+r^2\right)}{120 r^2 (r-2 M)^2},\nonumber \\
\sigma_{033}=\frac{M \left(56 M^3-54 M^2 r+12 M r^2+3 r^3\right)}{1680 r^3 (r-2 M)^2},\quad
\sigma_{040}=-\frac{M (M-8 r)}{24 r (r-2 M)^3},\quad
\sigma_{041}=\frac{M \left(5 M^2-3 M r-6 r^2\right)}{120 r^2 (r-2 M)^3},\nonumber \\
\sigma_{042}=\frac{M \left(42 M^3-70 M^2 r+39 M r^2-16 r^3\right)}{3360 r^3 (r-2 M)^3},\quad
\sigma_{050}=-\frac{M \left(M^2-2 M r+6 r^2\right)}{24 r^2 (r-2 M)^4},\nonumber \\
\sigma_{051}=\frac{M \left(20 M^3-31 M^2 r+12 M r^2+8 r^3\right)}{240 r^3 (r-2 M)^4},\quad
\sigma_{052}=\frac{M \left(7 M^4-21 M^3 r+23 M^2 r^2-11 M r^3+5 r^4\right)}{1680 r^4 (r-2 M)^4},\nonumber \\
\sigma_{060}=-\frac{M \left(35 M^3-86 M^2 r+86 M r^2-144 r^3\right)}{720 r^3 (r-2 M)^5},\quad
\sigma_{061}=\frac{M \left(245 M^4-532 M^3 r+421 M^2 r^2-120 M r^3-40 r^4\right)}{1680 r^4 (r-2 M)^5},\nonumber\\
\sigma_{070}=\frac{M \left(-15 M^4+44 M^3 r-54 M^2 r^2+36 M r^3-40 r^4\right)}{240 r^4 (r-2 M)^6},\nonumber\\
\sigma_{071}=\frac{M \left(280 M^5-763 M^4 r+832 M^3 r^2-444 M^2 r^3+100 M r^4+20 r^5\right)}{1120 r^5 (r-2 M)^6},\nonumber \\
\sigma_{080}=\frac{M \left(385 M^5-1316 M^4 r+1928 M^3 r^2-1576 M^2 r^3+788 M r^4-640 r^5\right)}{4480 r^5 (2 M-r)^7},\nonumber \\
\sigma_{090}=\frac{M \left(-5005 M^6+19558 M^5 r-33394 M^4 r^2+32584 M^3 r^3-19960 M^2 r^4+7984 M r^5-5040 r^6\right)}{40320 r^6 (r-2 M)^8}, \nonumber \\
\sigma_{200}=\frac{M}{r}-\frac{1}{2}, \quad
\sigma_{201}=\frac{M (r-2 M)}{6 r^2},\quad
\sigma_{202}=\frac{M \left(10 M^2-11 M r+3 r^2\right)}{60 r^3},\nonumber \\
\sigma_{203}=\frac{M \left(-92 M^3+142 M^2 r-78 M r^2+15 r^3\right)}{840 r^4},\quad
\sigma_{210}=-\frac{M}{2 r^2},\quad
\sigma_{211}=\frac{M (4 M-r)}{12 r^3},\nonumber \\
\sigma_{212}=-\frac{M \left(30 M^2-22 M r+3 r^2\right)}{120 r^4},\quad
\sigma_{213}=\frac{M \left(368 M^3-426 M^2 r+156 M r^2-15 r^3\right)}{1680 r^5},\quad
\sigma_{220}=\frac{M (5 M-4 r)}{12 r^3 (2 M-r)},\nonumber \\
\sigma_{221}=-\frac{M \left(23 M^2-20 M r+3 r^2\right)}{60 r^4 (2 M-r)},\quad
\sigma_{222}=\frac{M \left(686 M^3-802 M^2 r+281 M r^2-24 r^3\right)}{1680 r^5 (2 M-r)},\quad
\sigma_{230}=-\frac{M (M-r)^2}{4 r^4 (r-2 M)^2},\nonumber \\
\sigma_{231}=\frac{M \left(24 M^3-55 M^2 r+32 M r^2-4 r^3\right)}{120 r^5 (r-2 M)^2},\quad
\sigma_{232}=\frac{M \left(-630 M^4+1234 M^3 r-832 M^2 r^2+218 M r^3-15 r^4\right)}{1680 r^6 (r-2 M)^2},\nonumber \\
\sigma_{240}=-\frac{M \left(M^3-78 M^2 r+116 M r^2-48 r^3\right)}{240 r^5 (r-2 M)^3},\quad
\sigma_{241}=-\frac{M \left(553 M^4-269 M^3 r-444 M^2 r^2+322 M r^3-40 r^4\right)}{1680 r^6 (r-2 M)^3},\nonumber \\
\sigma_{250}=\frac{M \left(75 M^4-84 M^3 r-51 M^2 r^2+104 M r^3-40 r^4\right)}{240 r^6 (r-2 M)^4},\nonumber \\
\sigma_{251}=\frac{M \left(-4396 M^5+8227 M^4 r-4760 M^3 r^2+350 M^2 r^3+412 M r^4-60 r^5\right)}{3360 r^7 (r-2 M)^4},\nonumber \\
\sigma_{260}=\frac{M \left(2317 M^5-5560 M^4 r+4220 M^3 r^2-146 M^2 r^3-1254 M r^4+480 r^5\right)}{3360 r^7 (r-2 M)^5},\nonumber \\
\sigma_{270}=\frac{M \left(3759 M^6-12402 M^5 r+16015 M^4 r^2-9336 M^3 r^3+1347 M^2 r^4+1048 M r^5-420 r^6\right)}{3360 r^8 (r-2 M)^6},\nonumber \\
\sigma_{400}=\frac{M^2 (2 M-r)}{24 r^5}, \quad
\sigma_{401}=-\frac{M^2 \left(54 M^2-49 M r+11 r^2\right)}{360 r^6},\nonumber\\
\sigma_{402}=\frac{M^2 \left(1956 M^3-2702 M^2 r+1228 M r^2-183 r^3\right)}{10080 r^7},\quad
\sigma_{410}=\frac{M^2 (2 r-5 M)}{24 r^6},\nonumber \\
\sigma_{411}=\frac{M^2 \left(324 M^2-245 M r+44 r^2\right)}{720 r^7},\quad
\sigma_{412}=\frac{M^2 \left(-3423 M^3+4053 M^2 r-1535 M r^2+183 r^3\right)}{5040 r^8},\nonumber \\
\sigma_{420}=\frac{M^2 \left(429 M^2-394 M r+86 r^2\right)}{720 r^7 (2 M-r)},\quad
\sigma_{421}=\frac{M^2 \left(-7509 M^3+9170 M^2 r-3575 M r^2+438 r^3\right)}{5040 r^8 (2 M-r)},\nonumber \\
\sigma_{430}=\frac{M^2 \left(-301 M^3+460 M^2 r-228 M r^2+36 r^3\right)}{240 r^8 (r-2 M)^2},\nonumber \\
\sigma_{431}=\frac{M^2 \left(35472 M^4-63269 M^3 r+40912 M^2 r^2-11240 M r^3+1088 r^4\right)}{10080 r^9 (r-2 M)^2},\nonumber \\
\sigma_{440}=-\frac{M^2 \left(42507 M^4-94876 M^3 r+77976 M^2 r^2-27740 M r^3+3546 r^4\right)}{20160 r^9 (r-2 M)^3},\nonumber \\
\sigma_{450}=\frac{M^2 \left(-57987 M^5+178306 M^4 r-214952 M^3 r^2+126744 M^2 r^3-36328 M r^4+3992 r^5\right)}{20160 r^{10} (r-2 M)^4},\nonumber \\
\sigma_{600}=\frac{M^3 \left(26 M^2-25 M r+6 r^2\right)}{720 r^9},\quad
\sigma_{601}=\frac{M^3 \left(-1818 M^3+2475 M^2 r-1115 M r^2+166 r^3\right)}{15120 r^{10}},\nonumber \\
\sigma_{610}=-\frac{M^3 \left(117 M^2-100 M r+21 r^2\right)}{720 r^{10}},\quad
\sigma_{611}=\frac{M^3 \left(18180 M^3-22275 M^2 r+8920 M r^2-1162 r^3\right)}{30240 r^{11}},\nonumber \\
\sigma_{620}=\frac{M^3 \left(23931 M^3-31560 M^2 r+13652 M r^2-1930 r^3\right)}{30240 r^{11} (2 M-r)},\nonumber \\
\sigma_{630}=\frac{M^3 \left(-27687 M^4+51002 M^3 r-34745 M^2 r^2+10344 M r^3-1131 r^4\right)}{10080 r^{12} (r-2 M)^2},\nonumber \\
\sigma_{800}=\frac{M^4 \left(978 M^3-1393 M^2 r+660 M r^2-104 r^3\right)}{40320 r^{13}}, \quad
\sigma_{810}=\frac{M^4 \left(-6357 M^3+8358 M^2 r-3630 M r^2+520 r^3\right)}{40320 r^{14}}.
\end{gather}

\subsection*{Van Vleck determinant}
Inserting the above expansion for $\sigma(x,x')$ into the definition of the Van Vleck-Morette determinant,
Eq.~\eqref{eq:vanVleckDefinition}, gives
\begin{align}
\Delta^{1/2}(x,x')= 1 + \sum_{i,j,k=0}^{i+j+2 k\le7} \Delta^{1/2}_{ijk} (t'-t)^i (r'-r)^j (1-\cos \gamma)^k
 +\mathcal{O}(\epsilon^{10}),
\end{align}
where the non-zero coefficients are
\begin{gather}
\Delta^{1/2}_{002}=\frac{M^2}{15 r^2}, \quad
\Delta^{1/2}_{003}=\frac{M^2 (27 r-34 M)}{378 r^3}, \quad
\Delta^{1/2}_{012}=-\frac{M^2}{15 r^3}, \quad
\Delta^{1/2}_{013}=\frac{M^2 (17 M-9 r)}{126 r^4}, \quad
\Delta^{1/2}_{021}=\frac{M^2}{60 M r^3-30 r^4}, \nonumber \\
\Delta^{1/2}_{022}=\frac{M^2 (322 M-177 r)}{2520 r^4 (2 M-r)}, \quad
\Delta^{1/2}_{031}=\frac{M^2 (2 r-3 M)}{30 r^4 (r-2 M)^2}, \quad
\Delta^{1/2}_{032}=\frac{M^2 \left(-308 M^2+332 M r-93 r^2\right)}{1260 r^5 (r-2 M)^2}, \nonumber \\
\Delta^{1/2}_{040}=\frac{M^2}{60 r^4 (r-2 M)^2}, \quad
\Delta^{1/2}_{041}=-\frac{M^2 \left(910 M^2-1268 M r+459 r^2\right)}{5040 r^5 (r-2 M)^3}, \quad
\Delta^{1/2}_{050}=\frac{M^2 (4 M-3 r)}{60 r^5 (r-2 M)^3}, \nonumber \\
\Delta^{1/2}_{051}=\frac{M^2 \left(-1190 M^3+2664 M^2 r-2035 M r^2+537 r^3\right)}{5040 r^6 (r-2 M)^4}, \quad
\Delta^{1/2}_{060}=\frac{M^2 \left(5432 M^2-7720 M r+2943 r^2\right)}{30240 r^6 (r-2 M)^4}, \nonumber \\
\Delta^{1/2}_{070}=\frac{M^2 \left(1036 M^3-2120 M^2 r+1524 M r^2-393 r^3\right)}{2520 r^7 (r-2 M)^5}, \quad
\Delta^{1/2}_{201}=\frac{M^2 (r-2 M)}{30 r^5}, \nonumber \\
\Delta^{1/2}_{202}=\frac{M^2 \left(460 M^2-428 M r+99 r^2\right)}{2520 r^6}, \quad
\Delta^{1/2}_{211}=\frac{M^2 (5 M-2 r)}{30 r^6}, \quad
\Delta^{1/2}_{212}=-\frac{M^2 \left(690 M^2-535 M r+99 r^2\right)}{1260 r^7}, \nonumber \\
\Delta^{1/2}_{220}=-\frac{M^2}{30 r^6}, \quad
\Delta^{1/2}_{221}=-\frac{M^2 \left(443 M^2-412 M r+90 r^2\right)}{1260 r^7 (2 M-r)}, \quad
\Delta^{1/2}_{230}=\frac{M^2}{10 r^7}, \nonumber \\
\Delta^{1/2}_{231}=\frac{M^2 \left(161 M^3-464 M^2 r+317 M r^2-60 r^3\right)}{1260 r^8 (r-2 M)^2}, \quad
\Delta^{1/2}_{240}=\frac{M^2 \left(-3526 M^2+3746 M r-981 r^2\right)}{5040 r^8 (r-2 M)^2}, \nonumber \\
\Delta^{1/2}_{250}=-\frac{M^2 \left(9392 M^3-15628 M^2 r+8631 M r^2-1572 r^3\right)}{5040 r^9 (r-2 M)^3}, \quad
\Delta^{1/2}_{400}=\frac{M^2 (r-2 M)^2}{60 r^8}, \nonumber \\
\Delta^{1/2}_{401}=\frac{M^2 \left(-1300 M^3+1682 M^2 r-714 M r^2+99 r^3\right)}{5040 r^9}, \quad
\Delta^{1/2}_{410}=-\frac{M^2 \left(16 M^2-14 M r+3 r^2\right)}{60 r^9}, \nonumber \\
\Delta^{1/2}_{411}=\frac{M^2 \left(5850 M^3-6728 M^2 r+2499 M r^2-297 r^3\right)}{5040 r^{10}}, \quad
\Delta^{1/2}_{420}=\frac{M^2 \left(5648 M^2-4800 M r+981 r^2\right)}{10080 r^{10}}, \nonumber \\
\Delta^{1/2}_{430}=-\frac{M^2 \left(2020 M^2-1872 M r+393 r^2\right)}{2520 r^{11}}, \quad
\Delta^{1/2}_{600}=\frac{M^3 (199 M-89 r) (r-2 M)^2}{7560 r^{12}}, \nonumber \\
\Delta^{1/2}_{610}=\frac{M^3 \left(-3184 M^3+4224 M^2 r-1850 M r^2+267 r^3\right)}{5040 r^{13}}.
\end{gather}

\subsection*{Expansions of an arbitrary point on the world line about $x^\ab$}

The four velocity of a general geodesic orbit taken to lie in the equatorial plane is given by  the standard expressions~\cite{Chandrasekhar}
\begin{equation} \label{eqn: ualpha}
\dot{t}(\tau) = \frac{E r(\tau)}{r(\tau) - 2M}, \quad \quad
\dot{r}(\tau) = \sqrt{E^2 - \left(1 - \frac{2M}{r(\tau)}\right) \left(1 + \frac{L^2}{r(\tau)^2}\right)}, \quad \quad \dot{\theta}(\tau) = 0, \quad \quad
\dot{\phi}(\tau) = \frac{L}{r(\tau)^2}.
\end{equation}
It is straightforward to calculate the higher order proper time derivatives of these expressions and evaluate both the four velocity and its higher derivatives at $x^{\bar{a}}$, giving, for example,   
\begin{gather}
\dot{t}_0 = \frac{E r_0}{r_0 - 2M},
\quad \quad \dot{r}_0 = \sqrt{E^2 - \left(1 - \frac{2M}{r_0}\right) \left(1 + \frac{L^2}{r_0^2}\right)}, 
\quad \quad \dot{\theta}_0 = 0, 
\quad \quad \dot{\phi}_0 = \frac{L}{r_0^2}, 
\quad \nonumber\\ 
\ddot{t}_0 = -\frac{2 E M \dot{r}_0}{\left(r_0-2 M\right)^2}, 
\quad \quad \ddot{r}_0 = \frac{L^2 r_0-M r_0^2-3 L^2 M}{r_0^4}, 
\quad \quad \ddot{\theta}_0 = 0, 
\quad \quad \ddot{\phi}_0 = -\frac{2 L \dot{r}_0}{r_0^3}, 
\quad \nonumber \\
\dddot{t}_0 = \frac{2 E M \left[2 \left(E^2-1\right) r_0^4-r_0^2 \left(3 L^2+2 M^2\right)+9 L^2 M r_0+5 M r_0^3-6 L^2 M^2\right]}{r_0^4 \left(r_0-2 M\right)^3},
\quad \nonumber \\
\dddot{r}_0 = \frac{\dot{r}_0 \left(-3 L^2 r_0+2 M r_0^2 + 12 L^2 M\right)}{r_0^5},
\quad \nonumber \\
\dddot{\theta}_0 = 0,
\quad \quad \dddot{\phi}_0 = \frac{2 L \left[3 \left(E^2-1\right) r_0^3-4 L^2 r_0+7 M r_0^2+9 L^2 M\right]}{r_0^7}.
\label{eqn: Four Velocity Bar}
\end{gather}
Combining Eq.~\eqref{eqn: Four Velocity Bar} with Eq.~\eqref{eqn: xtilde}, we can express $x^{a'}$ in terms of $x^{\bar{a}}$ and $\Delta \tau$:
\begin{gather} 
t' = \frac{E r_0}{r_0-2 M} \Delta\tau -\frac{E M \dot{r}_0}{\left(r_0-2 M\right)^2}  \Delta\tau^2 + \cdots, \quad
r' = r_0+ \dot{r}_0  \Delta\tau - \frac{ \left(-L^2 r_0+M r_0^2+3 L^2 M\right)}{2 r_0^4} \Delta\tau ^2 + \cdots, 
\quad \nonumber \\
\theta' = \frac{\pi}{2}, 
\quad \quad \phi' = \frac{L}{r_0^2} \Delta\tau - \frac{L \dot{r}_0}{r_0^3} \Delta\tau^2 + \cdots.
\label{Anna eqn: xprimealpha}
\end{gather}
It is also straightforward to obtain $\delta x^{a'}$, in terms of $\Delta x^{a}$,
$x^{\bar{a}}$ and $\Delta \tau$ by noting that $\delta x^{a'} = x^{a'} - \Delta x^a - x^{\bar{a}}$.
Finally, we can calculate $u^{a'}$ in terms of $x^{\bar{a}}$ and $\Delta \tau$ by inserting $r'$ from Eq.~\eqref{Anna eqn: xprimealpha} into our equations for the four velocity:
\begin{align*}
u^{t'} =& \frac{E r_0}{r_0-2 M} - \frac{2 E M r  \rbdot}{\left(r_0-2 M\right)^2} \Delta \tau  \quad \nonumber \\
&+ \frac{E M \left\{6 L^2 M^2 - 9 L^2 M r_0 + r_0^2 \left[3 L^2+2 M^2 - 2 \left(E^2-1\right) r_0^2-5 M r_0\right]\right\}}{r_0^4 \left(2 M-r_0\right)^3} \Delta \tau ^2 + \cdots,
\end{align*}
\begin{gather}
u^{r'} = \rbdot + \frac{r_0 \left(L^2-M r_0\right)-3 L^2 M}{r_0^4} \Delta \tau  + \frac{\rbdot \left(12 L^2 M -3 L^2 r_0+2 M r_0^2\right)}{2 r_0^5} \Delta \tau ^2 + \cdots, \quad \nonumber \\
u^{\theta'} = 0, \quad \quad u^{\phi'} = \frac{L}{r_0^2} - \frac{2 L \rbdot}{r_0^3} \Delta \tau + \frac{L \left[3 \left(E^2-1\right) r_0^3-4 L^2 r_0+7 M r_0^2+9 L^2 M\right]}{r_0^7} \Delta \tau ^2 + \cdots. \quad
\label{eqn: ualphaprime}
\end{gather}

\subsection*{Expansions of retarded and advanced points}
Taking $\Delta \tau$ to have leading order $\epsilon$, the same leading order of our $\Delta x$ terms, we can further expand it in orders of $\epsilon$, giving
\begin{equation} \label{eqn: deltatau}
\Delta \tau = \tau_1 \epsilon + \tau_2 \epsilon^2 + \tau_3 \epsilon^3 + \tau_4 \epsilon^4 + \cdots.
\end{equation}
Substituting $\delta x^{a'}$ obtained from Eq.~\eqref{Anna eqn: xprimealpha} and $\Delta \tau$ from Eq.~\eqref{eqn: deltatau} into $\sigma (x, x')$, Eq.~\eqref{eqn: sigma}, gives $\sigma (x, x')$ as a function of $\Delta x^{a}$, $x^{\ab}$ and the $\tau_n$'s:
\begin{align}
\sigma(x,x') =& \frac{1}{2} \left[\frac{\left(2 M - \bar{r} \right) \Delta t^2}{\bar{r}}+\frac{\bar{r} \left(\Delta r-2 \rbdot \tau _1 \right) \Delta r}{\bar{r}-2 m}+\bar{r}^2 \left(\Delta \theta^2+\Delta \phi ^2\right) -\left(2 L \Delta \phi +\tau _1- 2 E \Delta t \right) \tau _1 \right] \nonumber \\
& + \frac{1}{2} \Bigg\{ \frac{2}{\bar{r}} \left(\frac{E M \Delta t}{\bar{r}-2 M}- L \Delta \phi \right) \tau _1 \Delta r +\frac{M \left(\rbdot \tau _1 -\Delta r\right) \Delta r^2}{\left(\bar{r}-2 M\right)^2}+\bar{r} \left[\left(\rbdot \tau _1 +\Delta r\right) \left(\Delta \theta ^2+\Delta \phi ^2\right)-\frac{2 \rbdot  \tau _2 \Delta r}{\bar{r}-2 M}\right] \nonumber \\
& -\frac{M \left(\rbdot \tau _1 +\Delta r \right) \Delta t^2 }{\bar{r}^2}-2 \left(L \Delta \phi -E \Delta t +\tau _1\right) \tau _2 \Bigg\} + \cdots.
\end{align}
If we now specify that $x^{a'}$ coincides with the point where the world line intersects with the light cone of $x$, we can use the equation $\sigma (x, x') = 0$ to solve for the $\tau_n$'s in terms of $\Delta x^{a}$ and $x^{\ab}$.  This gives us
\begin{align}
\tau_1 =& E \Delta t - L \Delta \phi + \frac{ \bar{r} \rbdot \Delta r}{2 M-\bar{r}} \pm \zrho, \nonumber \\
\tau_2 =& \frac{\pm 1}{8 \zrho } \Bigg( \frac{\left\{L^2 \left[\Delta r^2+4 M^2 \left(\Delta \theta ^2+3 \Delta \phi ^2\right)\right]-4  L M^2  (E \Delta t\pm 2 \zrho ) \Delta \phi - 2 E M^2 (3 E \Delta t \pm 2 \zrho ) \Delta t \right\} \Delta r}{M^2 \bar{r}} \nonumber \\
& +\frac{4 L M^2 \rbdot \Delta t^2 \Delta \phi -4 M^2 \left[\rbdot (E \Delta t\pm \zrho )+2 \Delta r \right] \Delta t^2 + L^2 \Delta r^3}{M \bar{r}^2} \nonumber \\
& +\frac{\left[4  L M^2 \left(E \Delta t-2 \rbdot \Delta r \right) \Delta \phi +8 E^2 M^3 \bar{r} \left(\Delta \theta ^2+\Delta \phi ^2\right) + L^2 \Delta r^2 - 2 E M^2 (3 E \Delta t\pm 2 \zrho ) \Delta t \right] \Delta r }{M^2 \left(2 M-\bar{r}\right)} \nonumber \\
& -\frac{ \left\{ \left(\bar{r}-2 M\right) \left[4 L M^2 \rbdot \Delta \phi + 4 M^2 \left(E \rbdot \Delta t \mp \rbdot \zrho + E^2 \Delta r\right)- L^2 \Delta r \right] + 8 E^2 M^3 \Delta r \right\} \Delta r^2}{M \left(\bar{r}-2 M\right)^3} - \frac{4 L^2 M \Delta r \Delta t^2}{\bar{r}^4} \nonumber \\
& -4 \bar{r} \left(\Delta \theta ^2+\Delta \phi ^2\right) \left[\left(E^2-2\right) \Delta r-\rbdot \left( E \Delta t- L \Delta \phi \pm \zrho \right)\right]\Bigg), 
\label{eqn: taus}
\end{align}
with the higher order terms following in the same manner.

Using our equations for $\Delta \tau$, Eqs.~\eqref{eqn: deltatau} and \eqref{eqn: taus}, we rewrite $x^{a'}$ (and consequently $\delta x^{a'}$), the four velocity, $u^{a'}$, $\Delta^{\frac{1}{2}} (x, x')$ and $\sigma_{a'}$  (Eqs.~\eqref{Anna eqn: xprimealpha}, \eqref{eqn: ualphaprime}, ~\eqref{eqn: Van Vleck 2} and \eqref{eqn: sigmaalphaprime} respectively), in terms of $\Delta x^{a}$ and $x^{\ab}$.

\subsection*{Bivector of Parallel Transport}

To calculate the bivector of parallel transport, $g^a{}_{b'} (x,x')$, we first write it
in terms of a coordinate expansion about $x$,
\begin{equation} \label{eqn: bivector expansion}
g^a{}_{b'}(x,x') = \delta^a{}_{b'} + G^a{}_{bc}(x) \delta x^{c'} + G^a{}_{bcd} (x) \delta x^{c'} \delta x^{d'} + G^a{}_{bcde}(x) \delta x^{c'} \delta x^{d'} \delta x^{e'} + \dots,
\end{equation}
where the coefficients $G^a{}_{b...}(x)$ are functions of $x^a$ written in terms of $\Delta x^a$ and $x^{\bar{a}}$.  Calculating $g^a{}_{b',c'} (x,x')$ is straight forward:
\begin{equation} \label{eqn: gabc}
g^a{}_{b',c'} (x,x') = G^a{}_{bc}(x) + 2 G^a{}_{bcd}(x) \delta x^{d'} + 3 G^a{}_{bcde} (x) \delta x^{d'} \delta x^{e'} + 4 G^a{}_{bcde}(x) \delta x^{d'} \delta x^{e'} \delta x^{f'} + \cdots.
\end{equation}
Using the identity $g^a{}_{b';c'} \sigma^{c'} = g^a{}_{b',c'} \sigma^{c'} - \Gamma^{d'}_{b'c'} g^{a}_{d'} \sigma^{c'} = 0 $ with Eqs.~\eqref{eqn: sigmaaprime}, \eqref{eqn: gabc}, \eqref{eqn: bivector expansion}, and our expression for $\delta x^{a'}$ (obtained from the previous section), one can calculate the above coefficients and hence obtain the bivector of parallel transport, $g^a{}_{b'} (x,x')$, in terms of $\Delta x^a$ and $x^{\bar{a}}$.

\subsection*{Scalar Singular Field}

Combining Eqs.~\eqref{eq:PhiS} and \eqref{eq: U}, the scalar singular field can be written as

\begin{equation} \label{eq: PhiS Simp}
\Phi^{\rm \sing}(x) =  \frac{q}{2} \Bigg[ \frac{\Delta^{\tfrac{1}{2}}(x,x')}{\sigma_{c'}(x,x') u^{c'}(x')} \Bigg]_{x'=x_{\rm \ret}}^{x'=x_{\rm \adv}} + \frac{q}{2} \int_{\tau_{\rm \ret}}^{\tau_{\rm \adv}} V(x,x(\tau')) d\tau.
\end{equation}
We already have everything required for the first term here, which gives the direct part of the
scalar singular field.  It should be noted that $x'=x_{\rm \ret}$ and $x'=x_{\rm \adv}$
are the equivalent of setting $\pm \rho = -\rho$ and $\pm \rho = +\rho$ respectively when
substituting the $\tau_n$, Eq.~\eqref{eqn: taus}, into $\Delta^{1/2}$, $\sigma_{a'}$ and $u^{a'}$.

In the scalar case, Eq.~\eqref{eq:V} for the scalar tail part becomes
\begin{equation} \label{eqn: VScalar}
V (x, x') = \sum^{\infty}_{\num=0} V_\num (x, x') \sigma^\num (x, x').
\end{equation}
To calculate coordinate expansions of the $V_\num$, first we require a coordinate expansion for $V_0$ about $x$ of the form,
\begin{equation}
V_0 (x, x') = v_{0}(x) + v_{0a} (x) \delta x^{a'} + v_{0ab} (x) \delta x^{a'} \delta x^{b'} + v_{0abc} (x) \delta x^{a'} \delta x^{b'} \delta x^{c'} + \cdots.
\end{equation}
The `initial condition' described by Eq.~\eqref{eq:recursionV0} in the scalar case then becomes
\begin{equation}
2 \sigma^{;a'} V_{0;a'} - 2 V_0 \Delta^{-\tfrac{1}{2}} \sigma^{;a'} \Delta^{\tfrac{1}{2}}{}_{;a'} + 2 V_0 + \left(\square' - m^2 -\xi R \right) \Delta^{\tfrac{1}{2}} = 0,
\end{equation}
and from this, it is quite simple to read off expressions for the coefficients $v_{0a...}$.  Once we have $V_0$ to the desired order, we compute a coordinate expansion for $V_\num$ ($\num > 0$) of the form 
\begin{equation}
V_\num (x, x') = v_{\num }(x) + v_{\num a} (x) \delta x^{a'} + v_{\num ab} (x) \delta x^{a'} \delta x^{b'} + v_{\num abc} (x) \delta x^{a'} \delta x^{b'} \delta x^{c'} + \cdots.
\end{equation}
The recursion relation for $V_\num$, Eq.~\eqref{eq:recursionVn}, in the scalar case is then
\begin{equation}
2 \num \sigma^{;a'} V_{\num ;a'} - 2 \num V_\num \Delta^{-\tfrac{1}{2}} \sigma^{;a'} \Delta^{\tfrac{1}{2}}{}_{;a'} + 2 \num \left( \num + 1\right) V_\num + \left(\square' - m^2 -\xi R \right) V_{\num-1} = 0,
\end{equation}
from which we can obtain expressions for the coefficients, $v_{\num a...}$. Here, the
number of terms which must be computed is determined by the accuracy to which
we require the singular field. For the present calculation, we require up to $v_{0\,abcde}$,
$v_{1\,abc}$ and $v_{2\,a}$.

Once we have $V_\num$ to the required $\num$, using Eqs.~\eqref{eqn: sigma} and \eqref{eqn: VScalar}, along with our expression for $\delta x^{a'}$ obtained from Eq.~\eqref{Anna eqn: xprimealpha}, we get $V(x,x')$ in terms of $\Delta \tau$, $\Delta x^a$ and $x^{\bar{a}}$.  This can be easily integrated over $\tau$ as required by Eq.~\eqref{eq: PhiS Simp}.  Our final expression for $\Phi^{\sing} (x)$ is then obtained by using Eqs.~\eqref{eqn: deltatau} and \eqref{eqn: taus} to remove the $\Delta \tau$ dependence.  As before $\tau_{\rm \ret}$ and $\tau_{\rm \adv}$ are obtained by allowing $\pm \rho = -\rho$ and $\pm \rho = +\rho$, respectively.

\subsection*{Electromagnetic Singular Field}

For the electromagnetic singular field, we use Eqs.~\eqref{eq:AS} and \eqref{eq: U} to give
\begin{equation} \label{eqn: A Simp}
A_{a}^{\rm \sing} = \frac{e}{2} \Bigg[\frac{\Delta^{\tfrac{1}{2}}(x,x') g_{aa'}(x,x')u^{a'}(x')}{\sigma_{c'}(x,x') u^{c'}(x')} \Bigg]_{x'=x_{\rm \ret}}^{x'=x_{\rm \adv}} + \frac{e}{2} \int_{\tau_{\rm \ret}}^{\tau_{\rm \adv}} V_{aa'}(x,z(\tau)) u^{a'}(x') d\tau,
\end{equation}
where $V_{aa'} (x, z(\tau'))$ is given by Eq.~\eqref{eq:V},
\begin{equation}
V^{aa'} (x, x') = \sum^{\infty}_{\num=0} V^{aa'}_\num (x, x') \sigma^\num (x, x'),
\end{equation}
and the relevant metrics at $x$ and $x'$ can be used to lower indices.  We require a coordinate expansion of $V^{aa'}_{0}$ of the form,
\begin{equation} \label{eqn: v0 EM}
V^{aa'}_0 (x, x') = v^{aa'}_{0}(x) + v^{aa'}_{0}{}_{b} (x) \delta x^{b'} + v^{aa'}_{0}{}_{bc} (x) \delta x^{b'} \delta x^{c'} + v^{aa'}_{0}{}_{bcd} (x) \delta x^{b'} \delta x^{c'} \delta x^{d'} + \cdots.
\end{equation}
Substituting this into the ``initial condition'' in Eq.~\eqref{eq:recursionV0}, which in the electromagnetic case is
\begin{equation}
2 \sigma^{;b'} V^{aa'}_{0}{}_{;b'} - 2 V^{aa'}_0 \Delta^{-\tfrac{1}{2}} \sigma^{;b'} \Delta^{\tfrac{1}{2}}{}_{;b'} + 2 V^{aa'}_0 + \left(\delta^{a'}{}_{b'}\square' - R^{a'}{}_{b'} \right) \left(\Delta^{\tfrac{1}{2}} g^{ab'}\right) = 0,
\end{equation}
the coefficients of Eq.~\eqref{eqn: v0 EM}, $v^{aa'}_{0}{}_{b \cdots}$, can easily be recursively obtained. It should be noted that the covariant derivatives do require the appropriate Christoffel symbols, which can be obtained from the suitable metric at $x'$.
Next, we construct coordinate expansions for the $V^{aa'}_{\num}$.  These have the form
\begin{equation}\label{eqn: vn EM}
V^{aa'}_\num (x, x') = v^{aa'}_{\num}(x) + v^{aa'}_{\num}{}_{b} (x) \delta x^{b'} + v^{aa'}_{\num}{}_{bc} (x) \delta x^{b'} \delta x^{c'} + v^{aa'}_{\num}{}_{bcd} (x) \delta x^{b'} \delta x^{c'} \delta x^{d'} + \cdots.
\end{equation}
Substituting Eq.~\eqref{eqn: vn EM} into the recursion relation \eqref{eq:recursionVn}, which for the electromagnetic case becomes
\begin{equation}
2 \num \sigma^{;b'} V^{aa'}_{\num}{}_{;b'} - 2 \num V^{aa'}_\num \Delta^{-\tfrac{1}{2}} \sigma^{;b'} \Delta^{\tfrac{1}{2}}{}_{;b'} + 2 \num \left( \num + 1\right) V^{aa'}_\num + \left(\delta^{a'}{}_{b'}\square' - R^{a'}{}_{b'} \right)V^{ab'}_{\num-1} = 0,
\end{equation}
we can recursively solve for the coefficients of Eq.~\eqref{eqn: vn EM}, $v^{aa'}_{\num}{}_{b \cdots}$.  Once we have $V^{aa'}_{\num}$ to the required $\num$, we carry out the same remaining steps as in the scalar case and use Eq.~\eqref{eqn: A Simp} to calculate the electromagnetic singular field.

\subsection*{Gravitational Singular Field}

In the gravitational case, Eqs.~\eqref{eq:hS} and \eqref{eq: U} give
\begin{equation} \label{eqn: h Simp}
\hb_{a b}^{\rm \sing} = 2 \mu \Bigg[\frac{\Delta^{\tfrac{1}{2}}(x,x') g_{a' (a} g_{b) b'}(x,x') u^{a'}(x') u^{b'}(x') }{\sigma_{c'}(x,x') u^{c'}(x')} \Bigg]_{x'=x_{\rm \ret}}^{x'=x_{\rm \adv}}
  + 2 \mu \int_{\tau_{\rm \ret}}^{\tau_{\rm \adv}} V_{aba'b'}(x,z(\tau')) u^{a'}(x') u^{b'}(x') d\tau,
\end{equation}
where $V_{aba'b'}(x,z(\tau'))$ is given by Eq.~\eqref{eq:V}.  For the gravitational case, this is
\begin{equation}
V^{aba'b'} (x, x') = \sum^{\infty}_{\num=0} V^{aba'b'}_\num (x, x') \sigma^\num (x, x'),
\end{equation}
where the appropriate metric at $x$ or $x'$ can be use to lower indices.  The coordinate expansion for $V^{aba'b'}_0 (x, x') $ is of the form
\begin{equation} \label{eqn: v0 G}
V^{aba'b'}_0 (x, x') = v^{aba'b'}_{0}(x) + v^{aba'b'}_{0}{}_{c} (x) \delta x^{c'} + v^{aba'b'}_{0}{}_{cd} (x) \delta x^{c'} \delta x^{d'} + v^{aba'b'}_{0}{}_{cde} (x) \delta x^{c'} \delta x^{d'} \delta x^{e'} + \cdots.
\end{equation}
We replace $V^{aba'b'}_0 (x, x')$ with Eq~\eqref{eqn: v0 G} in the ``initial condition'' described by Eq.~\eqref{eq:recursionV0}, which for the gravitational case is
\begin{equation}
2 \sigma^{;c'} V^{aba'b'}_{0}{}_{;c'} - 2 V^{aba'b'}_0 \Delta^{-\tfrac{1}{2}} \sigma^{;c'} \Delta^{\tfrac{1}{2}}{}_{;c'} + 2 V^{aba'b'}_0 + \left(\delta^{a'}{}_{c'}\delta^{b'}{}_{d'}\square' + 2 C^{a'}{}_{c'}{}^{b'}{}_{d'} \right) \left(\Delta^{\tfrac{1}{2}} g^{c'(a}g^{b) d'}\right) = 0.
\end{equation}
This equation may be used to recursively solve for the coefficients of Eq.~\eqref{eqn: v0 G}, $v^{aba'b'}_{0}{}_{c \cdots}$.
Next, the coordinate expansion of $V^{aba'b'}_\num (x, x')$ for $\num > 0$ has the form,
\begin{equation}\label{eqn: vn G}
V^{aba'b'}_\num (x, x') = v^{aba'b'}_{\num}{}_{0}(x) + v^{aba'b'}_{\num}{}_{c} (x) \delta x^{c'} + v^{aba'b'}_{\num}{}_{cd} (x) \delta x^{c'} \delta x^{d'} + v^{aba'b'}_{\num}{}_{cde} (x) \delta x^{c'} \delta x^{d'} \delta x^{e'} + \cdots.
\end{equation}
Substituting this into the recursion relation of Eq.~\eqref{eq:recursionVn}, which for the gravitational case has the form
\begin{equation}
2 \num \sigma^{;c'} V^{aba'b'}_{\num}{}_{;c'} - 2 \num V^{aba'b'}_\num \Delta^{-\tfrac{1}{2}} \sigma^{;c'} \Delta^{\tfrac{1}{2}}{}_{;c'} + 2 \num \left( \num + 1\right) V^{aba'b'}_\num + \left(\delta^{a'}{}_{c'}\delta^{b'}{}_{d'}\square' + 2 C^{a'}{}_{c'}{}^{b'}{}_{d'} \right)  V^{abc'd'}_{\num-1} = 0,
\end{equation}
we can recursively solve for the coefficiens of Eq.~\eqref{eqn: vn G}, $v^{aba'b'}_{\num}{}_{c\cdots}$.
As in the previous two cases, once we have $V^{aba'b'}_\num (x, x') $ for the required $\num$,
 it is straightforward to calculate the singular field using Eq.~\eqref{eqn: h Simp}.

\section{Covariant bitensor expansions} \label{sec:ExpansionCoefficients}
In this Appendix, we give covariant expansions for the beitensors appearing in the formal expression for the
singular field, Eq.~\eqref{eq:SingularField}. These are given in terms of the biscalars
$\sbar \equiv (g^{\alphab \betab} + u^{\alphab} u^{\betab}) \sigma_{\alphab} \sigma_{\betab}$,
(the projection of $\sigma_{\bar{a}}(x,\xb)$ orthogonal to the worldline), and
$\bar{r} = \sigma_{\alphab} u^{\alphab}$ (the projection of $\sigma_{\bar{a}}(x,\xb)$ along the worldline).
In writing the coefficients, we use the notation $[T_{a_1 \cdots a_n}]_{(k)}$ to denote the term of
order $\epsilon^k$ in the expansion of the tensor $T_{a_1 \cdots a_n}$, so that
\begin{equation}
T_{a_1 \cdots a_n} = \sum_{k=0}^\infty [T_{a_1 \cdots a_n}]_{(k)} \epsilon^k.
\end{equation}

\subsection*{Advanced and retarded points}
Eq.~\eqref{eq:SingularField} for the singular field includes bitensors at points $x'$ on the
world-line between the advanced and retarded points of $x$. We consolidate this dependance
to a single arbitrary point, $\xb$, on the world by expanding the dependence on $x'$ about $\xb$.
Denoting the proper distance along the world-line between $x_{\rm \adv}/x_{\rm \ret}$ and $\xb$ by
$\Delta \tau$, we may write the expansion of this distance in powers of $\epsilon$ as
\begin{gather*}
\Dtau{1} = \rbar \pm \sbar, \quad \Dtau{2} = 0, \quad
\Dtau{3} = \mp\frac{(\rbar \pm \sbar)^2 R_{u \sigma u \sigma}}{6\sbar}, \quad
\Dtau{4} = \mp\frac{(\rbar \pm \sbar)^2 \big((\rbar \pm \sbar) R_{u \sigma u \sigma ;u}- R_{u \sigma u \sigma ;\sigma}\big)}{24\sbar},
\end{gather*}
\begin{IEEEeqnarray}{rCl}
\Dtau{5} &=& \mp\frac{(\rbar \pm \sbar)^2}{360 \sbar^3} \Big\{
  5 R_{u \sigma u \sigma} R_{u \sigma u \sigma} (\rbar \mp 3 \sbar) (\rbar \pm \sbar)
  + \sbar^2 \Big[
      (\rbar \pm \sbar)^2 \Big(3 R_{u \sigma u \sigma; u u}
      + 4 R_{u \sigma u \ab} R_{u \sigma u}{}^{\ab}\Big)
\nonumber \\ && \qquad
    - (\rbar \pm \sbar) \Big(3 R_{u \sigma u \sigma; u \sigma}
      -\: 16 R_{u \sigma u \ab} R_{u \sigma \sigma}{}^{\ab}\Big)
    + 3 R_{u \sigma u \sigma; \sigma \sigma}
    + 4 R_{u \sigma \sigma \ab} R_{u \sigma \sigma}{}^{\ab}
  \Big]
\Big\},
\nonumber \\
\Dtau{6} &=& \pm\frac{(\rbar \pm \sbar)^2}{4320 \sbar^3} \Big\{
  30 R_{u \sigma u \sigma} \Big[R_{u \sigma u \sigma ; \sigma} (\rbar \mp 3 \sbar) (\rbar \pm \sbar)
        - R_{u \sigma u \sigma; u} (\rbar \mp 4 \sbar) (\rbar \pm \sbar)^2 \Big]
\nonumber \\ && \quad
  +\: \sbar^2 \Big[6 R_{u \sigma u \sigma; \sigma \sigma \sigma}
    + 36 R_{u \sigma \sigma \ab; \sigma} R_{u \sigma \sigma}{}^{\ab}
\nonumber \\ && \qquad
    - 2 (\rbar \pm \sbar) \Big(3 R_{u \sigma u \sigma; u \sigma \sigma}
      - 36 R_{u \sigma \sigma \ab; \sigma} R_{u \sigma u}{}^{\ab}
      - R_{u \sigma \sigma}{}^{\ab} (16 R_{u \sigma u \ab; \sigma}
        + 5 R_{u \sigma u \sigma; \ab}
        - 10 R_{u \sigma \sigma \ab; u})\Big)
\nonumber \\ && \qquad
      +\: 2 (\rbar \pm \sbar)^2 \Big(
          3 R_{u \sigma u \sigma; u u \sigma}
        - 30 R_{u \sigma u \ab; u} R_{u \sigma \sigma}{}^{\ab}
        + R_{u \sigma u}{}^{\ab} (13 R_{u \sigma u \ab; \sigma} + 5 R_{u \sigma u \sigma; \ab}
          - 25 R_{u \sigma \sigma \ab; u})
      \Big)
\nonumber \\ && \qquad
        -\: 6  (\rbar \pm \sbar)^3 (R_{u \sigma u \sigma; u u u}
        + 6 R_{u \sigma u \ab; u} R_{u \sigma u}{}^{\ab})
      \Big]
\Big\}.
\end{IEEEeqnarray}

\subsection*{Advanced and retarded distances}
Taking two derivatives of the world function, we obtain a bitensor that has the covariant expansion
\begin{IEEEeqnarray}{rCl}
\sigma_{ab} &=& g_{ab}
  - \tfrac13 R_{acbd} \sigma^c \sigma^d
  + \tfrac{1}{12} R_{acbd;e} \sigma^c \sigma^d \sigma^e
  + \Big(
      \tfrac{1}{45} R_{acpd} R^p{}_{ebf}
    + \tfrac{1}{60} R_{acbd;ef}
    \Big) \sigma^c \sigma^d \sigma^e \sigma^f
\nonumber \\ &&
  + \Big(
      \tfrac{1}{120} R_{acpd;e} R^p{}_{fbg}
    + \tfrac{1}{120} R_{acpd} R^p{}_{ebf;g}
    + \tfrac{1}{360} R_{acbd;efg}
    \Big) \sigma^c \sigma^d \sigma^e \sigma^f \sigma^g
\nonumber \\ &&
  - \Big(
      \tfrac{2}{945} R_{acpd} R^p{}_{eqf} R^q{}_{gbh}
    + \tfrac{1}{504} R_{acpd;ef} R^p{}_{gbh}
    + \tfrac{17}{5040} R_{acpd;e} R^p{}_{fbg;h}
\nonumber \\ && \quad
    + \tfrac{1}{504} R_{acpd} R^p{}_{ebf;gh}
    + \tfrac{1}{2520} R_{acbd;efgh}
    \Big) \sigma^c \sigma^d \sigma^e \sigma^f \sigma^g \sigma^h
\nonumber \\ &&
  + \Big(
      \tfrac{17}{20160} R_{acpd;e} R^p{}_{fqg} R^q{}_{hbi}
    + \tfrac{29}{30240} R_{acpd} R^p{}_{eqf;g} R^q{}_{hbi}
    + \tfrac{11}{30240} R_{acpd;efg} R^p{}_{hbi}
\nonumber \\ && \quad
    + \tfrac{17}{20160} R_{acpd;ef} R^p{}_{gbh;i}
    + \tfrac{17}{20160} R_{acpd} R^p{}_{eqf} R^q{}_{gbh;i}
    + \tfrac{17}{20160} R_{acpd;e} R^p{}_{fbg;hi}
\nonumber \\ && \quad
    + \tfrac{11}{30240} R_{acpd} R^p{}_{ebf;ghi}
    + \tfrac{1}{20160} R_{acbd;efghi}
    \Big) \sigma^c \sigma^d \sigma^e \sigma^f \sigma^g \sigma^h \sigma^i.
\end{IEEEeqnarray}
For the singular field, we require the expansion of $[\sigma_{a'}u^{a'}](z_\pm,x)$. Writing
$[\sigma_{a'}u^{a'}](\tau) = [\sigma_{a'}u^{a'}](z(\tau), x)$, expanding the dependence on $\tau$
about $\xb$ (using the method of Sec.~\ref{sec:Covariant} and making use of the above expansion
of $\sigma_{ab}$) and evaluating at $\tau = \tau_\pm$,
we obtain the coefficients of the expansion of
$[\sigma_{a'}u^{a'}]_\pm \equiv [\sigma_{a'}u^{a'}](z_\pm, x)$ about $\xb$.
They are:
\begin{gather*}
r_{(1)} = \sbar,\quad
r_{(3)} =
 - \frac{\rbar^2 - \sbar^2}{6\sbar} R_{u \sigma u \sigma},\quad
r_{(4)} =
 \frac{\rbar\pm\sbar}{24 \sbar} \Big[R_{u \sigma u \sigma \sigma} (\rbar \mp \sbar)
   - R_{u \sigma u \sigma u} (\rbar \pm \sbar) (\rbar \mp 2 \sbar)\bigg],
\end{gather*}
\begin{IEEEeqnarray}{rCl}
r_{(5)} &=& -\frac{1}{360 \sbar^3}
\Big\{
\sbar^2 \Big[(\rbar^2 - \sbar^2) \big(
    3 R_{u \sigma u \sigma \sigma \sigma}
    + 4 R_{u \sigma \sigma}{}^{\ab} R_{u \sigma \sigma \ab}
  \big)
  - (\rbar \pm \sbar)^2 (\rbar \mp 2 \sbar) \big(
      3  R_{u \sigma u \sigma \sigma u}
    - 16 R_{u \sigma \sigma}{}^\ab R_{u \sigma u \ab}
  \big)
\nonumber \\ && \qquad
  + (\rbar \pm \sbar)^3 (\rbar \mp 3 \sbar) \big(
      3 R_{u \sigma u \sigma u u}
    + 4 R_{u \sigma u}{}^\ab R_{u \sigma u \ab}
    \big)
  \Big]
+  5 \Big[(\rbar^2 - \sbar^2) R_{u \sigma u \sigma}\Big]^2
\Big\},
\nonumber \\
r_{(6)} &=& \frac{1}{4320 \sbar^3}
\Big\{
\sbar^2 \Big[
  6 (\rbar^2 - \sbar^2) \big(
      R_{u \sigma u \sigma \sigma \sigma \sigma}
    + 6 R_{u \sigma \sigma}{}^{\ab}{}_{\sigma} R_{u \sigma \sigma \ab}
  \big)
  -\: 6 (\rbar \pm \sbar)^4  (\rbar \mp 4 \sbar) \Big(
            R_{u \sigma u \sigma u u u}
        + 6 R_{u \sigma u \ab u} R_{u \sigma u}{}^{\ab}
      \Big)
\nonumber \\ && \qquad
  - 2 (\rbar \pm \sbar)^2 (\rbar \mp 2 \sbar)
    \Big(
        3 R_{u \sigma u \sigma u \sigma \sigma}
      - 36 R_{u \sigma \sigma \ab \sigma} R_{u \sigma u}{}^{\ab}
      - R_{u \sigma \sigma}{}^{\ab} (16 R_{u \sigma u \ab \sigma}
        + 5 R_{u \sigma u \sigma \ab}
        - 10 R_{u \sigma \sigma \ab u})
    \Big)
\nonumber \\ && \qquad
  +\: 2 (\rbar \pm \sbar)^3  (\rbar \mp 3 \sbar) \Big(
          3 R_{u \sigma u \sigma u u \sigma}
        - 30 R_{u \sigma u \ab u} R_{u \sigma \sigma}{}^{\ab}
        + R_{u \sigma u}{}^{\ab} (13 R_{u \sigma u \ab \sigma} + 5 R_{u \sigma u \sigma \ab}
          - 25 R_{u \sigma \sigma \ab u})
      \Big)
  \Big]
\nonumber \\ && \quad
  +  30 R_{u \sigma u \sigma} \Big[
   (\rbar^2 - \sbar^2)^2 R_{u \sigma u \sigma \sigma}
  -(\rbar\pm\sbar)^3(\rbar^2 \mp 3\rbar \sbar+4\sbar^2) R_{u \sigma u \sigma u}
  \Big]
\Big\}.
\end{IEEEeqnarray}

\subsection*{Van Vleck Determinant}
The Van Vleck determinant has the covariant expansion
\begin{IEEEeqnarray}{rCl}
\Delta^{1/2}(x,x') &=&
  1
  + \frac{1}{12} R_{ab} \sigma^a \sigma^b
  - \frac{1}{24} R_{ab;c} \sigma^a \sigma^b \sigma^c
  + \Big(
      \frac{1}{360} R_{paqb} R^p{}_c{}^q{}_d
    + \frac{1}{288} R_{ab} R_{cd}
    + \frac{1}{80}  R_{ab;cd}
    \Big)  \sigma^a \sigma^b \sigma^c \sigma^d
\nonumber \\ &&
  - \Big(
      \frac{1}{360} R_{paqb} R^p{}_c{}^q{}_{d;e}
    + \frac{1}{288} R_{ab} R_{cd;e}
    + \frac{1}{360}  R_{ab;cde}
    \Big)  \sigma^a \sigma^b \sigma^c \sigma^d \sigma^e
\nonumber \\ &&
  + \Big(
      \frac{1}{1260} R_{paqb} R^p{}_c{}^q{}_{d;ef}
    + \frac{1}{1344} R_{paqb;c} R^p{}_d{}^q{}_{e;f}
    + \frac{1}{5670} R_{paqb} R_{rc}{}^p{}_d R^q{}_e{}^r{}_f
    + \frac{1}{4320} R_{paqb} R^p{}_c{}^q{}_d R_{ef}
\nonumber \\ &&
    \qquad + \frac{1}{10368} R_{ab} R_{cd} R_{ef}
    + \frac{1}{1152} R_{ab;c} R_{de;f}
    + \frac{1}{960} R_{ab} R_{cd;ef}
    + \frac{1}{2016}  R_{ab;cdef}
    \Big)  \sigma^a \sigma^b \sigma^c \sigma^d \sigma^e \sigma^f
\nonumber \\ &&
  - \Big(
      \frac{1}{6048} R_{paqb} R^p{}_c{}^q{}_{d;efg}
    + \frac{1}{2240} R_{paqb;c} R^p{}_d{}^q{}_{e;fg}
    + \frac{1}{3780} R_{paqb} R_{rc}{}^p{}_d R^q{}_e{}^r{}_{f;g}
\nonumber \\ &&
    \qquad + \frac{1}{4320} R_{paqb} R^p{}_c{}^q{}_{d;e} R_{fg}
    + \frac{1}{8640} R_{paqb} R^p{}_c{}^q{}_d R_{ef;g}
    + \frac{1}{6912} R_{ab} R_{cd} R_{ef;g}
\nonumber \\ &&
    \qquad + \frac{1}{1920} R_{ab;c} R_{de;fg}
    + \frac{1}{4320} R_{ab} R_{cd;efg}
    + \frac{1}{13440}  R_{ab;cdefg}
    \Big)  \sigma^a \sigma^b \sigma^c \sigma^d \sigma^e \sigma^f \sigma^g.
\end{IEEEeqnarray}
Writing $\Delta^{1/2}(\tau) = \Delta^{1/2}(z(\tau), x)$, expanding the dependence on $\tau$
about $\xb$ (using the method of Sec.~\ref{sec:Covariant}) and evaluating at $\tau = \tau_\pm$,
we obtain the coefficients in the expansion of
$\Delta^{1/2}_\pm \equiv \Delta^{1/2}(z_\pm, x)$ about $\xb$. Specialized to the vacuum case,
they are:
\begin{equation*}
\Delta^{1/2}_{(0)} = 1,\quad
\Delta^{1/2}_{(1)} = 0,\quad
\Delta^{1/2}_{(2)} = 0,\quad
\Delta^{1/2}_{(3)} = 0,
\end{equation*}
\begin{IEEEeqnarray}{rCl}
\Delta^{1/2}_{(4)} &=& \frac{1}{360}
\Big[
    C_{\sigma \ab \sigma \bb}
  + 2 (\rbar \pm \sbar) C_{u (\ab |\sigma| \bb)}
  + (\rbar \pm \sbar)^2 C_{u \ab u \bb}
\Big]\Big[
    C_\sigma{}^\ab{}_\sigma{}^\bb
  + 2 (\rbar \pm \sbar) C_u{}^\ab{}_\sigma{}^\bb
  + (\rbar \pm \sbar)^2 C_u{}^\ab{}_u{}^\bb
\Big],
\nonumber \\
\Delta^{1/2}_{(5)} &=& \frac{1}{360}
\Big[
    C_{\sigma \ab \sigma \bb}
  + 2 (\rbar \pm \sbar) C_{u (\ab |\sigma| \bb)}
  + (\rbar \pm \sbar)^2 C_{u \ab u \bb}
\Big]
\Big[
  (\rbar \pm \sbar)
  \Big(
      C_\sigma{}^{\ab}{}_\sigma{}^{\bb}{}_u
    - 2 C_u{}^{\ab}{}_\sigma{}^{\bb}{}_\sigma
  \Big)
\nonumber \\ &&
\qquad
- (\rbar \pm \sbar)^2
  \Big(
      C_u{}^\ab{}_u{}^\bb{}_\sigma
    - 2 C_u{}^\ab{}_\sigma{}^\bb{}_u
  \Big)
+ (\rbar \pm \sbar)^3 C_u{}^\ab{}_u{}^\bb{}_u
- C_\sigma{}^\ab{}_\sigma{}^\bb{}_\sigma
\Big].
\end{IEEEeqnarray}

\subsection*{Bivector of parallel transport}
The derivative of the bivector of parallel transport has the covariant expansion
\begin{IEEEeqnarray}{rCl}
g_a{}^{a'} g_{a' b ; c}(x,x') &=&
  - \frac12 R_{bacd} \sigma^d
  + \frac16 R_{bacd;e} \sigma^d \sigma^e
  - \frac{1}{24} \Big(
      R_{bapd} R^p{}_{ecf}
    + R_{bacd;ef}
    \Big) \sigma^d \sigma^e \sigma^f
\nonumber \\ &&
  + \Big(
      \frac{1}{60} R_{bapd} R^p{}_{ecf;g}
    + \frac{7}{360} R_{bapd;e} R^p{}_{fcg}
    + \frac{1}{120} R_{bacd;efg}
    \Big) \sigma^d \sigma^e \sigma^f \sigma^g
\nonumber \\ &&
  - \Big(
      \frac{1}{240} R_{bapd} R^p{}_{ecf;gh}
      \frac{1}{120} R_{bapd;e} R^p{}_{fcg;h}
    + \frac{1}{180} R_{bapd;ef} R^p{}_{gch}
\nonumber \\ &&
    \qquad + \frac{1}{240} R_{bapd} R^p{}_{eqf} R^q{}_{gch}
    + \frac{1}{720} R_{bacd;efgh}
    \Big) \sigma^d \sigma^e \sigma^f \sigma^g \sigma^h.
\end{IEEEeqnarray}
For the singular field, we require the expansion of $g_{aa'}u^{a'}(z_\pm,x)$. Writing
$[g_{aa'}u^{a'}](\tau) = [g_{aa'}u^{a'}](z(\tau), x)$, expanding the dependence on $\tau$
about $\xb$ (using the method of Sec.~\ref{sec:Covariant} and making use of the above expansion
of the bivector of parallel transport) and evaluating at $\tau = \tau_\pm$,
we obtain the coefficients of the expansion of
$[g_{aa'}u^{a'}]_\pm \equiv g_{aa'}u^{a'}(z_\pm, x)$ about $\xb$. They are:
\begin{gather*}
\Big[g_{aa'} u^{a'}\Big]_{(0)} = g_{a\ab}u^\ab, \quad
\Big[g_{aa'} u^{a'}\Big]_{(1)} = 0, \quad
\Big[g_{aa'} u^{a'}\Big]_{(2)} = -\frac12 (\rbar \pm \sbar) g_{a}{}^\ab R_{u \sigma u \ab}, \nonumber \\
\Big[g_{aa'} u^{a'}\Big]_{(3)} = \frac16 (\rbar \pm \sbar) g_{a}{}^\ab (R_{u \sigma u \ab;\sigma}-(\rbar \pm \sbar) R_{u \sigma u \ab;u}), \nonumber \\
\end{gather*}
\begin{IEEEeqnarray}{rCl}
\Big[g_{aa'} u^{a'}\Big]_{(4)} &=& \pm \frac{1}{24 \sbar} g_{a}{}^\ab (\rbar \pm \sbar)
  \Big\{
    2 (\rbar \pm \sbar) R_{u \sigma u \ab} R_{u \sigma u \sigma}
    - \sbar \Big[
        R_{u \sigma u \ab ; \sigma \sigma}
      + R_{u \ab \sigma \bb} R_{u \sigma \sigma}{}^\bb
    + (\rbar \pm \sbar)^2 R_{u \sigma u a ; u u}
\nonumber \\ &&
\quad    + (\rbar \pm \sbar) \Big(
        R_{u \sigma u}{}^\bb ( 2 R_{u \ab \sigma \bb} + 3 R_{u \sigma \ab \bb})
      - R_{u \sigma u \ab ; u \sigma}
      + R_{u \ab u}{}^{\bb}(
        R_{u \sigma \sigma \bb}
      + (\rbar \pm \sbar) R_{u \sigma u \bb}
      )
    \Big)
    \Big]
  \Big\}, \nonumber \\
\Big[g_{aa'} u^{a'}\Big]_{(5)} &=& \pm \frac{1}{2160 \sbar} g_{a}{}^\ab (\rbar \pm \sbar)
  \Big\{
    \sbar \Big[
      18 \Big(
          R_{u \sigma u a ; \sigma \sigma  \sigma }
        -  R_{u \sigma u a ; u \sigma \sigma } (\rbar \pm \sbar)
        + R_{u \sigma u a ; u u \sigma } (\rbar \pm \sbar)^2
        - R_{u \sigma u a ; u u u} (\rbar \pm \sbar)^3 \Big)
\nonumber \\ &&
\quad    + 6 R_{u \sigma  \sigma b} \Big(
        7 R_{u a \sigma }{}^{b}{}_{ ;\sigma }
      + (\rbar \pm \sbar) ( 3 R_{u a u}{}^{b}{}_{ ;\sigma }
        - 4 R_{u a \sigma }{}^{b}{}_{;u}
        + 4 R_{u \sigma u a}{}^{;b})
      - 8 R_{u a u}{}^{b}{}_{;u}(\rbar \pm \sbar)^2\Big)
\nonumber \\ &&
\quad    + 9 R_{u a \sigma b} \Big(
        4 R_{u \sigma  \sigma }{}^{b}{}_{ ;\sigma }
      +  (\rbar \pm \sbar) (5 R_{u \sigma u}{}^{b}{}_{ ;\sigma }
        - 3 R_{u \sigma  \sigma }{}^{b}{}_{;u})
      - 6 R_{u \sigma u}{}^{b}{}_{;u} (\rbar \pm \sbar)^2\Big)
\nonumber \\ &&
\quad    + 2 R_{u \sigma u b} \Big(
        21  (\rbar \pm \sbar) (2 R_{u a \sigma }{}^{b}{}_{ ;\sigma }
        + 3 R_{u \sigma a}{}^{b}{}_{ ;\sigma } )
\nonumber \\ &&
\qquad      +  (\rbar \pm \sbar)^2 ( 17 R_{u a u}{}^{b}{}_{; \sigma }
        - 32 R_{u a \sigma }{}^{b}{}_{;u}
        - 36 R_{u \sigma a}{}^{b}{}_{;u}
        + 16 R_{u \sigma u a}{}^{;b})
      - 21 R_{u a u}{}^{b}{}_{;u} (\rbar \pm \sbar)^3 \Big)
\nonumber \\ &&
\quad    + 9 R_{u a u b} \Big(
        4 R_{u \sigma  \sigma }{}^{b}{}_{ ;\sigma } (\rbar \pm \sbar)
      + 3  (\rbar \pm \sbar)^2 (R_{u \sigma u}{}^{b}{}_{ ;\sigma }
        - R_{u \sigma  \sigma }{}^{b}{}_{;u})
      - 4 R_{u \sigma u}{}^{b}{}_{;u} (\rbar \pm \sbar)^3\Big)
\nonumber \\ &&
\quad    +  54 (\rbar \pm \sbar) R_{u \sigma a b} \Big(
         R_{u \sigma u}{}^{b}{}_{; \sigma }
     - 2 R_{u \sigma u}{}^{b}{}_{;u} (\rbar \pm \sbar)\Big)
    \Big]
\nonumber \\ &&
    + 15 (\rbar \pm \sbar) \Big[
       4 R_{u \sigma u \sigma } \Big(
          2  (\rbar \pm \sbar) R_{u \sigma u a; u}
        -   R_{u \sigma u a ;\sigma }\Big)
     + 3 R_{u \sigma u a} \Big(
            R_{u \sigma u \sigma ; u} (\rbar \pm \sbar)
        -   R_{u \sigma u \sigma ; \sigma }\Big)
     \Big]
  \Big\}.
\end{IEEEeqnarray}

\subsection*{Scalar tail}
The scalar tail bitensor, $V(x,x')$, may be expanded in a covariant series by writing it in the
form of a Hadamard series,
\begin{equation}
V(x,x') = V_0 (x,x') + V_1 (x,x') \sigma(x,x') + \cdots,
\end{equation}
and expanding each of the Hadamard coefficients $V_0(x,x'), V_1(x,x'), \cdots$ in a covariant Taylor
series,
\begin{align*}
V_{0} &= v_{0} - \tfrac{1}{2} v_{0\,;c} \sigma^c
+\tfrac{1}{2}   v_{0\,cd}  \sigma^c\sigma^d +\tfrac{1}{6} \left(  - \tfrac{3}{2} v_{0\,(cd;e)} + \tfrac{1}{4} v_{0\,;(cde)} \right) \sigma^c\sigma^d\sigma^e + \cdots,\\
V_{1} &= v_{1} - \tfrac{1}{2} v_{1\,;c}  \sigma^c + \cdots.
\end{align*}
The series coefficients required to obtain the expansion of the singular field to
$\mathcal{O}(\epsilon^4)$ [$V(x,x')$ to $\mathcal{O}(\epsilon^3)$] are given by
\begin{align*}
v_{0} &= \tfrac{1}{2} \left((\xi- \tfrac{1}{6}) R + m^2 \right),\\
v_{0\,}{}^{cd} &= -\tfrac{1}{180}R_{pqr}{}^{c}R^{pqrd}-\tfrac{1}{180}R^{c}{}_{p}{}^{d}{}_{q}R^{pq}+\tfrac{1}{90}R^{c}{}_{p}R^{dp}
-\tfrac{1}{120}\Box R^{cd}+\tfrac{1}{12}(\xi-\tfrac{1}{6})R R^{cd}+(\tfrac{1}{6}\xi-\tfrac{1}{40})R^{;cd}+\tfrac{1}{12}m^2 R^{cd},\\
v_1 &= \tfrac{1}{720}R_{pqrs}R^{pqrs} -\tfrac{1}{720}R_{pq}R^{pq} +\tfrac{1}{8}(\xi - \tfrac{1}{6})^2 R^2-\tfrac{1}{24}(\xi - \tfrac{1}{5})\Box R+\tfrac{1}{4}m^2(\xi - \tfrac{1}{6}) R + \tfrac{1}{8}m^4.
\end{align*}
For the singular field, we require the expansion of 
$\int_{\tau_{\rm \ret}}^{\tau_{\rm \adv}} V d\tau'$. Writing $V(\tau) = V(z(\tau), x)$ and expanding the dependence on $\tau$
about $\xb$ (using the method of Sec.~\ref{sec:Covariant} and making use of the above expansion
of $V$), we obtain an expansion in powers of $\Delta \tau$ that can be trivially integrated
between $\tau = \tau_-$ and $\tau = \tau_+$. Specialized to the vacuum case,
the required expansion coefficients are then:
\begin{gather*}
\Big[ \int_{\tau_{\rm \ret}}^{\tau_{\rm \adv}} V d\tau' \Big]_{(1)} = 0, \quad
\Big[ \int_{\tau_{\rm \ret}}^{\tau_{\rm \adv}} V d\tau' \Big]_{(2)} = 0, \quad
\Big[ \int_{\tau_{\rm \ret}}^{\tau_{\rm \adv}} V d\tau' \Big]_{(3)} = 0, \quad
\Big[ \int_{\tau_{\rm \ret}}^{\tau_{\rm \adv}} V d\tau' \Big]_{(4)} = 0.
\end{gather*}

\subsection*{Electromagnetic tail}
The electromagnetic tail bitensor, $V_{ab'}(x,x')$, may be expanded in a covariant series by
writing it in the form of a Hadamard series,
\begin{equation}
V_{ab'}(x,x') = g_{b'}{}^{b} [ V_{0 ab} (x,x') + V_{1 ab} (x,x') \sigma(x,x') + \cdots],
\end{equation}
and expanding each of the Hadamard coefficients, $V_{0 ab}(x,x'), V_{1 ab}(x,x'), \cdots$, in a
covariant Taylor series,
\begin{align*}
V_{0\,ab} &= v_{0\,(ab)} + \left( - \tfrac{1}{2} v_{0\,(ab);c} + v_{0\,[ab]c}\right) \sigma^c
+\tfrac{1}{2} \left(   v_{0\,(ab)cd} - v_{0\,[ab](c;d)}\right) \sigma^c\sigma^d\\
&\qquad +\tfrac{1}{6} \left(  - \tfrac{3}{2} v_{0\,(ab)(cd;e)} + \tfrac{1}{4} v_{0\,(ab);(cde)} + v_{0\,[ab]cde}\right) \sigma^c\sigma^d\sigma^e + \cdots,\\
V_{1\,ab} &= v_{1\,(ab)} + \left( - \tfrac{1}{2} v_{1\,(ab);c} + v_{1\,[ab]c}\right) \sigma^c + \cdots.
\end{align*}
The series coefficients required to obtain the expansion of the singular field to
$\mathcal{O}(\epsilon^4)$ [$V_{ab'}(x,x')$ to $\mathcal{O}(\epsilon^3)$] are given by
\begin{align*}
v_{0\,(ab)} &= \tfrac{1}{2} R_{ab}-\tfrac{1}{12}R g_{ab},\\
v_{0\,[ab]}{}^{c} &= \tfrac{1}{6} R^c{}_{[b;a]},\\
v_{0\,(ab)}{}^{cd} &= \tfrac{1}{6} R_{ab}{}^{;(cd)}+\tfrac{1}{12}R_{ab}R^{cd}+\tfrac{1}{12}R_{(a}{}^{pqc}R_{b)pq}{}^{d}\\
&\qquad + g_{ab}\left(-\tfrac{1}{180}R_{pqr}{}^{c}R^{pqrd}-\tfrac{1}{180}R^{c}{}_{p}{}^{d}{}_{q}R^{pq}+\tfrac{1}{90}R^{c}{}_{p}R^{dp}
-\tfrac{1}{72}RR^{cd}-\tfrac{1}{40}R^{;cd}-\tfrac{1}{120}\Box R^{cd}\right),\\
v_{0\,[ab]}{}^{cde} &= -\tfrac{3}{20}R^{(c}{}_{[a;b]}{}^{de)} -\tfrac{1}{12}R^{(c}{}_{[a;b]}R^{de)} -\tfrac{1}{20}R_{[a}{}^{pq(c}R_{b]pq}{}^{d;e)}
-\tfrac{1}{30}R_{ab}{}^{p(c}{}_{;q}R_{p}{}^{de)q}+\tfrac{1}{60}R_{abp}{}^{(c;d}R^{e)p}\\
&\qquad +\tfrac{1}{20}R_{abp}{}^{(c}R^{de);p}
-\tfrac{1}{20}R_{ab}{}^{p(c}R_{p}{}^{d;e)}\\
\end{align*}
and
\begin{align*}
v_{1\,(ab)} &= -\tfrac{1}{48}R_{apqr}R_{b}{}^{pqr} +\tfrac{1}{8}R_{ap}R_{b}{}^{p} -\tfrac{1}{24}RR_{ab}-\tfrac{1}{24}\Box R_{ab} \\
&\qquad + g_{ab} \left(\tfrac{1}{720}R_{pqrs}R^{pqrs} -\tfrac{1}{720}R_{pq}R^{pq} +\tfrac{1}{288}R^2+\tfrac{1}{120}\Box R\right),\\
v_{1\,[ab]}{}^{c} &=  \tfrac{1}{240}R_{[a}{}^{pqr}R_{b]pqr}{}^{;c} + \tfrac{1}{24}R_{[a}{}^{pqc}R_{b]p;q} + \tfrac{1}{120}R_{[a}{}^{pqc}R_{b]q;p} \\
&\qquad - \tfrac{1}{120}R^{c}{}_{pq[a}R^{pq}{}_{;b]}+ \tfrac{1}{24}R^{pc}{}_{;[a}R_{b]p}+ \tfrac{1}{24}R^{p}{}_{[a}R_{b]}{}^{c}{}_{;p}
- \tfrac{1}{360}R^{pc}R_{p[a;b]}- \tfrac{1}{24}R^{p}{}_{[a}R_{b]p}{}^{;c}+ \tfrac{1}{72}R R^c{}_{[a;b]}\\
&\qquad + \tfrac{1}{120}R^c{}_{[a;b]p}{}^p - \tfrac{1}{540}R_{abpq;r}R^{pqrc}- \tfrac{1}{540}R_{abpq;r}R^{rqpc}- \tfrac{1}{360}R_{abp}{}^{c}{}_{;q}R^{pq}
+ \tfrac{1}{120}R_{abpq}R^{cp;q}- \tfrac{1}{120}R_{ab}{}^{pc}R_{;p}.
\end{align*}
These are the same as those given by Brown and Ottewill \cite{Brown:Ottewill:1986} with the exception
of $v_{1\,[ab]}{}^{c}$, where we have corrected a sign error in one of their terms and
combined another two terms into a single term.

For the singular field, we require the expansion of 
$\int_{\tau_{\rm \ret}}^{\tau_{\rm \adv}} V_{ab'} u^{b'} d\tau'$. Writing
$[V_{ab'} u^{b'}](\tau) = [V_{ab'} u^{b'}](z(\tau), x)$ and expanding the dependence on $\tau$
about $\xb$ (using the method of Sec.~\ref{sec:Covariant} and making use of the above expansion
of $V_{ab'}$), we obtain an expansion in powers of $\Delta \tau$ that can be trivially integrated
between $\tau = \tau_-$ and $\tau = \tau_+$. Specialized to the vacuum case,
the required expansion coefficients are then:
\begin{gather*}
\Big[ \int_{\tau_{\rm \ret}}^{\tau_{\rm \adv}} V_{ab'} u^{b'} d\tau' \Big]_{(1)} = 0, \quad
\Big[ \int_{\tau_{\rm \ret}}^{\tau_{\rm \adv}} V_{ab'} u^{b'} d\tau' \Big]_{(2)} = 0,
\end{gather*}
\begin{IEEEeqnarray}{rCl}
\Big[ \int_{\tau_{\rm \ret}}^{\tau_{\rm \adv}} V_{ab'} u^{b'} d\tau' \Big]_{(3)} &=&
  \frac{1}{1152} (\bar{r} \pm \bar{s}) g_{a}{}^{\ab}
  \Big[
      16 R_{ubac} R_{u}{}^{b}{}_{u}{}^{c} (\rbar \pm \sbar)^2
    + 48 R_{u}{}^{b}{}_{\sigma}{}^{c} R_{\sigma c a b}
\nonumber \\ &&
    + 24  (\rbar \pm \sbar) \Big(
        R_{u}{}^{b}{}_{\sigma}{}^{c} R_{ucab}
      + R_{u}{}^{b}{}_{u}{}^{c} R_{\sigma b a c}
      + (\rbar \mp 2 \sbar) R_{bcde} R^{bcde}u_{a}
    \Big)
  \Big],
\nonumber \\
\Big[ \int_{\tau_{\rm \ret}}^{\tau_{\rm \adv}} V_{ab'} u^{b'} d\tau' \Big]_{(4)} &=&
  \frac{1}{51840} (\bar{r} \pm \bar{s}) g_{a}{}^{\ab}
  \Big\{
      144 \Big[
          2  R_{ua\sigma}{}^{bc} R_{\sigma b\sigma c}
        - 6  R_{\sigma}{}^{b}{}_{a}{}^{c}{}_{\sigma} R_{uc\sigma b}
        - 9  R_{u}{}^{b}{}_{\sigma}{}^{c}{}_{\sigma} R_{\sigma cab}
        +  R_{\sigma}{}^{bcd}{}_{\sigma} R_{\sigma bcd} u_{a}
        \Big]
\nonumber \\ &&
    -\:4 (\rbar \pm \sbar) \Big[
      18 \Big(
          6  R_{\sigma}{}^{b}{}_{a}{}^{c}{}_{\sigma} R_{ubuc}
        - 2  R_{ua\sigma}{}^{bc} R_{ub\sigma c}
        + 9  R_{u}{}^{b}{}_{\sigma}{}^{c}{}_{\sigma} R_{ucab}
        +  R_{uc\sigma b} (
            6  R_{u}{}^{b}{}_{a}{}^{c}{}_{\sigma}
\nonumber \\ &&
\qquad          - 2  R_{ua\sigma}{}^{bc}
          - 9  R_{\sigma}{}^{b}{}_{a}{}^{c}{}_{u})
        + 9  R_{u}{}^{b}{}_{u}{}^{c}{}_{\sigma} R_{\sigma bac}
        - 2  R_{uau}{}^{bc} R_{\sigma b\sigma c}
        - 6  R_{u}{}^{b}{}_{\sigma}{}^{c}{}_{u} R_{\sigma cab}
        \Big)
\nonumber \\ &&
\quad      +  (\rbar \mp 2 \sbar) \Big(27  R_{u}{}^{bcd}{}_{\sigma} R_{abcd}
        + 18  R_{a}{}^{bcd}{}_{\sigma} R_{ubcd}
        + 4  R_{ua}{}^{bcd} (R_{\sigma bcd}
        -  R_{\sigma dbc})\Big)
\nonumber \\ &&
\quad      - 3 \Big(6  R_{\sigma}{}^{bcd}{}_{\sigma} R_{ubcd}
        + 6 ( R_{u}{}^{bcd}{}_{\sigma}
        -   R_{\sigma}{}^{bcd}{}_{u}) R_{\sigma bcd}
        +  R^{bcde}{}_{\sigma} R_{bcde} (\rbar \mp 2 \sbar)
        \Big) u_{a}
      \Big]
\nonumber \\ &&
    + (\rbar \pm \sbar)^2 \Big[
        48 \Big(
            R_{ubuc} (
                2  R_{ua\sigma}{}^{bc}
              - 6  R_{u}{}^{b}{}_{a}{}^{c}{}_{\sigma}
              + 9  R_{\sigma}{}^{b}{}_{a}{}^{c}{}_{u}
              )
        + 6 R_{u}{}^{b}{}_{\sigma}{}^{c}{}_{u} R_{ucab}
        + 9  R_{u}{}^{b}{}_{a}{}^{c}{}_{u} R_{uc\sigma b}
\nonumber \\ &&
\qquad        + 2  R_{uau}{}^{bc} (
            R_{ub\sigma c}
          + R_{uc\sigma b}
          )
        + 6  R_{u}{}^{b}{}_{u}{}^{c}{}_{u} R_{\sigma bac}
        \Big)
\nonumber \\ &&
\quad      +  (\rbar \mp 3 \sbar) \Big(
            R_{ubcd} (
             27 R_{a}{}^{bcd}{}_{u}
           - 4  R_{ua}{}^{bcd}
           )
        + 4 R_{ua}{}^{bcd} R_{udbc}
        \Big)
\nonumber \\ &&
\quad      + 3 \Big(
          16 (
              R_{u}{}^{bcd}{}_{\sigma}
            - R_{\sigma}{}^{bcd}{}_{u}) R_{ubcd}
            - R^{bcde}{}_{u} R_{bcde} (\rbar \mp 3 \sbar)
        \Big) u_{a}
\nonumber \\ &&
\quad      + 6 R_{u}{}^{bcd}{}_{u} \Big(
          3 R_{abcd} (\rbar \mp 3 \sbar)
        - 8 R_{\sigma bcd} u_{a}
        \Big)
      - 432  R_{u}{}^{b}{}_{u}{}^{c}{}_{\sigma} R_{ubac}
      \Big]
\nonumber \\ &&
    + 36 (\rbar \pm \sbar)^3 \Big[
        6  R_{u}{}^{b}{}_{u}{}^{c}{}_{u} R_{ubac}
      + R_{ubuc} (2  R_{uau}{}^{bc} + 9  R_{u}{}^{b}{}_{a}{}^{c}{}_{u})
      - R_{u}{}^{bcd}{}_{u} R_{ubcd} u_{a}\Big]
  \Big\}.
\end{IEEEeqnarray}

\subsection*{Gravitational tail}
The gravitational tail bitensor, $V_{aa'bb'}(x,x')$, may be expanded in a covariant series by
writing it in the form of a Hadamard series,
\begin{equation}
V_{aa'bb'}(x,x') = g_{a'}{}^{c} g_{b'}{}^{d} [ V_{0 acbd} (x,x') + V_{1 acbd} (x,x') \sigma(x,x') + \cdots],
\end{equation}
and expanding each of the Hadamard coefficients, $V_{0 acbd}(x,x'), V_{1 acbd}(x,x'), \cdots$, in a
covariant Taylor series,
\begin{align*}
V_{0\,AB} &= v_{0\,(AB)} + \left( - \tfrac{1}{2} v_{0\,(AB);e} + v_{0\,[AB]e}\right) \sigma^e
+\tfrac{1}{2} \left(   v_{0\,(AB)ef} - v_{0\,[AB](e;f)}\right) \sigma^e\sigma^f\\
&\qquad +\tfrac{1}{6} \left(  - \tfrac{3}{2} v_{0\,(AB)(ef;g)} + \tfrac{1}{4} v_{0\,(AB);(efg)} + v_{0\,[AB]efg}\right) \sigma^e\sigma^f\sigma^g + \cdots,\\
V_{1\,AB} &= v_{1\,(AB)} + \left( - \tfrac{1}{2} v_{1\,(AB);c} + v_{1\,[AB]e}\right) \sigma^e+ \cdots.
\end{align*}
The required coefficients for $V_{0\,AB}$ are
\begin{eqnarray}
   v_{0\,(AB)}&=&v_{0\,(\overline{ab}\,\overline{cd})}=-C_{acbd},
\label{eq:v01}\\
   v_{0\,[AB]}{}^{e}&=&0,
\label{eq:v02}\\
   v_{0\,(AB)}{}^{ef}&=&v_{0\,(\overline{ab}\,\overline{cd})}{}^{ef}=
      -\frac{1}{3}C_{acbd}{}^{;(ef)}
      -\frac{1}{6}C_{ac}{}^{p(e}C_{bdp}{}^{f)}
      +\frac{1}{6}g_{ac}C_b{}^{pq(e}
         C_{dpq}{}^{f)}
      -\frac{1}{720} {\Pi}_{abcd} g^{ef} C^{pq{r}s}
         C_{pq{r}s},
\label{eq:v03}
\end{eqnarray}
and
\begin{eqnarray}
   v_{0\,[AB]}{}^{ef{g}}&=&v_{0\,[\overline{ab}\,\overline{cd}]}{}^{ef{g}}=
      \frac{1}{10}g_{ac}C_{[b}{^{pq(e;f}}
         C_{d]pq}{}^{g)}
      +\frac{1}{15}g_{ac}C_{bdpe;q}
         C^p{_f}{^q}{_{g}},
\label{eq:v04}
\end{eqnarray}
where
\begin{equation}
{\Pi}_{abcd} = {1 \over 2} g_{ac}
g_{bd} + {1 \over 2} g_{ad} g_{bc} +
{\kappa}g_{ab} g_{cd}.
\end{equation}
In Eqs.\ (\ref{eq:v01}) -- (\ref{eq:v04}), the right hand sides are
understood to be symmetrized on the index pairs $(ab)$ and $(cd)$.
The required coefficients for $V_{1\,AB}$ are
\begin{IEEEeqnarray}{rCl}
   v_{1\, (AB)} = v_{1\,(\overline{ab}\,\overline{cd})}&=&
      \frac{1}{12}{\Box}C_{acbd}
      +\frac{1}{2}C_a{^p}{_b}{^q}C_{cpdq}
      +\frac{1}{24}C_{ac}{^{pq}}C_{bdpq}
      -\frac{1}{96}g_{ac}g_{bd}C^{pqrs}C_{pqrs}
      +\frac{1}{720}{\Pi}_{abcd} C^{pqrs}
         C_{pqrs}
\IEEEeqnarraynumspace
\label{eq:v11}
\end{IEEEeqnarray}
and
\begin{align}
v_{1\,[AB]}{}^{e}&=v_{1\,[\overline{ab}\,\overline{cd}]}{}^{e} \nonumber \\
&=
\frac{1}{12} \left( C_{a}{}^{p}{}_b{}^q C_{cpdq}{}^{;e}-C_{c}{}^{p}{}_d{}^q C_{apbq}{}^{;e} \right)
+\frac{1}{12} \left( C_{a}{}^{p}{}^e{}^q C_{bcdp;q} - C_{c}{}^{p}{}^e{}^q C_{dabp;q} \right)
+\frac{1}{90} g_{ac} C^e{}^{pqr}C_{bdpq;r},
\label{eq:v12}
\end{align}
where again there is implicit symmetrization on the index pairs
($ab$) and ($cd$).
When ${\kappa}=-1/2$, Eq. (\ref{eq:v11}) agrees
with Eq.\ (A23) of Allen, Folacci and Ottewill \cite{Allen:1987bn} 
specialized to the vacuum case. Our expressions also agree with
Anderson, Flanagan and Ottewill \cite{Anderson:Flanagan:Ottewill:2004}, but we write
them here in a slightly more compact form. Note that the
expressions (\protect{\ref{eq:v01}}) -- (\protect{\ref{eq:v04}})
and (\ref{eq:v11}) -- (\ref{eq:v12})
are all traceless on the index pair $(cd)$, aside from the
terms involving the  
tensor ${\Pi}_{abcd}$.  This means that performing
a trace reversal on the index pair $(cd)$ is equivalent to
changing the value of ${\kappa}$ from $0$ to $-1/2$. For the calculation of the gravitational singular field, we require an
expansion of the trace reversed singular field and so we choose $\kappa = 0$.

\bibliography{references}{}
\bibliographystyle{apsrev}

\end{document}